\documentclass[aps,prd,11pt,a4paper,nofootinbib,oneside,superscriptaddress]{revtex4-2}

\usepackage{amsmath,amssymb,amsfonts,color}
\usepackage{tensor,slashed,paralist,cases,mathrsfs}
\usepackage{float,cancel,xcolor}
\usepackage{graphicx}
\usepackage{dcolumn}
\usepackage{bm}
\usepackage{tabularx}
\usepackage{multirow}
\usepackage{ulem}

\usepackage[colorlinks=true,urlcolor=blue,linkcolor=blue,citecolor=blue,linktocpage=true]{hyperref}

\usepackage{tikz}
\usetikzlibrary{matrix,decorations.pathreplacing,calc}

\pgfkeys{tikz/mymatrixenv/.style={decoration=brace,every left delimiter/.style={xshift=3pt},every right delimiter/.style={xshift=-3pt}}}
\pgfkeys{tikz/mymatrix/.style={matrix of math nodes,left delimiter=(,right delimiter={)},inner sep=2pt,column sep=1em,row sep=0.5em,nodes={inner sep=0pt}}}
\pgfkeys{tikz/mymatrixbrace/.style={decorate,thick}}

\newcommand\mymatrixbraceoffsetv{0.2em}

\newcommand*\mymatrixbracebottom[4][m]{
    \draw[mymatrixbrace] ($(#1.south west)!(#1-1-#3.south east)!(#1.south east)-(0,\mymatrixbraceoffsetv)$)
        -- node[below=2pt] {#4} 
        ($(#1.south west)!(#1-1-#2.south west)!(#1.south east)-(0,\mymatrixbraceoffsetv)$);
}

\tikzset{
    cheating dash/.code args={on #1 off #2}{
        \csname tikz@addoption\endcsname{%
            \pgfgetpath\currentpath%
            \pgfprocessround{\currentpath}{\currentpath}%
            \csname pgf@decorate@parsesoftpath\endcsname{\currentpath}{\currentpath}%
           \pgfmathparse{\csname pgf@decorate@totalpathlength\endcsname-#1}\let\rest=\pgfmathresult%
            \pgfmathparse{#1+#2}\let\onoff=\pgfmathresult%
            \pgfmathparse{max(floor(\rest/\onoff), 1)}\let\nfullonoff=\pgfmathresult%
            \pgfmathparse{max((\rest-\onoff*\nfullonoff)/\nfullonoff+#2, #2)}\let\offexpand=\pgfmathresult%
            \pgfsetdash{{#1}{\offexpand}}{0pt}}%
    }
}

\begin{document}


\vspace*{1.5em}

\title{p-Wave Dominant Majorana Dark Matter and Its Self-Interactions in a Gauged $U(1)_{L_\mu - L_\tau}$ Model}

\author{Kwei-Chou Yang}
\email{kcyang@cycu.edu.tw}

\affiliation{Department of Physics and Center for High Energy Physics, Chung Yuan Christian University,
200 Chung Pei Road, Taoyuan 32023, Taiwan}

\begin{abstract}
We propose a model of $p$-wave dominant Majorana dark matter (DM) featuring a macroscopic mass gap in a gauged $U(1)_{L_\mu - L_\tau}$ extension of the Standard Model. Spontaneous symmetry breaking generates a dominant Majorana mass via a strong Yukawa coupling, splitting the dark fermion into two distinct states. The lighter state ($\chi_-$) is the DM candidate, with an allowed mass of $\sim 10$ GeV to several hundred GeV. Crucially, within the $10 \sim 75$ GeV mass range and a scalar mediator at the tens of MeV scale, this strong coupling dictates the thermal relic abundance and induces significant elastic self-interactions. These self-interactions naturally resolve small-scale structure anomalies, such as the core-cusp problem, while satisfying massive cluster constraints. Furthermore, we place stringent joint constraints on the scalar mass and the Higgs mixing angle $\alpha$ using the latest LZ 2025 direct detection data. The model's gauge boson $Z^\prime$ maintains early-Universe thermal equilibrium and accommodates the muon anomalous magnetic moment $(g-2)_\mu$, remaining consistent with updated cosmological and experimental bounds.
\end{abstract}
\maketitle
\newpage

\section{Introduction}

The dark sector may possess a rich structure, potentially consisting of multiple particle states separated by a mass gap. In scenarios where the gauge interactions of these states are strictly off-diagonal, scattering processes inevitably induce transitions between them. Historically, frameworks featuring extremely small mass differences alongside such purely off-diagonal interactions were proposed as inelastic dark matter to reconcile discrepancies among early direct-detection experiments such as DAMA and CDMS \cite{Tucker-Smith:2001myb, Tucker-Smith:2004mxa}. Since then, research on multi-state dark sectors has expanded across a broader range of areas, including direct and indirect detection methods, particle physics experiments, and astronomical observations, as well as explanations and predictions of the Universe's evolution \cite{Batell:2009vb, Chang:2010en, Okada:2019sbb, DallaValleGarcia:2024zva, Wang:2025cth, Arkani-Hamed:2008hhe, Pospelov:2008jd, Finkbeiner:2009mi, Gustafson:2024aom, Berlin:2025fwx, Hooper:2025fda, Berlin:2023qco, Krnjaic:2025zjl, Izaguirre:2015zva, Berlin:2018jbm, Voronchikhin:2025eqm, Krnjaic:2024ols, Garcia:2024uwf, Abdullahi:2023tyk, CarrilloGonzalez:2021lxm, Heeba:2023bik, Brahma:2023psr, Roy:2025zvo}

A well-known framework realizing such mass-split dark sectors is pseudo-Dirac fermion dark matter \cite{DallaValleGarcia:2024zva, Gustafson:2024aom, Berlin:2023qco, Garcia:2024uwf, CarrilloGonzalez:2021lxm, Heeba:2023bik, Brahma:2023psr}. In a minimal $U(1)$ gauge extension, the dark sector typically comprises a Dirac fermion with a dominant mass $m_D$ and a complex scalar. A weak Yukawa coupling $f$ connects the dark fermions to the scalar. Upon spontaneous symmetry breaking (SSB) of the $U(1)$ symmetry, the scalar acquires a vacuum expectation value $v_S$, generating a secondary Majorana mass term $\sim f v_S$. Consequently, the original Dirac fermion splits into two Majorana eigenstates with masses $m_D \pm f v_S$. However, in this pseudo-Dirac scenario, preserving a small mass splitting generally restricts the coupling $f$ to be extremely small, severely limiting the phenomenological richness of the dark sector, particularly regarding DM self-interactions.

In this work, we propose a novel and phenomenologically rich model of $p$-wave dominant Majorana DM featuring a macroscopic mass gap that naturally connects key astrophysical observations with various direct, indirect, and particle experimental constraints. In contrast to the pseudo-Dirac paradigm, our dark sector features fermions ($\chi, \chi^c$) with a subdominant Dirac mass $m_D$ and a complex scalar $S$. Here, we introduce a strong Yukawa coupling $f$ between the dark fermions and the scalar. After SSB, this large coupling generates a dominant Majorana mass $f v_S$, splitting the dark fermions into two distinct states separated by a macroscopic mass gap, with masses $f v_S \pm m_D$. Due to the macroscopic mass gap, the primordial abundance of the heavier state ($\chi_+$) is exponentially suppressed in the early Universe, leaving the lighter Majorana eigenstate ($\chi_-$) as the sole dark matter candidate today.

In our framework, the dark matter is significantly heavier than the light scalar mediator ($m_\chi \gg m_S$). By introducing a macroscopic mass gap, the primordial abundance of the excited state $\chi_+$ is exponentially suppressed, safely neutralizing the powerful $s$-wave coannihilation channel ($\chi_- \chi_+ \to Z^\prime S$). Consequently, driven by the strong Yukawa coupling $f$, strictly $p$-wave dominated annihilations of the ground state into the light $S$ and $Z^\prime$ final states dominate the freeze-out process and dictate the relic abundance. Moreover, the light scalar induces strong dark matter self-interactions that transfer heat inward within galactic halos, effectively alleviating core density problems \cite{Tulin:2017ara, Adhikari:2022sbh, Kaplinghat:2015aga}. 

Before symmetry breaking, our dark sector respects a $U(1)_{L_{\mu}-L_{\tau}}$ gauge symmetry. This represents the simplest Standard Model (SM) extension accommodating $p$-wave dominant Majorana dark matter while remaining anomaly-free. Since the SM muon and tau doublets carry opposite charges under this symmetry, the gauge boson $Z^{\prime}$ couples to them \cite{He:1990pn, He:1991qd, Altmannshofer:2016jzy, Kamada:2018zxi, Foldenauer:2018zrz, Escudero:2019gzq, Asai:2020qlp, Borah:2021jzu, Holst:2021lzm, Drees:2021rsg, Hapitas:2021ilr, Figueroa:2024tmn, Asai:2021wzx, Zu:2021odn}.
Historically, this interaction was proposed to address the $(g-2)_{\mu}$ anomaly. However, recent updates to the SM prediction \cite{Aliberti:2025beg} align well with experimental data \cite{Muong-2:2023cdq, Muong-2:2024hpx, Muong-2:2006rrc, Aoyama:2020ynm, Davies:2025pmx}, thereby imposing stringent upper bounds on the $Z^{\prime}$ mass and its coupling to muons. 
Consequently, we use these robust new limits to determine the allowed parameter space for our model, and we will review the current experimental constraints on this space in detail.

Crucially, despite these restricted, tiny gauge couplings for the dark sector and leptons ($\lesssim 10^{-3}$), $S$ and $Z^{\prime}$ successfully remain in thermal equilibrium with the SM bath via $S \leftrightarrow Z^{\prime} Z^{\prime}$ and $Z^{\prime} \leftrightarrow \nu_{\mu,\tau} \bar{\nu}_{\mu,\tau}$ until dark matter freeze-out. Consequently, thermalization does not require a large scalar mixing angle $\alpha$ with the SM Higgs. This naturally accommodates extremely small $\alpha$ values, which we evaluate against severe joint constraints on the $(m_S, \alpha)$ parameter space from recent LZ 2025 direct detection data \cite{LZ:2024zvo}. Furthermore, high-scale kinetic mixing between the $Z^{\prime}$ and the photon ensures that, even after standard weak interactions decouple, the electromagnetic plasma ($\gamma, e^\pm$) temporarily maintains thermal contact with the neutrino sector via $e^+ e^- \leftrightarrow Z^{\prime}$. This extended equilibrium subsequently alters the effective number of relativistic species ($N_{\rm eff}$), offering a potential pathway to alleviate the cosmological Hubble tension.

While both our framework and standard pseudo-Dirac models generate two Majorana states after symmetry breaking, their underlying mass mechanisms and gauge interactions are fundamentally different. In pseudo-Dirac models, the dark matter mass is dominated by an initial Dirac term. This leads to gauge interactions governed by an inelastic vector current ($\bar\chi_+ \gamma_\mu \chi_-$), which largely preserves macroscopic Dirac-like phenomenology. In contrast, the eigenmasses in our model are predominantly generated by the Majorana mass term following spontaneous symmetry breaking (SSB). Consequently, the $Z^\prime$ boson couples to the dark sector via an inelastic axial-vector current ($\bar\chi_+ \gamma_\mu \gamma_5 \chi_-$), a distinct signature of its true Majorana nature. To accurately reflect this dominant mass origin and the strictly off-diagonal axial-vector interactions, we refer to our framework as the $p$-wave dominant Majorana dark matter model.

Finally, our macroscopic mass gap exponentially suppresses the primordial abundance of the excited state $\chi_+$ during the early Universe. This thermodynamical depletion effectively ensures a pure ground-state relic ($r \to 0$) in the late Universe, which inherently acts as a strict kinematic closure against endothermic processes such as spin-dependent inelastic scattering with nucleons. Furthermore, we analyze the impact of gauge kinetic mixing generated at high energy scales on the effective number of relativistic species ($N_{\rm eff}$), offering a potential pathway to alleviate the Hubble tension.

The rest of this paper is organized as follows. We first establish the theoretical framework and the general experimental constraints of the model. In Sec.~\ref{sec:model}, we detail the $p$-wave dominant Majorana $U(1)_{L_\mu - L_\tau}$ model, with eigenstate derivations deferred to Appendix~\ref{app:pseudo-M}. Sec.~\ref{subsec:gauge} addresses the gauge mixing between the $Z^\prime$ and SM neutral bosons at both high-energy and one-loop levels, providing detailed interaction Lagrangian calculations in Appendix~\ref{app:kinetic-to-mass}. Sec.~\ref{sec:parameters} defines the independent parameters of our framework. Sec.~\ref{sec:g-2} explores the parameter space constrained by the $(g-2)_\mu$ measurement and other relevant experiments. Sec.~\ref{sec:neff} investigates the model's predictions for $N_{\rm eff}$ and its potential to alleviate the Hubble tension. We then analyze the model-independent particle physics bounds on the scalar mixing angle $\alpha$ in Sec.~\ref{sec:pp-expt}.

Next, we turn our main focus to the dark matter phenomenology. In Sec.~\ref{sec:boltz}, we examine the thermal evolution of the dark sector, demonstrating how the macroscopic mass gap safely renders the $\chi_- \chi_+ \to S Z^\prime$ coannihilation inactive, thereby allowing the purely $p$-wave annihilations into $SS$ and $Z^\prime Z^\prime$ to exclusively dictate the freeze-out process.
Utilizing the dynamically determined coupling required for the correct relic abundance, Sec.~\ref{sec:dd-expt} evaluates the stringent constraints from dark matter direct detection. Sec.~\ref{sec:mass-gap} analyzes how the macroscopic mass gap exponentially suppresses the excited state and ensures a pure ground-state relic. Sec.~\ref{sec:sommerfeld-relic} incorporates the Sommerfeld enhancement to evaluate its impact on the relic abundance and indirect detection signatures. Sec.~\ref{sec:self-scatterings} discusses dark matter self-scattering and its role in addressing small-scale structure challenges. In Sec.~\ref{sec:discussions}, we review additional experimental probes capable of testing this scenario, and we conclude with a summary in Sec.~\ref{sec:summary}. Further details and relevant formulas are included in the Appendices.

\section{Description of the $p$-wave dominant Majorana dark matter model}\label{sec:model}

We consider a minimal extension of the $p$-wave dominant Majorana DM model for which the dark sector is gauged under a local $U(1)_{L_\mu -L_\tau}$ symmetry with a gauge boson $Z^\prime$.  Meanwhile, the SM $\mu$ and $\tau$ sectors can interact with the dark sector by exchanging the gauge boson $Z^\prime$ associated with the $L_\mu -L_\tau$ symmetry,
\begin{align}
{\cal L} \supset - g_{\mu\tau} Z^{\prime}_{\mu} 
 J_{L_\mu-L_\tau}^\mu \,,
\end{align}
with
\begin{equation}
J_{L_\mu-L_\tau}^\mu = \bar\mu \gamma^\mu \mu - \bar\tau \gamma^\mu \tau
+ \overline{\nu_{\mu_L}} \gamma^\mu \nu_{\mu_L} - \overline{\nu_{\tau_L}} \gamma^\mu \nu_{\tau_L}  \,, 
\end{equation}
where $g_{\mu\tau}$ is the gauge coupling constant associated with the $Z^{\prime}$ boson. Under the $U(1)_{L_{\mu}-L_{\tau}}$ symmetry, the muon and tau sectors carry charges of $+1$ and $-1$, respectively.
We assume the dark sector contains a vector-like Dirac fermion $\chi$ and a Higgs-like complex scalar $\phi_S$ with the  U(1)$_{L_{\mu} - L_{\tau}}$ charge assigned to be $+q_d$ and $+2q_d$, respectively. The charge assignments for all relevant particles are summarized in Table~\ref{tab:list-charge}.

\begin{table}[t!]
  \centering
  \begin{tabular}{|c|ccc|}\hline
    & \hskip0.2cm SU(2)$_L$  \hskip0.35cm & \hskip0.35cm U(1)$_{Y}$   \hskip0.35cm & \hskip0.35cm U(1)$_{L_{\mu} - L_{\tau}}$:  $q_{L_d}$ \hskip0.35cm \\ [0ex] \hline
      $L_{\mu} = (\nu_{\mu},\ \mu_L)$  & $2$                          & $-1/2$                & \ \ $1$\\
      $L_{\tau} = (\nu_{\tau},\ \tau_L)$  & $2$                          & $-1/2$                & $-1$\\
      $\mu_R$                                       & $1$                         & $-1$                   &\, \ $1$\\
      $\tau_R$                                       & $1$                         & $-1$                   & $-1$\\
      $\phi_H$                                       & $2$                         & $1/2$                  & $0$\\ \hline
      $\chi$                                           & $1$                          & $0$                     & $q_d$ \\
      ${\phi_S}$                                     & $1$                         & $0$                    & $2q_d$ \\ \hline
  \end{tabular}
 \caption{ \small Charge assignments, where the generic ${\rm U(1)_{L_\mu-L_\tau}}$ charge is denoted as  $q_{L_d}$.
 }
  \label{tab:list-charge}
\end{table}

 In the following subsections, we will separately discuss (i) the mixing of the dark scalar and SM Higgs, (ii) the dark Majorana mass eigenstates and their interactions, and (iii) the possible kinetic mixing between $Z^\prime$ and SM neutral gauge bosons, $Z$ and $A$ (photon).

\subsection{The Higgs sector}\label{sec:Higgs-sector}

The Higgs portal may hint at the dark sector interacting with SM particles owing to mixing between the dark scalar $\phi_S$ and the SM Higgs doublet $\phi_H$.
The relevant Lagrangian is described by
\begin{align}
{\cal L}_{\rm Higgs} = (D_\mu \phi_H)^\dag D^\mu \phi_H + (D_\mu \phi_S)^\dag D^\mu \phi_S -V(\phi_H, \phi_S) \,,
\end{align}
where $D_\mu \phi_{S,H} =(\partial_\mu + i q_{S,H} g_{\mu\tau} Z_\mu^\prime)\phi_{S,H}$ with $U(1)_{L_\mu -L_\tau}$ charges $q_H=0$ and $q_S=2 q_d$, and with $q_S g_{\mu\tau} \equiv 2 g_\chi $. Here, the potential is given by
\begin{align}
V (\phi_H, \phi_S) =  -\mu_H^2  \phi_H^\dag \phi_H + \lambda_H (\phi_H^\dag \phi_H)^2 - \mu_S^2 \phi_S^\dag \phi_S + \lambda_S  (\phi_S^\dag \phi_S)^2 
                        + \lambda_{SH}  (\phi_S^\dag \phi_S)(\phi_H^\dag \phi_H) \,.
\end{align}
We consider $SU(2)_L\times U(1)_Y \times U(1)_{L_\mu -L_\tau}$ is spontaneously broken into $U(1)_{\rm QED} \times Z_2$, where the $Z_2$ symmetry, under which $\chi_\pm \to - \chi_\pm$, stabilizes the lightest dark-sector Majorana fermion which becomes the dark matter candidate (which will be further discussed in the following subsection). The parameter requirements are as follows. The spontaneous symmetry breaking (SSB) requires $\mu_H^2 > 0$ and $\mu_S^2 >0$.
The vacuum stability needs $\lambda_H >0, \lambda_S>0$ and $\lambda_{SH} >  -2 \sqrt{\lambda_H \lambda_S}$. The conditions with scalar eigenmasses $m_S^2, m_H^2 >0$ give $\lambda_{SH}^2 <  4 \lambda_H \lambda_S$.

Writing  $\phi_H = \exp (i \tau^a \zeta^a /v_H) \cdot (\, 0, (\phi_h + v_H)/\sqrt{2} \, )^T$ and $\phi_S =\exp(i \sigma/v_S)  (\phi_s +v_S)/\sqrt{2} $, the three fields $\zeta^a$ and the field $\sigma$ can be set as zero respectively through the unitarity transformation; the former three ($\zeta^a$) have then become the longitudinal degrees of freedom (DoFs) of $W^\pm$ and $Z$, while the latter one ($\sigma$) the longitudinal DoF of $Z^\prime$. The vacuum expectation value $v_S$, for which its additional phase (or sign) can be absorbed by field redefinitions of the dark sector particles, is taken to be positive.
Thus, $(D_\mu \phi_S)^\dag D^\mu \phi_S$ will give rise to the $Z^\prime$'s mass term, $ m_{Z^\prime}^2 Z^\prime_\mu Z^{\prime\mu} /2$ with $m_{Z^\prime} =2g_\chi v_S$.

We can then represent the post-SSB Higgs-sector Lagrangian in terms of the neutral fields $(\phi_s, \phi_h)$, 
where the mass term is
\begin{align}
\frac{1}{2}
(\phi_s  \,, \  \phi_h)
{\cal M}^2
\Bigg(\begin{array}{c}
\phi_s \\  \phi_h
\end{array}\Bigg) 
\end{align}
with 
\begin{equation}
{\cal M}^2=
\left(
\begin{array}{cc}
 2 \lambda_S v_S^2 & \lambda_{SH} v_S v_H \\ 
\lambda_{SH} v_S v_H &  2 \lambda_H v_H^2
\end{array}
\right)  \,.
\label{eq:mass_matrix}
\end{equation}
Diagonalizing the mass matrix, the two neutral fields can be further expressed by the corresponding mass eigenstates $(S, H)$,
\begin{align}
\left(\begin{array}{c}
\phi_s \\  \phi_h
\end{array}\right) 
&=
\left(\begin{array}{cc}
\cos\alpha & \sin\alpha \\
-\sin\alpha & \cos\alpha
\end{array}\right)
\left(\begin{array}{c}
S  \\  H
\end{array}\right),  
\end{align}
where the mixing angle $\alpha$ satisfies the condition,
\begin{align}
\sin 2\alpha = \frac{v_S v_H \lambda_{SH}}{\sqrt{(\lambda_H v_H^2 - \lambda_S v_S^2)^2 + (\lambda_{SH} v_S v_H)^2 }} 
     =  \frac{2 v_S v_H \lambda_{SH}}{m_H^2 -m_S^2}\,,
\end{align}
and the eigenmasses squared ($m_S^2, m_H^2$) are given by
\begin{align}
m_S^2 &= \lambda_H v_H^2 + \lambda_S v_S^2 
             - \sqrt{(\lambda_H v_H^2 - \lambda_S v_S^2)^2 + (\lambda_{SH} v_S v_H)^2}
       \,,  \nonumber\\
m_H^2 &= \lambda_H v_H^2 + \lambda_S v_S^2 
             + \sqrt{(\lambda_H v_H^2 - \lambda_S v_S^2)^2 + (\lambda_{SH} v_S v_H)^2} \,.
 \label{eq:eigen-mass-Higgs}                          
\end{align}
In the paper, we use $H$ to be the SM-like Higgs with mass $m_H=125.25$~GeV \cite{pdg2022}.
The Lagrangian for $S$, which includes its self-interaction terms and interactions with the dark gauge field, is given by
\begin{align}
{\cal L} \supset 
&\frac{1}{2} \partial_\mu S \partial^\mu S -\frac{1}{2} m_S^2 S^2 
- \Big( \frac{g_\chi m_S^2 c_\alpha^3}{m_{Z^\prime}} + \frac{m_S^2  s_\alpha^3}{2 v_H}  \Big) S^3  
+ \frac{g_{\chi}^2}{2 m_{Z^\prime}^2} (c_\alpha^2 m_S^2 +s_\alpha^2 m_H^2) c_\alpha^4 S^4
\nonumber\\
&+ 2 c_\alpha^2 g_\chi^2 S^2 Z^{\prime\mu} Z^\prime_\mu + 2 c_\alpha g_\chi m_{Z^\prime} S Z^{\prime\mu} Z^\prime_\mu + \dots \,,
  \label{eq:L-DS-S}
\end{align}
where $c_\alpha \equiv \cos\alpha$ and $s_\alpha \equiv \sin\alpha$.

\subsection{The dark sector part involving dark fermions}\label{subsec:dark-sector}

The interaction containing the dark sector fermion is given by
\begin{align}
{\cal L}_\chi =i \bar\chi \gamma^\mu D_\mu \chi -m_D \bar\chi \chi
                     -\frac{f}{\sqrt{2}} \overline{\chi^c} \chi \phi_S^* - \frac{f^*}{\sqrt{2}} \overline{\chi} \chi^c \phi_S \,,
  \label{eq:L-DS-DM}
\end{align}
where $m_D >0$ corressponds to a Dirac mass, and $\chi^c \equiv C \bar{\chi}^T$ is the charge conjugate of $\chi$.
In Eq.~(\ref{eq:L-DS-DM}), $D_\mu = \partial_\mu + i g_\chi Z^\prime_\mu$ is the covariant derivative, where the dark gauge coupling is defined as $g_\chi = g_{\mu\tau} \times q_d$. In this study, we treat $g_\chi$ as a free parameter to account for its relative strength compared to the lepton sector coupling $g_{\mu\tau}$.
We have assumed a single Yukawa coupling $f$ for both the left- and right-handed chiral components of the dark fermion $\chi$. Although a Dirac fermion generally allows for two independent couplings, $y_N$ and $y_{\bar{N}}$ (corresponding to the $\overline{\chi^c_R} \chi_L$ and $\overline{\chi^c_L} \chi_R$ terms, respectively) as discussed in Ref.~\cite{Kamada:2018zxi}, we explicitly invoke a discrete parity symmetry, $\chi_L \leftrightarrow \chi_R$. This symmetry enforces $y_N = y_{\bar{N}} \equiv f$, which remains robust against radiative corrections due to the vector-like nature of the gauge interactions in this model.

After SSB, $\phi_S$ develops a vacuum expectation value given by $\langle \phi_S \rangle =v_S /\sqrt{2}$. The mass term for the dark fermion in the Lagrangian is given by 
\begin{align}
{\cal L}_m= {\cal L}_m^D +{\cal L}_m^M \,,
\end{align}
where the Dirac mass part is 
\begin{align}
-{\cal L}_m^D&=  \frac{m_D}{2} (\bar\chi \chi + \overline{\chi^c} \chi^c) \nonumber\\
                      &= \frac{m_D}{2} ( \overline{\chi_R} \chi_L + \overline{\chi_L} \chi_R + \overline{\chi^c_R} \chi^c_L + \overline{\chi^c_L} \chi^c_R  ) \,,
\end{align}
and the Majorana mass part is 
\begin{align}                    
-{\cal L}_m^M&=  \frac{f v_S}{2}  (\overline{\chi^c} \chi + \overline{\chi} \chi^c ) \nonumber\\
                      &=  \frac{f v_S}{2}( \overline{\chi^c_R} \chi_L + \overline{\chi_L} \chi^c_R + \overline{\chi^c_L} \chi_R + \overline{\chi_R} \chi^c_L  ) \,.
\end{align}
Here we take $f$ to be positive and $f v_S \gg m_D$. Note that through this paper, we adopt the notations\footnote{In the literature, someones instead use $\chi^c_L \equiv (\chi_L)^c$ and $\chi^c_R \equiv (\chi_R)^c$.} $\chi^c_L \equiv (\chi^c)_L  =(\chi_R)^c$ and  $\chi^c_R \equiv (\chi^c)_R  =(\chi_L)^c$.
When the Majorana mass dominates ($f v_S \gg m_D$), the system splits into two distinct Majorana eigenstates with masses,
\begin{align}
m_\pm = f v_S \pm m_D \,,
\end{align}
which correspond to the physical Majorana fermions, 
\begin{align}
\chi_+ &= \frac{1}{\sqrt{2}} ( \chi_L + \chi_R + \chi^c_R + \chi^c_L)
\hskip0.5cm {\rm and}   
\hskip0.5cm
\chi_- = \frac{1}{\sqrt{2}} (-\chi_L + \chi_R - \chi^c_R + \chi^c_L)  \,,
\end{align}
respectively. Appendix~\ref{app:pseudo-M} contains the relevant derivation.  Rewriting Eq.~(\ref{eq:L-DS-DM}) in the mass basis, the Lagrangian ${\cal L}_\chi$ is given by
\begin{align}
{\cal L}_\chi = &\frac{i}{2} (\bar\chi_+ \gamma_\mu \partial^\mu \chi_+  + \bar\chi_- \gamma_\mu \partial^\mu \chi_-)
                       - \frac{1}{2} g_\chi Z^{\prime \mu} (\bar\chi_+ \gamma_\mu \gamma_5 \chi_-  +  \bar\chi_- \gamma_\mu \gamma_5 \chi_+) \nonumber\\
                       & -\frac{1}{2} [ m_+ \bar\chi_+ \chi_+  +   m_-  \bar\chi_-  \chi_-  ] 
                         -\frac{f}{2}  (S \cos\alpha + H \sin\alpha) (\bar\chi_+ \chi_+  +  \bar\chi_-  \chi_- )  \,.
 \label{eq:L-DS-DMr}
\end{align}

It is important to distinguish our framework from other $U(1)$ gauge extensions. In widely studied pseudo-Dirac scenarios \cite{Kamada:2018zxi}, the dark matter mass is dominated by a large Dirac term ($m_D \gg f v_S$), leading to gauge interactions via an inelastic vector current. Conversely, recent literature on sub-GeV dark matter \cite{Balan:2025uke} explores a distinct phenomenological limit, where degenerate Majorana states interact with the dark photon solely via an elastic axial-vector current.

Our model operates in the inverse hierarchy: the dominant mass originates from the Majorana term generated by the SSB of $\phi_S$, while a tiny Dirac mass induces a slight splitting ($f v_S \gg m_D \neq 0$). To accurately reflect this physics, we refer to our candidate as $p$-wave dominant Majorana dark matter, highlighting its dominant mass origin, the non-zero mass splitting, and the strictly off-diagonal nature of its axial-vector gauge interactions. This distinct mass origin drives two crucial phenomenological features. First, as shown in Eq.~(\ref{eq:L-DS-DMr}), the states couple to the $Z^\prime$ exclusively via a strictly off-diagonal, inelastic axial-vector current ($\bar{\chi}_+ \gamma_\mu \gamma_5 \chi_- + \text{h.c.}$). Because standard elastic gauge scattering is absent, the scalar portal becomes the primary avenue for coherent elastic direct detection, allowing us to place stringent constraints on the mixing angle $\alpha$ (Sec.~\ref{sec:dd-expt}). Second, while pseudo-Dirac setups require highly suppressed couplings to maintain a tiny splitting---thereby hindering large self-interactions---our large Majorana mass is governed by an unsuppressed Yukawa coupling $f$. This naturally accommodates robust dark matter self-interactions (Sec.~\ref{sec:self-scatterings}) without fine-tuning.

\subsection{The gauge boson sector}\label{subsec:gauge}

Assuming a generic kinetic mixing between the $Z^\prime$ and the SM hypercharge gauge boson $\hat{B}$ at a high-energy scale, electroweak symmetry breaking induces effective mixings with the neutral SM bosons. The relevant gauge Lagrangian reads
\begin{align}
{\cal L}_{\rm gauge} =& - \frac{1}{4} \hat{Z}_{\mu\nu} \hat{Z}^{\mu\nu}  - \frac{1}{4} \hat{F}_{\mu\nu} \hat{F}^{\mu\nu} - \frac{1}{4} \hat{Z}^\prime_{\mu\nu} \hat{Z}^{\prime \mu\nu} 
+ \frac{1}{2} m_Z^2 \hat{Z}_{\mu} \hat{Z}^{\mu} +  \frac{1}{2} m_{Z^\prime}^2 \hat{Z}'_{\mu} \hat{Z}'^{\mu} \nonumber\\
 &  - \frac{1}{2} \epsilon_0^A \hat{Z}^\prime_{\mu\nu} \hat{F}^{\mu\nu}  - \frac{1}{2} \epsilon_0^Z \hat{Z}^\prime_{\mu\nu} \hat{Z}^{\mu\nu}  \,,
\label{eq:gaugeL}
\end{align}
where hatted fields denote interaction eigenstates, $m_{Z^\prime} =2g_\chi v_S$, and $\hat{V}_{\mu\nu} \equiv \partial_\mu \hat{V}_\nu - \partial_\nu \hat{V}_\mu$ for $\hat{V} \in \{\hat{Z}, \hat{A}, \hat{Z}^\prime \}$. Originating from the gauge-invariant operator $-\frac{\epsilon}{2}\hat{Z}^{\prime}_{\mu\nu}\hat{B}^{\mu\nu}$ \cite{Holdom:1985ag}, the bare parameters are related by $\epsilon_0^A = \epsilon \cos\theta_W$ and $\epsilon_0^Z = -\epsilon \sin\theta_W$. Consequently, $\epsilon_0^Z = -\epsilon_0^A \tan\theta_W$, leaving only one independent mixing parameter.

In addition to the residual mixing among the gauge fields, allowing the dark gauge boson to directly couple to $\mu$ and $\tau$ lepton sectors, these neutral gauge bosons coupled to SM and dark fermions are described by the Lagrangian,
\begin{align}
{\cal L} \supset  & - (g_Z J_Z^\mu, \ e J_\gamma^\mu, \  g_{\mu\tau} J_{L_\mu-L_\tau}^\mu +g_{\chi} J_D^\mu) {\bf \hat V}_\mu \,,
  \label{eq:current-coupling}
\end{align}
where ${\bf \hat V}_\mu= (\hat Z_\mu,\,  \hat A_\mu,\,  \hat Z^\prime_\mu)^T$ is a vector column formed by gauge fields,  $J_Z^\mu = J_3^\mu -  \sin^2\theta_WJ_\gamma^\mu $ is the neutral current with $J_3^\mu = \bar\psi \gamma^\mu P_L T^3 \psi$,
$J_\gamma^\mu =\bar\psi \gamma^\mu Q \psi$ is the electromagnetic current with $\psi$ being the SM fermion and $Q$ being the corresponding electromagnetic charge operator, $g_Z \equiv e/(\sin\theta_W \cos\theta_W)$, and dark Majorana current $J_{D}^\mu =(1/2) (\bar\chi_+ \gamma_\mu\gamma_5 \chi_-  + \bar\chi_- \gamma_\mu\gamma_5 \chi_+)$. Here,  $P_L=(1-\gamma_5)/2$ is the left-handed projector and the eigenvalue of $T^3$ is $\pm 1/2$ for a left-handed doublet SM fermion or $0$ for an isosinglet SM fermion.

However, $\hat{Z}'$ can still mix with the photon $\hat{A}$ and with the $\hat{Z}$ boson at a one-loop level due to $m_\mu \not= m_\tau$ at low energy with the scale $\mu \lesssim m_\mu$.
Thus, the Lagrangian relevant to the gauge mixing, involving the one-loop corrections, is described by
\begin{align}
{\cal L}_{\rm mix} = - \frac{1}{2}  \epsilon_A  \hat{Z}^\prime_{\mu\nu} \hat{F}^{\mu\nu} - \frac{1}{2} \epsilon_Z \hat{Z}^\prime_{\mu\nu}  \hat{Z}^{\mu\nu} \,,
\label{eq:gaugeMixing}
\end{align}
with $\epsilon_A = \epsilon_0^A + \epsilon_{\text{loop}}^A$ and $\epsilon_Z = \epsilon_0^Z + \epsilon_{\text{loop}}^Z$, where the one-loop mixing parameters below the $m_{\mu}$ scale are given by  \cite{Holdom:1985ag, Schmitz:2009, Gherghetta:2019coi, Kamada:2018zxi}
\begin{align}
 \epsilon_{\text{loop}}^A = - \frac{e g_{\mu\tau} }{12 \pi^2}  \ln \left( \frac{m_\tau^2}{m_\mu^2} \right)
 \simeq  - \frac{g_{\mu\tau} }{69}
 \,, \ \ \
 \epsilon_{\text{loop}}^Z =  \left( -\frac{1}{4} + \sin^2\theta_W  \right) \frac{1}{ \sin\theta_W  \cos\theta_W }\epsilon_A^\prime  \,,
\end{align}
with $\theta_W$ being the Weinberg angle which is satisfied with $\sin\theta_W =0.231$ at the scale $m_Z$ in the $\overline{\rm MS}$ scheme \cite{pdg2022}.

To obtain the physical mass eigenstates, we diagonalize the kinetic and mass terms of the gauge fields. Following the procedure outlined in Appendix~\ref{app:kinetic-to-mass}, the interaction eigenstates $\hat{\bf V}_{\mu}$ are related to the physical states ${\bf V}_{\mu}=(Z_{\mu},A_{\mu},Z_{\mu}^{\prime})^{T}$ via the transformation $\hat{\bf V}_{\mu} = {\bf U}^{-1}{\bf V}_{\mu}$. 
Substituting this relation into Eq.~(\ref{eq:current-coupling}), the resultant Lagrangian describing the interactions of the physical gauge bosons coupled to the SM and dark currents is given by
\begin{align}
{\cal L} & \supset   - (g_Z J_Z^\mu, \ e J_\gamma^\mu, \  g_{\mu\tau} J_{L_\mu-L_\tau}^\mu + g_\chi J_D^\mu)  {\bf U}^{-1} {\bf V}_\mu \nonumber \\
& = - e J_\gamma^\mu A_\mu -  g_Z J_Z^\mu Z_\mu -  g_{\mu\tau} J_{L_\mu-L_\tau}^\mu  Z^\prime_\mu  -  g_\chi  J_{D}^\mu Z^\prime_\mu \nonumber\\
     &\  \  + \epsilon_A \, e J_\gamma^\mu Z^\prime_\mu 
      + \epsilon_Z \frac{m_Z^2}{m_Z^2-m_{Z^\prime}^2} (g_{\mu\tau} J_{L_\mu-L_\tau}^\mu + g_\chi J_D^\mu) Z_\mu 
      - \epsilon_Z \frac{m_{Z^\prime}^2}{m_Z^2-m_{Z^\prime}^2} g_Z J_Z^\mu Z^\prime_\mu + {\cal O} (\epsilon_{A,Z}^2)  \,.
  \label{eq:amu-L}
\end{align}

\section{The parameters}\label{sec:parameters}

We adopt $H$ to be the observed Higgs resonance with mass $m_H = 125.25$~GeV and vacuum expectation value $v_H = 246.22$~GeV \cite{pdg2022}. Compared with the SM, the present model comprises 8 independent parameters:
\begin{equation}
m_{-}, \quad m_+ \equiv m_- + \delta m, \quad m_{Z^\prime}, \quad m_S, \quad g_{\mu\tau}, \quad f, \quad \alpha, \quad \text{and} \quad \epsilon_0^A \,,
\label{eq:parameters}
\end{equation}
with $\delta m =2 m_D$.
Note that the dark gauge coupling $g_\chi$ is not an independent parameter in this setup. From the mass relations in Eq.~(15) and Eq.~(18), it is determined by the Yukawa coupling $f$, the gauge boson mass $m_{Z'}$, and the mass splitting $\delta m$ via the exact relation:
\begin{equation}
g_\chi = \frac{f m_{Z^\prime}}{m_+ + m_-} = \frac{f m_{Z^\prime}}{2m_- + \delta m}.
\label{eq:gchi_relation}
\end{equation}

Our analysis of the parameter space is guided by the following strict considerations:
(i) We require the mass splitting $\delta m$ to satisfy the thermodynamic condition $(\delta m / m_-) x_f \gg 1$, where $x_f = m_- / T_{\rm fo}$ is the freeze-out parameter. Equivalently, this dictates $\delta m \gg T_{\rm fo}$. This strict kinematic hierarchy ensures that the primordial abundance of the excited state $\chi_+$ is exponentially suppressed by the Boltzmann factor $e^{-\delta m / T_{\rm fo}}$ during the freeze-out epoch, safely neutralizing the coannihilation channel while maintaining the splitting subdominant to the base mass ($\delta m < m_-$).
(ii) The kinetic mixing parameter $\epsilon_A$ is defined as $\epsilon_A \simeq \epsilon_0^A + \epsilon_{\text{loop}}^A$, where $\epsilon_0^A$ represents the primordial kinetic mixing generated at high scales. Its phenomenological impact will be discussed in Sec.~\ref{sec:g-2}.
(iii) To ensure dark matter remains in thermal equilibrium with the SM bath before freeze-out, we assume $m_S > 2m_{Z^\prime}$.
(iv) Due to the thermodynamical depletion of $\chi_+$, the relic abundance is exclusively dictated by the purely $p$-wave annihilation processes of the ground state, $\chi_{-} \chi_{-} \to SS$ and $Z^\prime Z^\prime$. The macroscopic mass gap safely renders the $s$-wave $\chi_- \chi_+ \to S Z^\prime$ coannihilation inactive. This requirement uniquely fixes the coupling $f$ for a given dark matter mass $m_-$. Consequently, $g_\chi$ is derived using Eq.~(\ref{eq:gchi_relation}).
(v) We assume $m_S \ll m_{\pm}$ to properly enhance the self-interaction cross-section for solving small-scale structure problems.
(vi) The scalar mixing angle $\alpha$ is constrained by various experimental searches, as analyzed in Secs.~\ref{sec:pp-expt} and \ref{sec:dd-expt}.

\section{\texorpdfstring{Constraints from $(g-2)_\mu$ and kinetic mixing}{} }\label{sec:g-2}

Recently, the prediction for the muon anomalous magnetic moment has been updated by the Muon $g-2$ Theory Initiative \cite{Aliberti:2025beg}. This new assessment incorporates lattice QCD results for the hadronic vacuum polarization, leading to a Standard Model prediction that is in better agreement with experimental data, unlike previous assessments  \cite{Muong-2:2023cdq, Muong-2:2024hpx, Muong-2:2006rrc, Aoyama:2020ynm, Davies:2025pmx} which favored a significant new physics contribution. Consequently, the difference between the experimental value and the Standard Model prediction is now given by
\begin{equation}
\Delta a_\mu \equiv a_\mu^{\text{exp}} - a_\mu^{\text{SM}} = (38 \pm 63) \times 10^{-11}.
\label{eq:delta_amu_new}
\end{equation}

In the $m_{Z^\prime} \ll m_Z$ limit, the leading contribution to the muon anomalous magnetic moment is well approximated by \cite{Jegerlehner:2009ry}
\begin{align}
\Delta a_\mu  & \simeq \frac{ (g_{\mu\tau} + e \, \epsilon_A )^2 }{ 4 \pi^2} \int_0^1 dx \frac{m_{\mu}^2 \, x^2 (1-x)}{m_{Z^\prime}^2 (1-x) + m_\mu^2 \, x^2} \,,
\label{eq:Delta_a}
\end{align}where the effective kinetic mixing $\epsilon_A$ incorporates both the primordial $\epsilon_0^A$ and an irreducible $\mu/\tau$ loop-induced shift, yielding $e \epsilon_A \simeq e\epsilon_0^A - 0.0044 g_{\mu\tau}$ (see Appendix~\ref{app:gminus2} for full expressions).

\begin{figure}[t!]
\begin{center}
\includegraphics[width=0.48\textwidth]{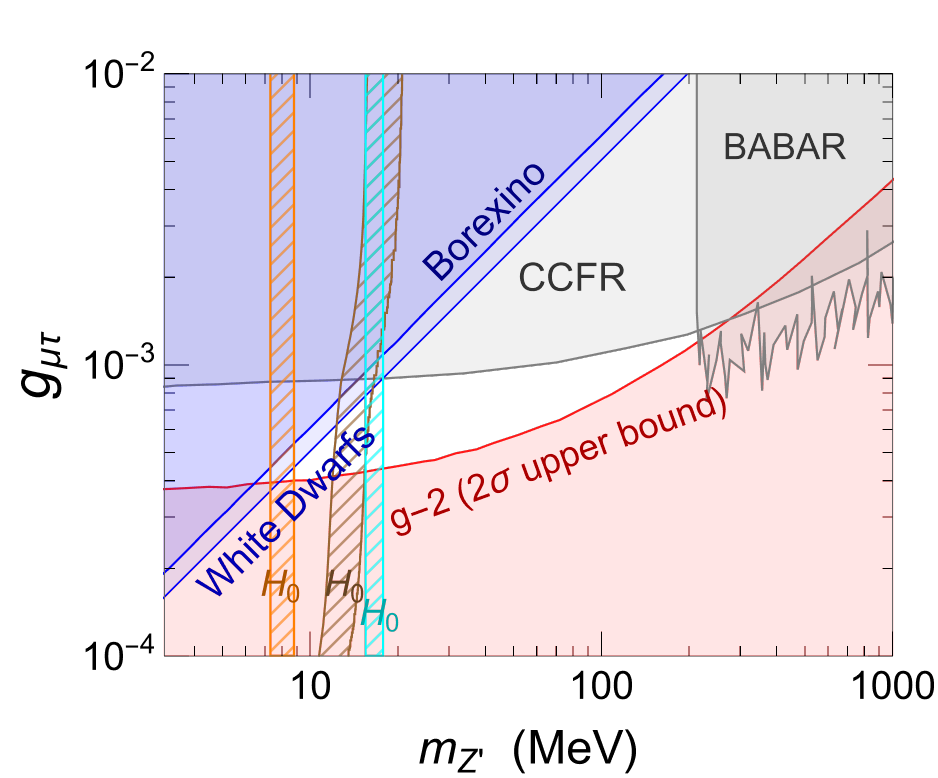}
\caption{The parameter space of the gauged $U(1)_{L_\mu - L_\tau}$ model. Gray shaded regions indicate exclusions from CCFR~\cite{CCFR:1991lpl, Altmannshofer:2014pba} and BABAR~\cite{BaBar:2016sci}, while constraints from WD cooling~\cite{Dreiner:2013tja, Bauer:2018onh} and Borexino~\cite{Bellini:2011rx, Kamada:2018zxi} are indicated by blue boundaries (assuming vanishing primordial kinetic mixing, $\epsilon_0^A = 0$). The solid red line represents the updated $(g-2)_\mu$ $2\sigma$ upper bound \cite{Aliberti:2025beg}; the region below this line is allowed by the new data. The hatched bands indicate the Hubble-tension preferred region ($3.12 \lesssim N_{\text{eff}} \lesssim 3.33$) for kinetic mixings $|\epsilon_A| = 10^{-7}$ (orange), $g_{\mu\tau}/69$ (brown), and $0$ (cyan), appearing in that order from left to right at $g_{\mu\tau}=10^{-4}$. The domain to the left of these bands yields $N_{\text{eff}} > 3.33$ and is disfavored by Planck 2018 constraints~\cite{Planck:2018vyg}.
}
\label{fig:g-2}
\end{center}
\end{figure}

Recent theoretical updates \cite{Aliberti:2025beg} align the SM prediction of $(g-2)_{\mu}$ with experimental data, imposing a strict upper bound on $\Delta a_\mu$. As summarized in Fig.~\ref{fig:g-2}, this new $2\sigma$ limit (solid red line) leaves a broad viable parameter space. Specifically, the region $m_{Z^\prime} \lesssim 200$ MeV is bounded primarily by neutrino trident production at CCFR~\cite{CCFR:1991lpl, Altmannshofer:2014pba} (gray shaded region), whereas BABAR searches for $Z^\prime \to \mu^+\mu^-$~\cite{BaBar:2016sci} dominate at heavier masses. Crucially, the resolution of the $(g-2)_\mu$ anomaly reopens the $m_{Z^\prime} > 200$ MeV regime, which was previously ruled out by the tension between the old $(g-2)_\mu$ favored band and BABAR exclusions.

We now turn to the constraints arising from the kinetic mixing between the dark sector and the photon. As introduced in Sec.~\ref{subsec:gauge}, the mixing parameter receives contributions from both the irreducible one-loop vacuum polarization, $\epsilon_{\text{loop}}^A \simeq -g_{\mu\tau}/69$, and any primordial kinetic mixing $\epsilon_0^A$ generated at high scales. 
While often assumed to be negligible, a non-zero $\epsilon_0^A$ can interfere with the loop-induced mixing, significantly altering the phenomenological bounds.

The constraints from White Dwarf (WD) cooling~\cite{Dreiner:2013tja, Bauer:2018onh} and Borexino~\cite{Bellini:2011rx, Kamada:2018zxi}, arising essentially from the effective kinetic mixing between the $Z^\prime$ and the photon, are depicted as blue boundaries in Fig.~\ref{fig:g-2} under the assumption of vanishing primordial kinetic mixing ($\epsilon_0^A = 0$).
Taking the interference between these two contributions into account, the standard bounds on $g_{\mu\tau}$ can be analytically generalized as:
\begin{align}
g_{\mu\tau} \times \left| 1 - \frac{69 \epsilon_0^A}{g_{\mu\tau}} \right|^{1/2} \le 6.1\times 10^{-4} \times \frac{m_{Z^\prime}}{12~\text{MeV}} \,.
\label{eq:WD_limit_new}
\end{align}
This inequality shows that if the primordial mixing $\epsilon_0^A$ cancels the loop contribution (i.e., $\epsilon_0^A \approx g_{\mu\tau}/69$), the effective mixing is suppressed, allowing for a much larger $g_{\mu\tau}$ than the standard bound suggests. Crucially, even if the condition $e \epsilon_A \ll g_{\mu\tau}$ is satisfied, small kinetic mixing effects from high-energy scales can significantly alter the constraints derived from both WD cooling and Borexino data. This interference effect effectively relaxes the constraints compared to the standard bounds ($\epsilon_0^A = 0$) shown in Fig.~\ref{fig:g-2}.

In Fig.~\ref{fig:g-2}, the hatched areas, corresponding to $3.12 \lesssim N_{\text{eff}} \lesssim 3.33$, represent the parameter space that helps alleviate the disagreement between cosmological and local Hubble constant measurements, known as the Hubble tension ($H_{0}$). This preferred range is supported by analyses combining Planck data with local $H_0$ measurements \cite{Planck:2018vyg}. 
 the orange, brown, and cyan hatched bands represent the scenarios with $|\epsilon_{A}|=10^{-7}$, $g_{\mu\tau}/69$, and $0$, respectively (appearing in this order from left to right at $g_{\mu\tau}=10^{-4}$). The parameter space to the left of these bands yields $N_{\text{eff}} > 3.33$, which is disfavored by Planck 2018 constraints (TT, TE, EE+lowE+lensing+BAO at 95\% CL) \cite{Planck:2018vyg}. A detailed analysis of $N_{\text{eff}}$ will be given in Sec.~\ref{sec:neff}.

\section{Constraint from effective number of relativistic species and Hubble tension}\label{sec:neff}
\begin{figure}[t!]
\begin{center}
\includegraphics[width=0.41\textwidth]{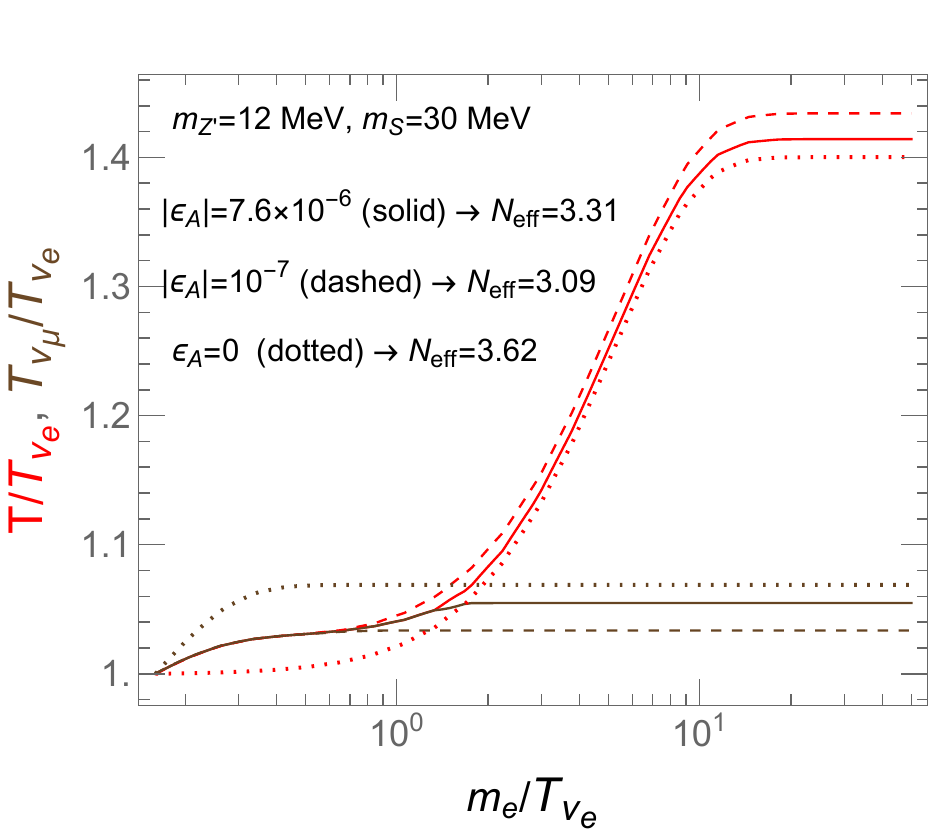}\hskip0.6cm
\includegraphics[width=0.41\textwidth]{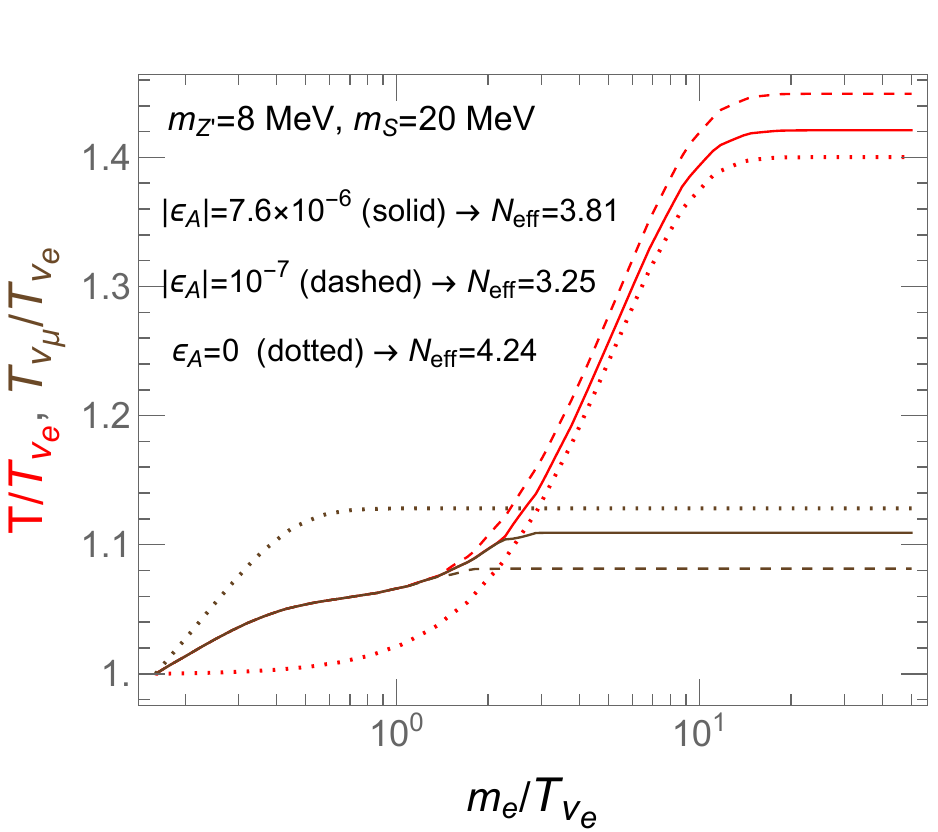}
\caption{The evolution of the ratios $T/T_{\nu_e}$ (red) and $T_{\nu_\mu} / T_{\nu_e}$ (brown) for $|\epsilon_A| =7.6\times 10^{-6}$ (solid line), $10^{-7}$ (dashed line), and $0$ (dotted line), with $m_{Z^\prime}=12$~MeV and $m_S=30$~MeV in the left panel, and $m_{Z^\prime}=8$~MeV and $m_S=20$~MeV in the right panel. The resulting $N_{\rm eff}$ value is also given.
}
\label{fig:temp-ratio}
\end{center}
\end{figure}

\begin{figure}[t!]
\begin{center}
\includegraphics[width=0.46\textwidth]{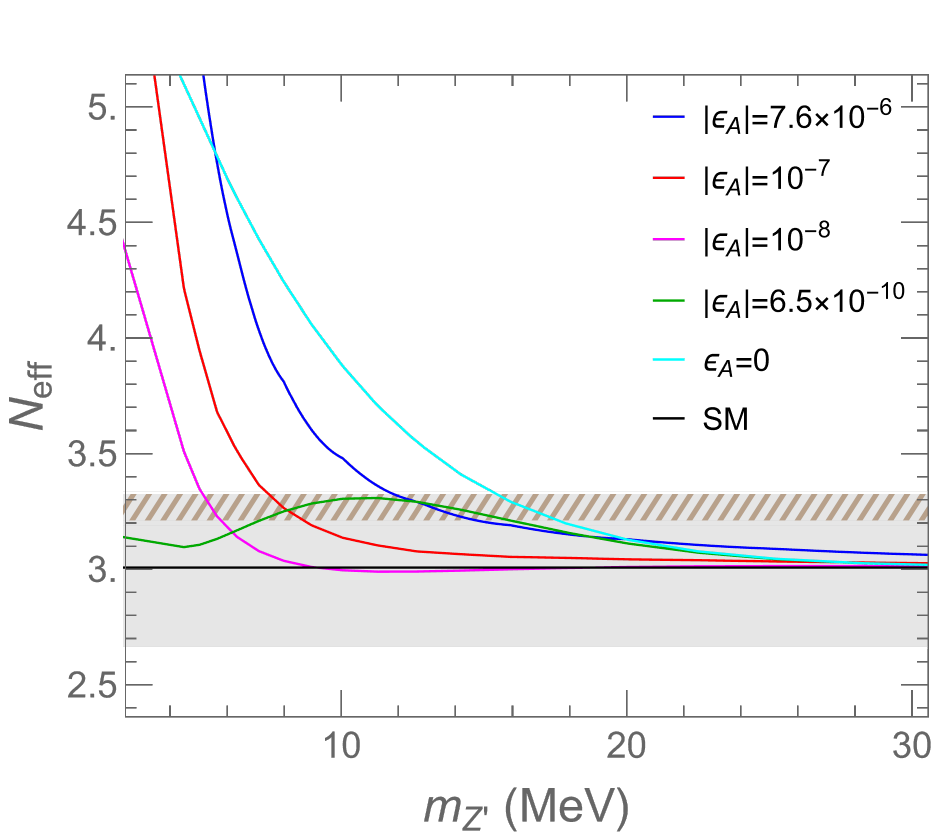}
\caption{ $N_{\text{eff}}$ as a function of $m_{Z^\prime}$. The solid curves correspond to model predictions with $|\epsilon_A| = 0, 7.6\times 10^{-6}, 10^{-7}, 6.5\times 10^{-10}$, and $10^{-8}$ (from top to bottom at $m_{Z^\prime}=7$~MeV). The horizontal line indicates the SM value. The gray and hatched regions show the 95\% CL allowed range ($N_{\text{eff}} = 2.99^{+0.34}_{-0.33}$) from Planck 2018~\cite{Planck:2018vyg}. The hatched brown sub-region ($3.12 \lesssim N_{\text{eff}} \lesssim 3.33$) is favored by local $H_0$ measurements to alleviate the Hubble tension, with the lower bound corresponding to the $1\sigma$ preference ($N_{\text{eff}} \approx 3.27 \pm 0.15$~\cite{Planck:2018vyg}).}
\label{fig:neff}
\end{center}
\end{figure}

During thermal evolution, the decay rates of $S \to  Z^\prime  Z^\prime$ and $Z^\prime \to \nu_\mu \bar{\nu}_\mu, \nu_\tau \bar{\nu}_\tau$, are greater than the Hubble expansion rate. As a result, $S$ and $ Z^\prime$ remain in thermal equilibrium with $\mu$- and $\tau$-neutrinos, maintaining a zero chemical potential even after neutrinos decouple from photons. However, due to residual kinetic mixing between $ Z^\prime$ and the electromagnetic field, $ Z^\prime$ can decay into an $e^+ e^-$ pair, with the decay width,
\begin{align}
\Gamma_{ Z^\prime\to e^+ e^-} = \frac{\alpha \epsilon_A^2 m_{Z^\prime}}{3} \left(1+ \frac{2 m_e^2}{m_{Z^\prime}^2} \right)  \left(1- \frac{4 m_e^2}{m_{Z^\prime}^2} \right)^{1/2} \,.
\end{align}
Since the DoFs of $S$ and $ Z^\prime$ are ultimately integrated into the SM bath, these effects can impact the effective number of relativistic species, $N_{\rm eff}$. This parameter is defined based on the total radiation energy density, $\rho_t$, and the photon density, $\rho_\gamma$, as
\begin{align}
\frac{\rho_t}{\rho_\gamma} = 1+ N_{\rm eff}\, \frac{7}{8}  \bigg(\frac{4}{11} \bigg)^{4/3} \,.
\end{align} 

Before electron neutrino-photon decoupling, all neutrinos stay in thermal equilibrium with photon-coupled particles through weak interactions with $e^\pm$ and via $Z^\prime \leftrightarrow e^+ e^-$. About one second after the Big Bang, when the weak interaction rate falls below the expansion rate of the universe, its corresponding temperature is estimated to be around $T\approx 3.2$~MeV \cite{Escudero:2019gzq}. From this point forward, interactions between the two groups of particles---$(\gamma, e^\pm)$ and $(\nu_\mu,\nu_\tau, S, Z^\prime)$---mainly originate from $Z^\prime \leftrightarrow e^+ e^-$, which results in the transfer of entropy between them. To clarify, we define three temperatures: $T$, $T_{\nu_e}$, and $T_{\nu_\mu}$, corresponding to the sets $(\gamma, e^\pm)$, $(\nu_e)$, and $(\nu_\mu,\nu_\tau, S, Z^\prime)$, respectively. When the temperature drops below approximately 3.2 MeV, we can estimate the neutrino-to-photon temperature ratio and $N_{\rm eff}$ by analyzing the separate evolution of the comoving entropy densities, as described in Ref.~\cite{Kamada:2018zxi}: 
\begin{align}
 \frac{1}{a^3} \frac{d}{dt} \Big( s_\gamma (T) a^3 + 2 s_e(T) a^3  \Big) 
& = \frac{1}{T} \Big( \langle \Gamma_{{Z^\prime}\to e^+ e^-} \rangle_{T_{\nu_\mu}} \, \rho_{Z^\prime} (T_{\nu_\mu}) - \langle \Gamma_{{Z^\prime}\to e^+ e^-} \rangle_{T} \, \rho_{Z^\prime} (T)  \Big)\,,  
\label{eq:co-entropy-1} \\
 \frac{1}{a^3} \frac{d}{dt} \Big( 2 s_{\nu_\mu}  (T_{\nu_\mu}) a^3  + 2 s_{\nu_\tau}  (T_{\nu_\mu}) a^3 
& + s_S  (T_{\nu_\mu}) a^3 + s_{Z^\prime}  (T_{\nu_\mu}) a^3\Big) \nonumber\\
& = - \frac{1}{T_{\nu_\mu}} \Big( \langle \Gamma_{{Z^\prime} \to e^+ e^-} \rangle_{T_{\nu_\mu}} \, \rho_{Z^\prime} (T_{\nu_\mu}) - \langle \Gamma_{{Z^\prime} \to e^+ e^-} \rangle_{T} \, \rho_{Z^\prime} (T)  \Big)\,,  
\label{eq:co-entropy-2} \\
a^3 \frac{d}{dt} \Big(2 s_{\nu_e} (T_{\nu_e}) a^3 \Big) & =0 \,,
\label{eq:co-entropy-3}
\end{align}
 where the evolution of the scale factor $a$ is determined by the Hubble rate,
 \begin{eqnarray}
 \frac{\dot{a}}{a} 
 &=&  \frac{
\sqrt{\rho_\gamma (T) + 2 \big(\rho_e(T) + \rho_{\nu_e}(T_{\nu_e}) 
   + \rho_{\nu_\mu}(T_{\nu_\mu}) + \rho_{\nu_\tau} (T_{\nu_\mu}) \big)
  + \rho_S (T_{\nu_\mu}) + \rho_{Z^\prime} (T_{\nu_\mu}) 
  } }{\sqrt{3} M_{\rm pl}} \,,
\end{eqnarray}
and  the thermally averaged width of ${Z^\prime}$ at temperature $T_i$ is given by
 \begin{equation}
  \langle \Gamma_{{Z^\prime}\to e^+ e^-} \rangle_{T_i}  = \Gamma_{Z^\prime}  \frac{K_1 (m_{Z^\prime}/T_i) }{K_2(m_{Z^\prime}/T_i) } \,,
 \end{equation}
with $K_{1,2}$ the modified Bessel functions of the second kind.
Here, $s_i$ and $\rho_i$ denote the entropy density and thermal equilibrium energy density (at zero chemical potential) of species $i$, respectively, both of which account for the internal spin degrees of freedom. For details of their forms, please refer to Appendix B of Ref.~\cite{Kamada:2018zxi}. The factor of "2" preceding them accounts for its antiparticle. In Eqs.~(\ref{eq:co-entropy-1}) and (\ref{eq:co-entropy-2}), detailed balance \cite{Kolb:1990} is invoked to parameterize the energy transfer rate of the inverse decay $e^+ e^- \to {Z^\prime}$ in terms of the thermally averaged decay rate of $Z^\prime$ evaluated at temperature $T$.

Figure~\ref{fig:temp-ratio} shows the behavior of temperature ratios $T/T_{\nu_e}$ and $T_{\nu_\mu}/T_{\nu_e}$ using different values for the kinetic mixing parameter $|\epsilon_A| =7.6 \times 10^{-6}$ (which corresponds to using $\epsilon_A = - [e g_{\mu\tau} /(12 \pi^2] \ln (m_\tau^2 / m_\mu^2 )$ with $g_{\mu\tau}=5.2 \times 10^{-4}$), $10^{-7}$, and $0$, with $m_{Z^\prime}=12$~MeV and $m_S=30$~MeV, while in the right panel, $m_{Z^\prime}=8$~MeV and $m_S=20$~MeV. For a ${Z^\prime}$-$\gamma$ kinetic mixing parameter $|\epsilon_A|  > 10^{-8}$, the $Z^\prime  \leftrightarrow e^+ e^-$ interaction helps equilibrate the ($\gamma, e^\pm$) and ($\nu_\mu, \nu_\tau, S, Z^\prime $) baths, maintaining $T =T_{\nu_\mu}$ until $T< m_e$. As the temperature drops, entropy from $S$ and $Z^\prime $ is shared between these two baths. A larger $|\epsilon_A|$ extends the equilibrium period, causing $T_{\nu_\mu} / T_{\nu_e}$ to increase and $T/T_{\nu_e}$ to decrease.

On the other hand, for $|\epsilon_A|  < 10^{-8}$, the baths decouple at a temperature of approximately 3.2 MeV, below which weak interactions cannot maintain temperature equilibrium among them. In this condition, an even smaller $|\epsilon_A| $ results in less entropy being transferred to the ($\gamma, e^\pm$) bath, while the eaten $S$ and $Z^\prime $ further increase $T_{\nu_\mu}$ when $T<m_e$.

In Fig.~\ref{fig:neff}, we plot the resulting $N_{\rm eff}$ versus $m_{Z^\prime }$ during the later stage of the Big Bang Nucleosynthesis (BBN) epoch, where we have adopted $m_S=2 m_{Z^\prime }+5$~MeV. The obtained $N_{\rm eff}$ is insensitive to $m_S$ for our consideration of $m_S>2m_{Z^\prime }$. 
In Figs.~\ref{fig:g-2} and \ref{fig:neff}, the hatched region corresponds to $3.12 \lesssim N_{\text{eff}} \lesssim 3.33$. While standard $\Lambda$CDM predicts $N_{\text{eff}}^{\text{SM}} \simeq 3.046$, an elevated value of $N_{\text{eff}}$ is known to be positively correlated with the Hubble constant $H_0$ in CMB analyses, thereby reducing the tension with local measurements  \cite{Vagnozzi:2019ezj, DiValentino:2021izs}. Specifically, including a prior from local $H_0$ measurements favors $N_{\text{eff}} \approx 3.27 \pm 0.15$.
However, we note that increasing $N_{\text{eff}}$ is not a complete solution to all cosmological tensions. As pointed out in Ref. \cite{Planck:2018vyg} (see Fig. 35 therein), an increase in $N_{\text{eff}}$ tends to increase the amplitude of matter fluctuations, $\sigma_8$, which may exacerbate the tension with weak lensing surveys that prefer a lower $S_8$ (or $\sigma_8$). Therefore, the parameter space highlighted here should be interpreted as alleviating the $H_0$ tension, with the caveat that the $\sigma_8$ tension requires a more comprehensive discussion beyond the scope of this minimal model.

 Under the constraint of $N_{\rm eff}<3.33$ from Planck (TT, TE, EE + lowE + lensing + BAO at 95\% CL) \cite{Planck:2018vyg}, we find $m_{Z^\prime } \gtrsim 12$~MeV for $|\epsilon_A| = 7.6\times 10^{-6}$, and $m_{Z^\prime } \gtrsim 7.5$~MeV for $|\epsilon_A| =10^{-7}$. For much smaller $|\epsilon_A|$, {\it e.g.}, $6.5\times10^{-10}$, an even wider mass range, $6.9~{\rm MeV}<m_{Z^\prime }<16.9~{\rm MeV}$, can help relax the Hubble tension, whereas for $\epsilon_A=0$, the favored value shifts to higher masses. Before proceeding, it is worth noting that neutrino oscillations imply rapid changes in neutrino flavor, which may equilibrate the temperatures $T_{\nu_e} \simeq T_{\nu_\mu}$. Such effects might further slightly alter the $N_{\rm eff}$ \cite{Escudero:2019gzq}.

\section{\texorpdfstring{Particle physics experiments on the mixing angle $\alpha$}{}}\label{sec:pp-expt}

While a vanishing scalar-Higgs mixing angle ($\alpha=0$) was assumed in Ref.~\cite{Kamada:2018zxi} for simplicity, our study explores the more general scenario where $\alpha \neq 0$. In this section, we establish the model-independent phenomenological constraints on the $(m_S, \alpha)$ parameter space arising strictly from particle physics experiments. The complementary constraints from dark matter direct detection---which intricately depend on the DM mass and the coupling $f$ determined by the thermal relic abundance---will be addressed later in Sec.~\ref{sec:dd-expt}.
\begin{figure}[t!]
\begin{center}
\includegraphics[width=0.46\textwidth]{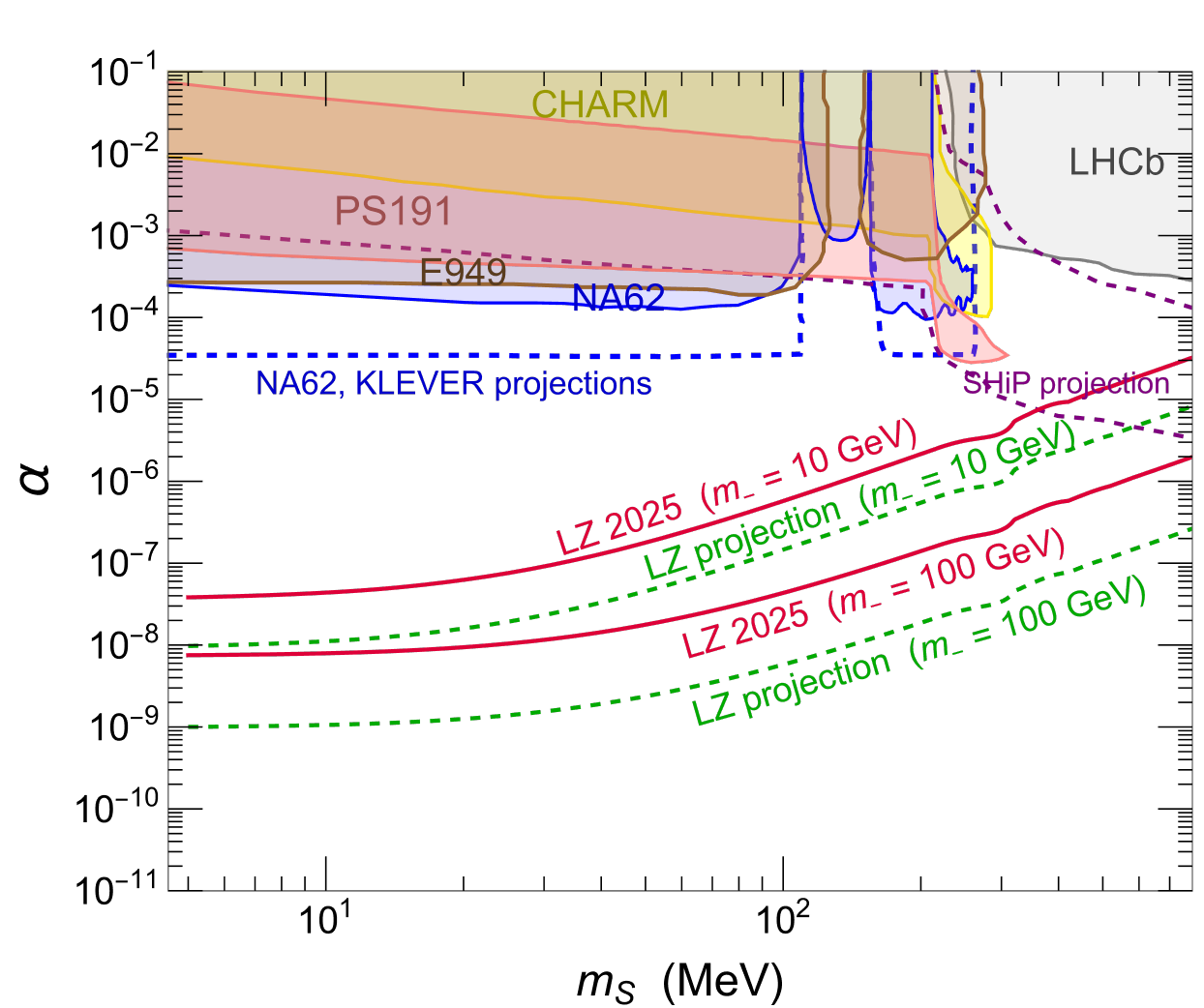}
\caption{Current 90\% CL exclusion limits and future projections for the Higgs mixing angle $\alpha$ as a function of the scalar mass $m_S$ in the sub-GeV region. Solid-colored areas with boundaries indicate parameter space excluded by particle physics experiments \cite{CHARM:1985anb, Winkler:2018qyg, Bernardi:1985ny, Bernardi:1986hs, Bernardi:1987ek, Gorbunov:2021ccu, BNL-E949:2009dza, NA62:2021zjw, NA62:2020pwi}, while dashed boundaries show expected sensitivity projections \cite{Winkler:2018qyg, Beacham:2019nyx, NA62:2017rwk, Alekhin:2015byh, SHiP:2018yqc, Bondarenko:2019vrb}. The robust joint constraints derived from the latest LZ 2025 direct detection results \cite{LZ:2024zvo}, as well as the future LZ projections \cite{LZ:2018qzl}, are denoted for two representative dark matter masses: $m_- = 10 \text{ GeV}$ and $m_- \ge 100 \text{ GeV}$ (detailed in Sec.~\ref{sec:dd-expt}).} 
\label{fig:alpha-cons}
\end{center}
\end{figure}

Experimental signatures involving a light $S$ in the final state are potentially detectable. Consequently, searches for missing energy from the long-lived scalar $S$ escaping detectors, or displaced vertices in beam-dump experiments, are highly relevant \cite{ CHARM:1985anb, Winkler:2018qyg, Bernardi:1985ny, Bernardi:1986hs, Bernardi:1987ek, Gorbunov:2021ccu, BNL-E949:2009dza, NA62:2021zjw, NA62:2020pwi, Beacham:2019nyx, NA62:2017rwk, Alekhin:2015byh, SHiP:2018yqc, Bondarenko:2019vrb}. In our model, this sub-GeV scalar $S$ can mix with the SM Higgs boson, enabling its production in rare meson decays at colliders and fixed-target experiments. The main search for a sub-GeV scalar focuses on $K\to \pi S$, with the branching ratio depending on $\sin^2 \alpha$.

Fig.~\ref{fig:alpha-cons} shows the current experimental limits on the mixing angle in relation to the sub-GeV $m_S$, including reinterpretations of earlier beam-dump experiments, CHARM \cite{Winkler:2018qyg, CHARM:1985anb} and PS191 \cite{Bernardi:1985ny, Bernardi:1986hs, Bernardi:1987ek, Gorbunov:2021ccu}.

The E949 \cite{BNL-E949:2009dza} and NA62 \cite{NA62:2021zjw} experiments have placed limits on a long-lived scalar $S$ by measuring $K^+ \to \pi^+ \bar{\nu} \nu$. In this model, a weakly interacting $S$ can be produced in the decay $K^+ \to \pi^+ S$, sharing similar experimental characteristics with the main background process $K^+ \to \pi^+ \bar{\nu} \nu$. In the 110–155 MeV range, the relevant decay chain is $K^+ \rightarrow \pi^+ \pi^0 (\gamma)$ followed by $\pi^0 \rightarrow \text{invisible}$, which involves radiative photons in the final state. Assuming $S$ escapes detection, NA62 has established constraints on $\mathcal{B}(K^+ \rightarrow \pi^+ S)$ based on the 90\% upper limit for the $\pi^0 \rightarrow \text{invisible}$ channel \cite{NA62:2020pwi}.

The KLEVER project aims to measure the branching ratio of the rare decay $K_L \to \pi^0 \nu \nu$ with 20\% accuracy \cite{Beacham:2019nyx}.  It complements the NA62 experiment, which is expected to measure the charged mode $K^+ \to \pi^+ \bar{\nu} \nu$ with about 10\% precision by the end of LHC Run 3.  \cite{Beacham:2019nyx, NA62:2017rwk}.  In Fig.~\ref{fig:alpha-cons}, these projections are shown assuming ${\cal B}(K^+ \to \pi^+ S) \sim 2\times 10^{-12}$. The SHiP experiment can reconstruct final states with two charged tracks. We also present the projected sensitivity of SHiP for a scalar decaying into two charged particles \cite{Alekhin:2015byh, SHiP:2018yqc}, as reanalyzed by Winkler \cite{Winkler:2018qyg}.

\section{ Boltzmann equations and thermal freeze-out}\label{sec:boltz}

In this section, we will estimate the freeze-out parameter $x_f = m_{-}/T_{\rm fo}$ and the related DM annihilation cross section that can result in the present-day relic abundance, where $T_{\rm fo}$ is the DM freeze-out temperature. 
To guarantee a sufficiently large dark fine-structure constant ($\alpha_f$) for resolving small-scale structure anomalies, we focus on the kinematic regime where $(\delta m / m_-) x_f \gg 1$. In this scenario, the heavier state $\chi_+$ is heavily Boltzmann-suppressed during the freeze-out epoch, rendering its number density negligible ($n_{\chi_+} \ll n_{\chi_-}$). Consequently, the dark matter relic is almost entirely composed of the ground state $\chi_-$ from the early Universe. The final dark matter abundance is thus accurately described by the number density of $\chi_-$ alone,
\begin{equation}
n \simeq  n_{\chi_-} \,,
\end{equation}
of which the evolution follows the Boltzmann equation,
\begin{align}
\frac{dn}{dt}  + 3 H n  \simeq 
- \sum_{X}  \langle \sigma v \rangle_{\chi_- \chi_- \to X} 
     \left( n^2 - (n^{\text{eq}})^2 
      \right) \,,
\label{eq:boltz-primary} 
\end{align}
with $X \equiv$ all possible final states. Addressing small-scale issues requires significant self-interactions among dark matter particles, facilitated by light $S$ particles that are strongly coupled to dark matter via a large coupling constant. Therefore, we consider that the DM freeze-out process is strictly governed by the combined annihilation channels into scalars ($\chi_{-}\chi_{-} \to SS$) and gauge bosons ($\chi_{-}\chi_{-} \to Z^\prime Z^\prime $). The coannihilation channel $\chi_{-}\chi_{+} \to Z^\prime S$, despite containing an $s$-wave component, is safely negligible due to the exponential Boltzmann suppression of the $\chi_+$ population.
During this epoch, $S$ maintains its thermal equilibrium density at the same temperature as the SM bath primarily via the reactions $S \leftrightarrow Z^\prime Z^\prime$ and $Z^\prime \leftrightarrow \text{SM SM}$, meaning $n_S = n_S^{\text{eq}}$, until after $\chi_{-}$ freeze-out\footnote{The decay widths of $ Z^\prime$ and $S$, including the Higgs decaying into dark sector particles, are listed in Appendix~\ref{app:widths}}. This property has been applied in Eq.~(\ref{eq:boltz-primary}).

In our analysis, the requirement of reproducing the observed relic abundance fixes the Yukawa coupling $f$ (and consequently determines $g_\chi$ via Eq.~(\ref{eq:gchi_relation})) for a given dark matter mass. Although our numerical calculation incorporates all relevant channels, we will show that the scalar channel $\chi_{-}\chi_{-} \to SS$ plays the primary role, while the $Z^\prime Z^\prime$ final state provides a subdominant but non-negligible contribution. Defining the effective thermally averaged cross section as $\langle \sigma_{\rm eff} v \rangle \simeq \langle \sigma v \rangle_{\chi_- \chi_- \to X}$, we rewrite Eq.~(\ref{eq:boltz-primary}) in terms of the yield $Y = n/s$, which represents the ratio of $n$ to the entropy density of the Universe,
\begin{align}
\frac{d Y}{dx} =
   & - \frac{m_-}{x^2} \bigg( \frac{\pi}{45 G} \bigg)^{1/2} g_*^{1/2} \langle \sigma_{\rm eff} v \rangle
  \left(   Y^2 - \big(Y^{\rm eq} \big)^2 \right) \,,
  \label{eq:boltz-approx}
 \end{align}
where $Y^{\rm eq} =n^{\rm eq}/s$, and  $g_*^{1/2} \equiv \tilde{h}_{\rm eff}  /g_{\rm eff}^{1/2}$  with
$\tilde{h}_{\rm eff}  \equiv h_{\rm eff} [1+(1/3) (d\ln h_{\rm eff} / d\ln T)]$. Here, $h_{\rm eff}$ and $g_{\rm eff}$ are the effective degrees of freedom (DoFs) for the Universe's total entropy and energy density, respectively, with
\begin{equation}
s = h_{\rm eff}(T) \frac{2 \pi^2}{45}\, T^3, \quad \rho = g_{\rm eff}(T) \frac{\pi^2}{30}T^4 \,.
\end{equation}
In our analysis, we adopt the effective degrees of freedom associated with a QCD phase transition temperature of $T_{\rm QCD}=150$ MeV \cite{Cerdeno:2011tf}.

We follow the method outlined in Refs.~\cite{Gondolo:1990dk, Yang:2022zlh} to determine the freeze-out parameter $x_f$ and the appropriate coupling constant $f$ related to the DM annihilation cross section that yields the current relic abundance. We define $T_{\rm fo} (=m_-/x_f)$ as the temperature where $Y - Y^{\rm eq} = \delta\cdot Y^{\rm eq}$, ensuring that $d\delta/dx \ll 1$ and that $\delta$ is approximately 1. Next, we derive the approximation from Eq.~(\ref{eq:boltz-approx}), given by
\begin{align}
\left( 1 -\frac{27}{8x} \right)  \approx
x^{-1/2} e^{-x} 
\delta (\delta+2) \sqrt{\frac{45}{\pi^5}} \frac{g _*^{1/2} M_{\rm pl} m_-}{2 h_{\rm eff}  } g_{-}  \langle \sigma_{\rm eff} v\rangle \,,
\label{eq:xf}
\end{align}
 at $x=x_f$. Here, $g_- = 2$ denotes the internal degrees of freedom of the Majorana fermion $\chi_-$, and $M_{\rm pl} \equiv (8\pi G)^{-1/2}$ signifies the reduced Planck mass. Since $x_f$ logarithmically depends on $\delta$, we set $\delta=1$ for the analysis. 
 
 The freeze-out process is strictly dominated by the $p$-wave annihilations $\chi_{-}\chi_{-} \to SS$ and $Z^\prime Z^\prime$, yielding the effective cross section,
\begin{align}
 \langle \sigma_{\rm eff} v\rangle 
 & \simeq  \langle \sigma v\rangle_{\chi_- \chi_- \to SS} +  \langle \sigma v\rangle_{\chi_- \chi_- \to Z^\prime Z^\prime}  \,.
\end{align}
Assuming $m_{Z^\prime} \ll m_-$,  $m_S \ll m_-$, and neglecting terms proportional to $x^{-n}$ for $n \geq 3$, the thermally averaged cross sections are approximately given by
\begin{align}
& \langle \sigma v\rangle_{\chi_- \chi_- \to SS}  
 \simeq \frac{9 f^4}{ 64 \pi m_-^2} x^{-1} - \frac{51 f^4}{ 64 \pi m_-^2} x^{-2} \,, 
\label{eq:ss}
\\
& \langle \sigma v\rangle_{\chi_- \chi_- \to  Z^\prime  Z^\prime}  \nonumber\\
&\ \ \simeq 
     \frac{g_\chi^4}{16 \pi m_-^2}
     \Bigg\{  \left(1- \frac{2 \delta{m}}{ m_-}\right) + 3 \frac{ f^2 m_-^2}{ g_\chi^2 m_{Z^\prime}^2} 
    \bigg[ \left( 1 - \frac{8}{3} d  \Big(1- \frac{\delta{m}}{4 m_-}\Big) + \frac{8}{3} d^2  \Big(1- \frac{\delta{m}}{m_-}\Big)  \right)  \left( \frac{1}{x} +\frac{3}{x^2} \right) \nonumber\\
&\ \ \ \ \ \ - \left(5- 12 d\Big(1- \frac{\delta{m}}{2 m_-}\Big) + 12 d^2 \Big(1- \frac{17 \delta{m}}{9 m_-}\Big) \right)  \frac{1}{x^2} \bigg] \Bigg\} \nonumber\\
&\ \ 
    \simeq \frac{ f^4 } {64 \pi m_-^2}  \big(x^{-1}  -3 x^{-2} \big) w(\delta{m}/m_-) \,, 
\label{eq:zpzp}
\end{align}
where $d= g_\chi m_-/ (f m_{Z^\prime})$. In the present $p$-wave dominant Majorana DM model, we have $2d = (1+\delta{m}/(2m_-))^{-1}$ due to $g_\chi (m_+ + m_-) = f m_{Z^\prime}$. All of these active thermally averaged cross sections are $p$-wave dominant.
It is important to note that the channel with $Z^\prime$ final states (Eq.~(\ref{eq:zpzp})) receives significant enhancement factor proportional to $(f m_-/ g_\chi m_{Z^\prime})^2$ in its $p$-wave term. 

In deriving Eq.~(\ref{eq:zpzp}), we performed an expansion assuming $(\delta{m} /m_-)^2 \ll 1$ while maintaining the thermodynamical requirement $(\delta m / m_-) x_f \gg 1$. However, evaluating the full cross section for $\chi_- \chi_- \to  Z^\prime  Z^\prime$ (see Eq.~(\ref{app:zpzp})) introduces a mild suppression factor $\delta{m}/m_-$ for increasing mass splittings (e.g., $w(0)=1$ versus $w(2)\approx 0.64$).
Throughout our numerical analysis, we adopt a representative value of $w=0.9$. We emphasize that the required value of $f$ is highly insensitive to the exact choice of $w$, as the $SS$ final state overwhelmingly dominates the total annihilation rate.

Furthermore, we neglect annihilations into SM fermions ($\chi_- \chi_- \to \mu^+ \mu^-, \nu_\mu \bar{\nu}_\mu, \tau^+\tau^- , \nu_\tau \bar{\nu}_\tau$), as their cross sections are severely suppressed compared to the dark sector channels in the parameter space of interest. The detailed expressions for all relevant annihilation cross sections, including the analytical derivation of the $\chi_- \chi_+ \to Z^\prime S$ channel, which explicitly demonstrates its unsuppressed $s$-wave nature and thereby necessitates the aforementioned Boltzmann suppression, are provided in Appendix~\ref{appsec:xs}.

It is noteworthy that in our framework, the freeze-out hierarchy differs fundamentally from the pseudo-Dirac DM model \cite{Kamada:2018zxi} in its theoretical origin, even though both ultimately rely on the population suppression of the heavier state. In the pseudo-Dirac model, the mass splitting is directly proportional to the coupling ($\delta m \sim f v_S$); requiring a large $f$ for self-interactions inherently produces a macroscopic mass splitting, which exponentially suppresses the $\chi_+$ population and consequentially deactivates the $\chi_- \chi_+ \to Z^\prime S$ channel.
In contrast, in our $p$-wave dominant Majorana framework, the mass splitting between the two states is governed by the independent Dirac mass term ($\delta m = 2m_D$). Crucially, as detailed in Appendix~\ref{appsec:xs}, the $s$-wave amplitude for the $\chi_- \chi_+ \to Z^\prime S$ coannihilation is essentially unsuppressed in the degenerate limit. If both states were equally populated, this $s$-wave channel would dominate the freeze-out process, severely restricting $f$ to phenomenologically small values. To elegantly bypass this and accommodate the self-interacting DM scenario, we specifically require a sufficiently large mass splitting ($(\delta m/m_-) x_f \gg 1$) to exponentially suppress the primordial $\chi_+$ abundance. With the $\chi_+$ population heavily depleted, this unsuppressed $s$-wave channel is effectively neutralized, and the freeze-out is robustly driven by the $p$-wave $\chi_- \chi_- \to SS$ annihilation. Consequently, a much larger Yukawa coupling $f$ is mathematically necessitated, which naturally provides the massive self-interaction cross sections required to resolve small-scale structure anomalies.

We integrate Eq.~(\ref{eq:boltz-approx}) from freeze-out $x_f$ to $x_0$, with yields $Y_f$ and $Y_0$, respectively, considering that $Y^{\rm eq}$ is insignificant compared to $Y$. 
Here $x_0 = m_-/T_0$, with $T_0$ indicating the photon bath temperature below which DM self-interactions become too weak to maintain thermal equilibrium among themselves. The potential corrections arising from the Sommerfeld effect will be addressed separately in Sec.~\ref{sec:sommerfeld-relic}. For $x >x_0$, the DM particles become free-streaming and non-thermal. However, they still follow a Maxwell-Boltzmann distribution, $f_{\rm MB} \propto \exp[ - p^2/(2m_{\rm DM} T_{\rm DM})]$, because the momentum redshifts as $p \propto a^{-1}$, and the DM temperature scales as $T_{\rm DM}=T^2/T_{kd} \propto a^{-2}$.

 Consequently, we get
\begin{align}
-\int^{Y_{0}}_{Y_f}  \frac{d Y}{Y^2}
 & \approx  m_- \bigg( \frac{\pi}{45 G} \bigg)^{1/2} 
  \Bigg( \int^{x_{kd}}_{x_f}   g_*^{1/2}  \langle \sigma_{\rm eff} v\rangle  \frac{dx}{ x^2}
  +   \int^{x_0}_{x_{kd}}   g_*^{1/2}  \langle \sigma_{\rm eff} v\rangle  \frac{dx}{ x^2} 
\Bigg) \,, 
\label{eq:yield-1}
\end{align}
where $x_{kd} (=m_-/ T_{kd})$ represents the value of $x$ at the temperature when DM kinetically decouples from the bath. 
Between $x_f$ and $x_{kd}$, DM particles remain in kinetic equilibrium with the SM bath through elastic collisions, especially $\chi_- \, S \leftrightarrow \chi_- \, S$ even after freeze-out (refer to Appendix~\ref{app:chiS-el} for a detailed discussion); in other words, we have $T_{\rm DM} = T$. Both $S$ and $ Z^\prime$ stay in thermal equilibrium with the $(\nu_\mu, \nu_\tau)$ bath through interactions such as $S \leftrightarrow  Z^\prime  Z^\prime$ and $ Z^\prime \leftrightarrow \nu_\mu \bar\nu_\mu, \nu_\tau \bar\nu_\tau$. Additionally, as explained in Sec.~\ref{sec:neff} and shown in Fig.~\ref{fig:temp-ratio}, the process $ Z^\prime \leftrightarrow e^+ e^-$ maintains $T = T_{\nu_\mu}$ until the temperature drops below $m_e$. 
It is observed that $\chi_-$, with a mass ranging from a few GeV to several hundred GeV, kinetically decouples from the scalar field $S$ at a temperature ($T_{kd}$) roughly between 210 and 250 MeV. All relevant details are available in Appendix~\ref{app:chiS-el}. When $T< T_{kd}$, the dark matter's temperature $T_{\rm DM}$ instead decreases as $a^{-2} \propto T^{2}$, i.e., $T_{\rm DM}=T^2/T_{kd}$.

Since the annihilation rate of dark matter is nearly negligible at $x=x_0$, we can use $Y_0$ as its current value. The left side of Eq.~(\ref{eq:yield-1}) is roughly $Y_0^{-1}$ because $Y_f \gg Y_0$, while the integral on the right side is mainly influenced by the region near $x_f$, making it insensitive to the value of $x_0$.
The DM relic abundance \cite{pdg2022} is related to $Y_0$ with
\begin{align}
\Omega_{\rm DM} \simeq  \frac{m_- Y_0  s_0}{\rho_c} \simeq 0.1200/h^2 \,,
\label{eq:relic}
\end{align}
where $\rho_c = 1.0537\times 10^{-5} h^2 ({\rm GeV}/c^2) {\rm cm}^{-3}$ is the critical density with $h\simeq 0.674$, and $s_0=2891.2~\text{cm}^{-3}$ is the entropy density today.  
\begin{figure}[t!]
\begin{center}
\includegraphics[width=0.41\textwidth]{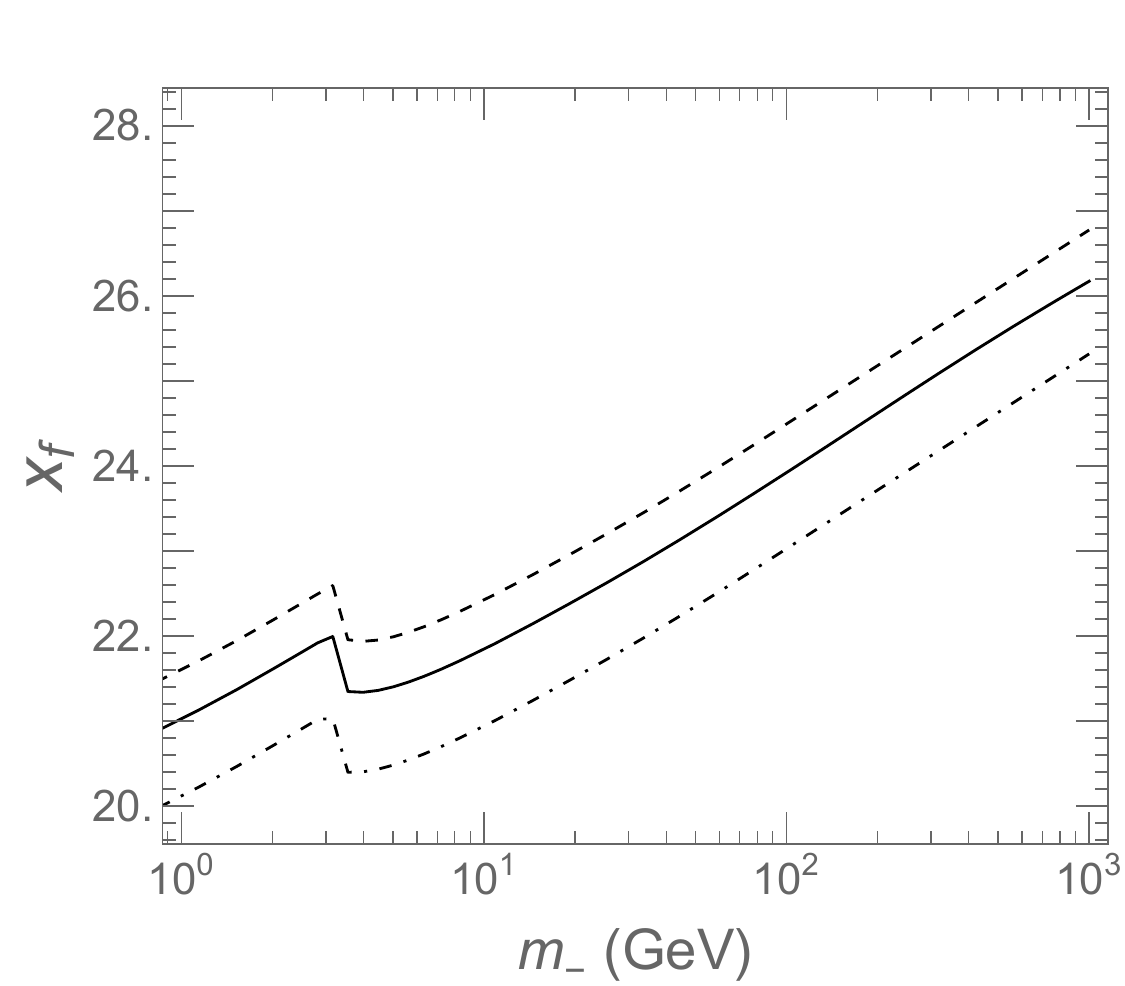}\hskip0.6cm
\includegraphics[width=0.415\textwidth]{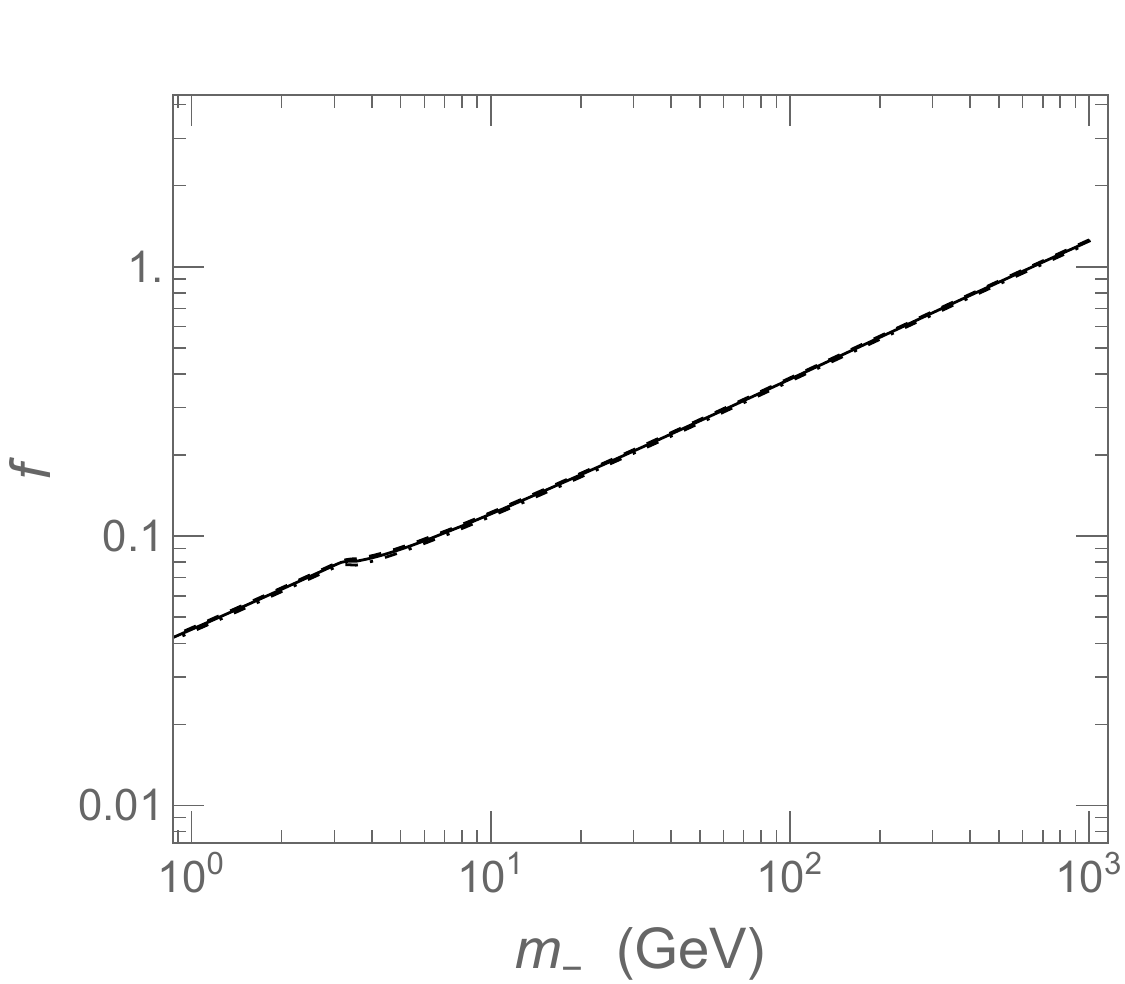}
\caption{Left panel: $x_f$ as a function of $m_-$. Right panel: The coupling constant $f$ that can produce the correct DM relic density as a function of $m_-$.  The solid curve corresponds to  $\delta=1$, while the dashed and dotdashed curves are for $\delta =1.5$ and $0.5$, respectively. The unsmooth part of the curves is due to the QCD phase transition at $T_{\rm QCD}=150$~MeV.
}
\label{fig:xf-xs-mx}
\end{center}
\end{figure}
From Eqs.~(\ref{eq:xf}) and (\ref{eq:yield-1}), we numerically compute $x_f$ and the coupling constant $f$ as functions of $m_-$ using the DM relic abundance detailed in Eq.~(\ref{eq:relic}). We examine the scenario where $m_{-}\gg m_{Z^\prime}$ and the scalar $S$ is located in the range of $12~\text{MeV} \lesssim m_S \lesssim 400~\text{MeV}$, while also fulfilling the condition $m_{S}\gtrsim 2 m_{Z^\prime}$. The outcomes are illustrated in Fig.~\ref{fig:xf-xs-mx} for $m_- \in (1, 1000)~\text{GeV}$. For a comparative analysis, when $\delta$ is $\pm 0.5$, we find that the uncertainty in $f$ is roughly less than 1\%, meaning the corresponding error of $\langle \sigma v\rangle_{\chi_- \chi_- \to SS} $ is below 5\%. 

In our analysis, the coupling $f$ is fixed by the relic density requirement, $\Omega_{\rm DM} h^2 \simeq 0.12$. Since the annihilation cross section scales as $\langle\sigma_{\rm eff} v\rangle \propto f^4/m_-^2$, a constant relic density implies $f \propto m_-^{1/2}$. Using the relation in Eq.~(\ref{eq:gchi_relation}), the derived dark gauge coupling behaves as $g_\chi \propto m_{Z^\prime} m_-^{-1/2}$.

\begin{figure}[t!]
\begin{center}
\includegraphics[width=0.41\textwidth]{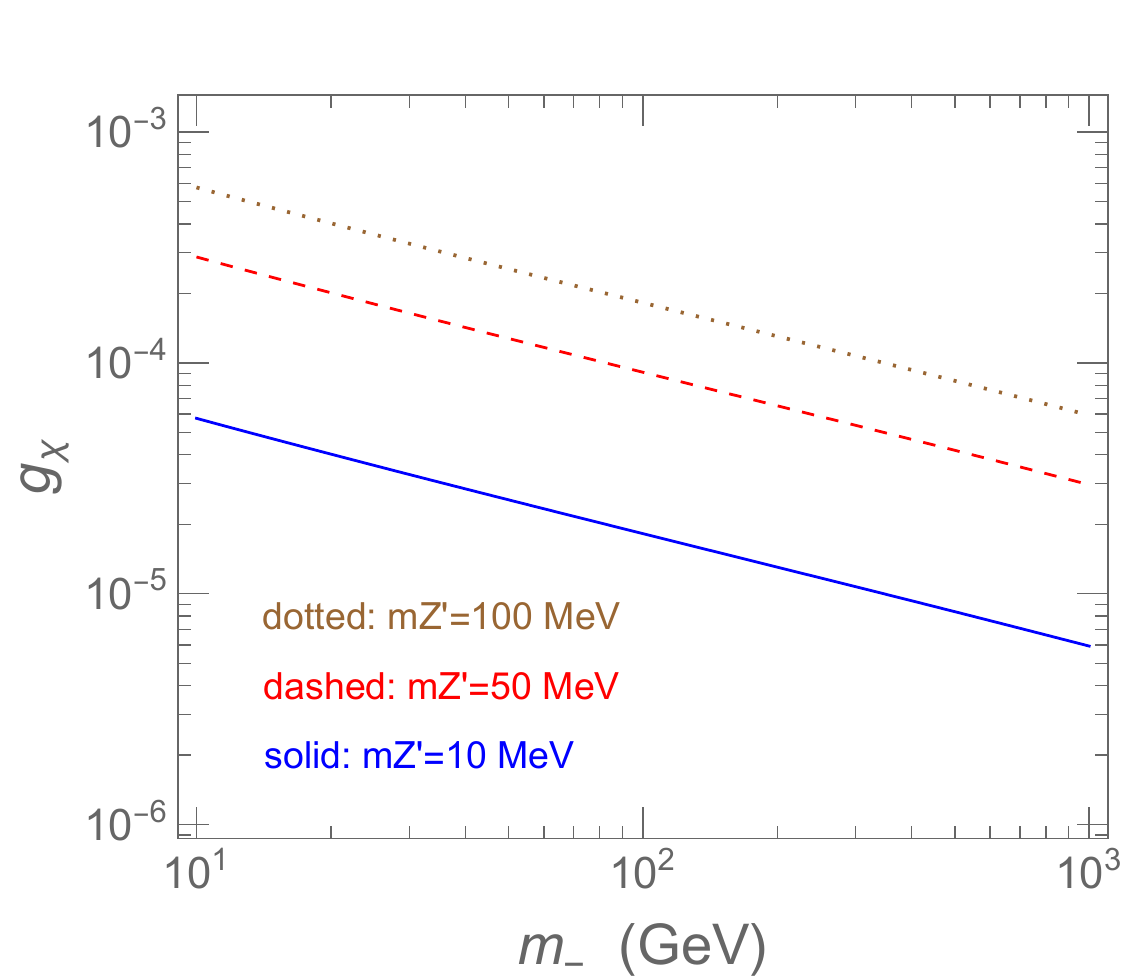}\hskip0.6cm
\includegraphics[width=0.41\textwidth]{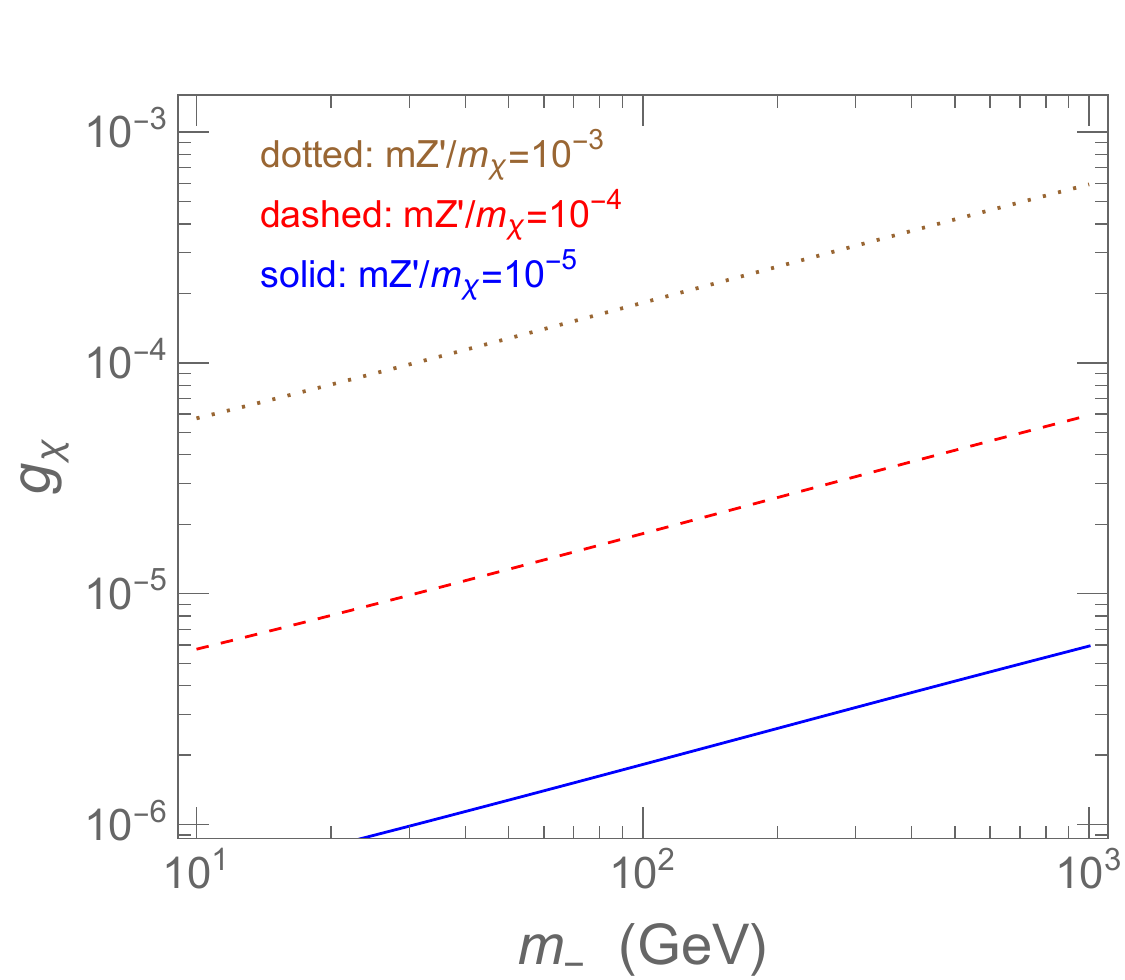}
\caption{
The dynamically derived dark gauge coupling $g_\chi = f m_{Z^\prime} / [(2+\kappa)m_-]$ as a function of the DM mass $m_-$. 
To satisfy the macroscopic mass gap condition uniformly across the mass range, we set the mass splitting as $\delta m = \kappa m_-$ with $\kappa = 0.1$.
The \textbf{left panel} shows the results for fixed gauge boson masses $m_{Z^\prime} = 10$ (solid blue), $50$ (dashed red), and $100$~MeV (dotted brown). 
The \textbf{right panel} assumes fixed mass ratios of $m_{Z^\prime}/m_- = 10^{-5}$ (solid blue), $10^{-4}$ (dashed red), and $10^{-3}$ (dotted brown). 
In both cases, the Yukawa coupling $f$ is determined by the relic density constraint (with $\delta=1$) as shown in Fig.~\ref{fig:xf-xs-mx}.}
\label{fig:g_chi_m_minus}
\end{center}
\end{figure}
In Fig.~\ref{fig:g_chi_m_minus}, we present the numerical results for the dynamically derived $g_\chi$ as a function of the DM mass $m_-$. To strictly satisfy the thermodynamical decoupling condition $(\delta m / m_-) x_f \gg 1$ without fine-tuning, we parameterize the splitting as a fixed subdominant fraction of the dark matter mass, setting $\delta m = \kappa m_-$ with $\kappa = 0.1$. Consequently, the exact relation simplifies to $g_\chi = f m_{Z^\prime} / [(2+\kappa)m_-]$.
The left panel shows the case for fixed gauge boson masses $m_{Z^\prime} = 10, 50, 100$ MeV. Here, $g_\chi$ decreases as $m_-$ increases. By maintaining a constant macroscopic ratio $\kappa = \delta m / m_-$, the exact gauge coupling intrinsically strictly follows the $m_-^{-1/2}$ scaling. It also shows that for a lighter $Z^\prime$, a smaller gauge coupling is required.
The right panel illustrates the case where the mass ratio $m_{Z^\prime}/m_-$ is kept constant. In this scenario, substituting the constant ratio $m_{Z^\prime}/m_-$ into the exact relation reveals that $g_\chi \propto f / (2+\kappa)$, and thus strictly grows as $m_-^{1/2}$.
The results show that for the parameter space of interest, the derived gauge coupling $g_\chi$ generally remains small ($g_\chi \lesssim 10^{-3}$), while the Yukawa coupling $f$ is adjusted to satisfy the observed relic density.

\section{Direct detection constraints}\label{sec:dd-expt}

With the dark matter relic density and the required Yukawa coupling $f$ dynamically determined in Sec.~\ref{sec:boltz}, we now evaluate the constraints from direct detection experiments. In our model, direct detection imposes stringent limits on the mixing angle $\alpha$, which can easily surpass the aforementioned beam-dump bounds. Here, we focus on coherent elastic scattering.

In our model, direct detection imposes stringent limits on the mixing angle $\alpha$, which can surpass beam-dump bounds. 
In this section, we focus on the coherent elastic scattering between dark matter and target nuclei. This process is dominated by the $t$-channel exchange of the light scalar mediator $S$. We defer the discussion of inelastic $Z^\prime$ interactions to Sec.~\ref{sec:discussions}.

The differential event rate can be expressed as
\begin{equation}
\label{eq:Rate-1}
\frac{\mathrm{d} R_T}{\mathrm{d} E_R} = N_T \frac{\rho_\odot}{m_-} \frac{f^2}{32 \pi} \frac{m_T}{v_H^2} 
\frac{  m_N^2 \sin^2 2\alpha}{(\vec{q}^2 + m_S^2)^2} \sum_{N, N' = p, n} f_N f_{N'} 
\int_{v_{\text{min}}(E_R)} \mathrm{d}^3 v \, \frac{1}{v} f_\oplus(\vec{v},t) F_{M}^{(N, N')}(y, T) \ , 
\end{equation}
where $E_R$ is the nuclear recoil energy, $N_T$ is the number of target atoms per unit mass, and $m_T$ is the mass of the target nucleus $T$. The momentum transferred during the collision is given by $|\vec{q}| = \sqrt{2m_T E_R}$. To accurately model energy-dependent effects during experimental fitting, this relationship must be explicitly substituted into the propagator denominator. Furthermore, $m_N$ represents the nucleon mass, $v_H$ is the Higgs vacuum expectation value, and $f_N \simeq 0.3$ is the effective nucleon coupling \cite{Cirelli:2024ssz}. The nuclear form factors $F_{M}^{(N,N')}(y, T)$, which account for the structure of the nuclei \cite{Anand:2013yka}, are functions of $y = (|\vec{q}|b/2)^2$, with $b \simeq \sqrt{41.467 / (45A^{-1/3} - 25A^{-2/3})}$ being the harmonic oscillator parameter for a nucleus with mass number $A$. Finally, we adopt a local dark matter density of $\rho_\odot \simeq 0.35 \text{ GeV/cm}^3$ in our analysis, and $f_\oplus(\vec{v},t)$ denotes the dark matter velocity distribution in the Earth's rest frame (see Ref.~\cite{Yang:2017zor} for details).

The goodness-of-fit test statistic (TS) follows an asymptotic $\chi^2$ distribution and can be expressed as \cite{pdg2022}
\begin{equation}\label{eq:TS-dd}
{\rm TS} (\alpha, m_S; m_-) = 2 \left[ n^{\rm th}(\alpha, m_S; m_-)  + n^b - \hat{n}  - n^{\text{obs}}  \ln \left( \frac{n^\text{th}(\alpha, m_S; m_-) + n^{\rm b}}{\hat{n}} \right)   \right] \,,
\end{equation}
where $n^{\rm th}$ and $n^b$ denote the theoretically predicted and expected background event numbers, respectively; $n^{\rm obs}$ is the observed number of events; and $\hat{n}$ is the maximum likelihood estimate (MLE) of the total expected events $n = n^{\rm th} + n^b$. Consequently, setting $\text{TS} = 2.71$ yields a one-sided 95\% confidence level (C.L.) upper limit, allowing us to derive exclusion contours in the $(m_S, \alpha)$ parameter space for a fixed dark matter mass $m_-$. Furthermore, the theoretically expected number of events at the LZ experiment is calculated as
\begin{equation}
n^{\text{th}}= \mathcal{E} \int_0^\infty dE_R \sum_T \epsilon_T (E_R) \frac{dR_T}{dE_R} \,, \nonumber\\
\end{equation}
where $\mathcal{E}$ represents the total experimental exposure, and $\epsilon_T(E_R)$ accounts for the detection efficiency and signal acceptance for a target nucleus $T$ with recoil energy $E_R$.

Figure~\ref{fig:alpha-cons} displays these stringent upper bounds on $\alpha$ as a function of $m_S$, based on the recent results from the LUX-ZEPLIN (LZ) experiment (LZ 2025) \cite{LZ:2024zvo}. While results from PandaX-4T \cite{PandaX-4T:2021bab} and XENONnT \cite{XENON:2023cxc} provide competitive but slightly less stringent constraints, LZ 2025 currently serves as the leading representative benchmark. 
Since the $10$--$100 \text{ GeV}$ mass range is favored by small-scale anomalies (see Sec.~\ref{sec:self-scatterings}), we select $m_- = 10 \text{ GeV}$ and $m_- \ge 100 \text{ GeV}$ as benchmarks. For $m_- \gtrsim 100 \text{ GeV}$, the thermal relic density requirement $f \propto m_-^{1/2}$ makes the limits on the mixing angle $\alpha$ essentially DM mass-independent. As shown, LZ 2025 strictly surpasses colliders and fixed-target experiments across the sub-GeV region, bounding $\alpha \lesssim 2\times 10^{-8}$ for $m_S = 50 \text{ MeV}$ in the heavy mass regime. Future LZ upgrade projections \cite{LZ:2018qzl} are shown as dashed lines.

\section{Excited state fraction and cosmological safety}\label{sec:mass-gap}

For the $p$-wave dominant Majorana DM, the mass splitting $\delta m$ between the excited state $\chi_+$ and the ground state $\chi_-$ plays a crucial role in dictating the cosmological history. As established in Sec.~\ref{sec:boltz}, shutting down the $s$-wave coannihilation channel ($\chi_- \chi_+ \to Z^\prime S$) to preserve a maximally large Yukawa coupling $f$ necessitates a sufficiently large mass splitting such that $(\delta m / m_-) x_f \gg 1$.

Under this thermodynamical condition, the heavier state $\chi_+$ is subject to a severe exponential Boltzmann suppression during the freeze-out epoch. Consequently, the comoving number density of the excited state is entirely negligible compared to the ground state ($n_{\chi+} \ll n_{\chi-}$). The primordial fraction of the excited state, defined as $r \equiv n_{\chi_+}/(n_{\chi_+} + n_{\chi_-})$, inherently approaches zero right from the early Universe.

This profound thermodynamical depletion leads to an exceptionally clean cosmological history. Since the dark matter relic is virtually 100\% composed of the ground state $\chi_-$ immediately after freeze-out, the subsequent late-time decay process ($\chi_+ \rightarrow \chi_- \nu_{\mu,\tau} \overline{\nu}_{\mu,\tau}$) is rendered cosmologically inconsequential. We therefore naturally circumvent any stringent constraints from Big Bang Nucleosynthesis (BBN) \cite{Kawasaki:2000en} or the effective number of relativistic species ($N_{\rm eff}$) in the Cosmic Microwave Background (CMB) that typically restrict models with late-decaying dark sector particles.

In the present-day Universe, the dark matter halo consists purely of $\chi_-$. As we will explore in Sec.~\ref{sec:self-scatterings}, this pure $\chi_-$ composition ($r \to 0$) simplifies the evaluation of the velocity-dependent self-interactions, which are dynamically driven by the $p$-wave resonant scattering of the ground state to resolve the small-scale structure anomalies.

\section{Sommerfeld effects on thermal relic abundance and dark matter annihilation}\label{sec:sommerfeld-relic}

The Sommerfeld enhancement can significantly affect dark matter annihilation rates, potentially altering both the thermal relic density and indirect detection signals. Because dark matter particles can exchange a ladder of light scalar mediators ($S$) with a relatively large Yukawa coupling $f$ prior to annihilation, the cross section can be substantially boosted. Here, we evaluate the impact of these non-perturbative effects on the relic density and indirect detection individually.

\subsection{Dark matter relic density}\label{subsec:sommerfeld-relic}

In Sec.~\ref{sec:boltz}, we estimated the dark matter yield $Y$ neglecting Sommerfeld effects. However, as the dark matter particles become non-relativistic, the exchange of light scalars may induce a Sommerfeld enhancement. The thermal average of the annihilation cross-section is given by
\begin{align}
 \langle \sigma v\rangle_{\chi_-\chi_-}
 = \int S_1 \cdot (\sigma v)_{\chi_-\chi_-} f_r (v) dv \,,
 \label{eq:thermal-xs-sommer}
\end{align}
where, from Appendix~\ref{appsec:xs}, the tree-level $p$-wave annihilation cross section for the pure ground state ($\chi_- \chi_- \to SS, Z^\prime Z^\prime$) is given by
 \begin{align}
(\sigma v)_{\chi_-\chi_-}  \simeq 
\frac{9.9 f^4}{ 384 \pi m_-^2} v^2  \,.
\label{eq:xs-rel-v}
\end{align}
Here, we fix the parameter $w=0.9$ as defined in Eq.~(\ref{eq:zpzp}), noting that our subsequent analysis remains insensitive to its precise value. Additionally, $S_1$ denotes the Sommerfeld enhancement factor associated with the $\ell=1$ ($p$-wave) partial wave. 
The relative velocity distribution of the dark matter particles in the center-of-mass frame, $f_r(v)$, follows the Maxwell-Boltzmann distribution
\begin{align}
f_r(v) = \sqrt{\frac{2}{\pi}} \frac{v^2}{v_0^3} e^{-\frac{v^2} {2 v_0^2}}
      = \sqrt{\frac{2}{\pi}} \left(\frac{x_-}{2}\right)^{3/2} v^2 e^{-\frac{x_- v^2} {4}} \,,
\label{eq:dist-rel-v}
\end{align}
where $v_0 =\sqrt{ 2T_{\rm DM} /m_-} $ is the most probable velocity and $x_- =m_-/ T_{\rm DM}$. While the $p$-wave cross section is velocity-suppressed ($\sigma v \propto v^2$) at low temperatures, the Sommerfeld factor $S_1$ could partially lift this suppression, potentially requiring a readjustment of the coupling $f$ to satisfy the observed relic density.

Summing the ladder diagrams in the calculation increases the square of the scattering amplitude by a so-called Sommerfeld factor.
Considering the DM annihilation process in the center-of-mass frame, the initial DM particles are $\vec{p}$ and $ -\vec{p}$, while the final $S$ particles possess momenta of $\vec{p'}$ and $ -\vec{p'}$.
Generally, a full amplitude, $A$, that includes the Sommerfeld effect, is represented by an integral involving the wave function $\phi^*_{\vec{p}} (\vec{r})$ and the bare scattering amplitude, $A_0$ \cite{Iengo:2009ni, Iengo:2009xf},
\begin{align}
A (\vec{p}, \vec{p'} ) =
\int d^3 r \,  \phi^*_{\vec{p}} (\vec{r})  \int \frac{d^3 q}{ (2\pi)^3 }  e^{i \vec{q}\cdot \vec{r}}A_0 (\vec{q}, \vec{p'} ) \,,
\label{eq:sommer-amp}
\end{align}
where we can further expand $A, A_0$, and $\phi^*_{\vec{p}} (\vec{r})$ in partial waves,
\begin{align}
A (\vec{p}, \vec{p'} )  &=   \sum_\ell A_\ell (\vec{p}, \vec{p'} ) = 
   \sum_\ell p^\ell p^{\prime\ell} P_\ell (\hat{p} \cdot \hat{p'}) a_\ell\,, \\
A_0 (\vec{q}, \vec{p'} )  &=  \sum_\ell A_0 (\vec{q}, \vec{p'} ) =
   \sum_\ell q^\ell p^{\prime\ell} P_\ell (\hat{q} \cdot \hat{p'})  a_{0,\ell}\,, \\
\phi^*_{\vec{p}} (\vec{r}) &= \frac{ 1}{ p} \sum_\ell i^\ell (2\ell+1) e^{i \delta_\ell} P_\ell (\hat{p}\cdot \hat{r}) \frac{\chi_{p,\ell}(\alpha_f m_- r) }{r \, C_\ell} \,,
\label{eq:sommer-amp-fns}
\end{align}
with $\alpha_f = f^2/(4\pi)$ and the symbols with hats denoting the corresponding unit vectors.
Here, the wave function $\chi_{p,\ell}$ satisfies the differential equation,
\begin{equation}
\frac{1}{m_-}\frac{d^2}{d r^2} \chi_{p,\ell} -\frac{\ell (\ell+1)}{m_- r^2} \chi_{p, \ell} + \frac{p^2}{m_-} \chi_{p, \ell} - V(r) \chi_{p, \ell}=0 \,,
\label{eq:sommer-diff-o}
\end{equation}
where in the coordinate space the DM interaction through the exchange of $S$ can be described by the Yukawa potential,
 \begin{equation}
 V(r)=-\frac{\alpha_f e^{-m_S r}}{r} .\\
 \label{eq:yukawa-1}
 \end{equation}
 Using the dimensionless variable $x= \alpha_f m_- r$, we can express Eq.~(\ref{eq:sommer-diff-o}) as
\begin{equation}
\frac{d^2}{d x^2} \chi_{p,\ell} -\frac{\ell (\ell+1)}{x^2} \chi_{p, \ell} + a^2 \chi_{p, \ell} + \frac{e^{-x/b}}{x} \chi_{p, \ell}=0 \,,
\label{eq:sommer-diff}
\end{equation}
where 
\begin{align}
a = \frac{v}{2 \alpha_f} \,, \quad  b=\frac{\alpha_f m_-}{m_S} \,.
\label{eq:parameter-ab}
\end{align}
Following a similar derivation as in Refs.~\cite{Iengo:2009ni, Iengo:2009xf}, we obtain the $\ell$-wave partial amplitude from Eq.~(\ref{eq:sommer-amp}), 
\begin{equation}
A_\ell (\vec{p}, \vec{p'} ) \simeq \sqrt{S_\ell} \,  A_0 (\vec{p}, \vec{p'} ) \,,
\label{eq:sommer-amp-1}
\end{equation}
where the Sommerfeld enhancement is given by
\begin{align}
S_\ell = \left|  \frac{(2\ell+1)!!}{C_\ell} \left( \frac{2\alpha_f}{v} \right)^{\ell +1}  \right|^2 \,.
\label{eq:sommer-sl}
\end{align}
The constant $C_\ell$ can be determined by
\begin{align}
C_\ell^2 = [\chi_{p, \ell} (x)^2 + \chi_{p, \ell}(x-\pi/2) ^2]  |_{x\to \infty} \,.
\label{eq:sommer-cl}
\end{align}
In this paper, we consider that the DM relic density is dominated by the $p$-wave annihilations (see Eqs.~(\ref{eq:ss})-(\ref{eq:zpzp})). Thus, the relevant Sommerfeld factor is given by 
\begin{align}
S_1 = \left|  \frac{3}{C_1} \left( \frac{2\alpha_f}{v} \right)^{2}  \right|^2 \,.
\label{eq:sommer-cl-1}
\end{align}

In Sec.~\ref{sec:boltz}, we neglected the Sommerfeld effect during freeze-out as it is only significant at low velocities. However, resonant enhancement can potentially lead to a ``reannihilation'' catastrophe post-kinetic-decoupling ($x > x_{kd}$), where the late-time integration measure diverges \cite{Binder:2017lkj}. To verify our $p$-wave dominated model's immunity to this catastrophe, we analyze the integrated effect post-kinetic-decoupling. Throughout the post-decoupling era---encompassing both the briefly self-thermalized and subsequent free-streaming phases---the DM temperature scales as $T_{\rm DM} = T^2 / T_{kd} \propto x^{-2}$, strictly yielding $\langle v^2 \rangle \propto x^{-2}$.

By examining the asymptotic limits of the Sommerfeld factor $S_1$ via the Hulth\'{e}n approximation \cite{Kahlhoefer:2017umn}, we identify three kinematic regimes as follows: 
(i) A high-velocity unenhanced regime ($v \gg \alpha_f$) yields $S_1 \simeq 1$ and $\langle \sigma v \rangle_{\chi_-\chi_-} \propto v^2$. 
(ii) A narrow Coulomb regime ($m_S/m_- < v < \alpha_f$), which is highly restricted or non-existent due to our relic density requirement $f \propto m_-^{1/2}$, gives $S_1 \sim v^{-3}$ and $\langle \sigma v \rangle_{\chi_-\chi_-} \propto S_1 v^2 \sim v^{-1} \sim x$. This results in a logarithmically divergent integral, $\int x \cdot x^{-2} dx \sim \ln(x)$. Assuming $m_S=60$~MeV, for instance, this regime ceases to exist entirely for $m_- \lesssim 16$ GeV, while for a heavy mass like $m_- = 300$ GeV, it is constrained within the narrow window $0.0002 < v < 0.0437$. 
(iii) In the deep saturation regime ($v < m_S/m_-$), the finite mass of the mediator fundamentally alters the long-range nature of the potential. For generic parameter space (away from exact resonances), this finite force range truncates the enhancement, causing $S_1$ to strictly saturate to a constant limit \cite{Kahlhoefer:2017umn}. Consequently, $\langle \sigma v \rangle_{\chi_-\chi_-} \propto v^2 \sim x^{-2}$, making the integration measure $\int x^{-2} \cdot x^{-2} dx = \int x^{-4} dx \sim -x^{-3}$ absolutely convergent. The only exception occurs strictly on a resonance pole (corresponding to a zero-energy bound state \cite{Kamada:2020buc, Kamada:2023iol}), where the quantum mechanical resonance overcomes the finite-range truncation, yielding $S_1 \sim v^{-4}$ and leading to a linear reannihilation divergence ($\int x^2 \cdot x^{-2} dx = \int dx \sim x$).
   
To explicitly quantify this and prove the absence of reannihilation, we define the integrated enhancement ratio $\mathcal{R}$ over the integration measure as:
\begin{align}
  \mathcal{R} \equiv \frac{\int^{x_{\rm max}}_{x_f}   g_*^{1/2}  \langle \sigma v\rangle_{\chi_-\chi_-,S}  \frac{dx}{ x^2} }
  {   \int^{x_{\rm max}}_{x_{f}}   g_*^{1/2}  \langle \sigma v\rangle_{\chi_-\chi_-,0}  \frac{dx}{ x^2} }
 \,, 
\label{eq:SEeffect}
\end{align}
where $\langle \sigma v\rangle_{\chi_-\chi_-,0}$ ($\langle \sigma v\rangle_{\chi_-\chi_-,S}$) denotes the cross section without (with) the Sommerfeld effect.

\begin{figure}[t!]
\begin{center}
\includegraphics[width=0.47\textwidth]{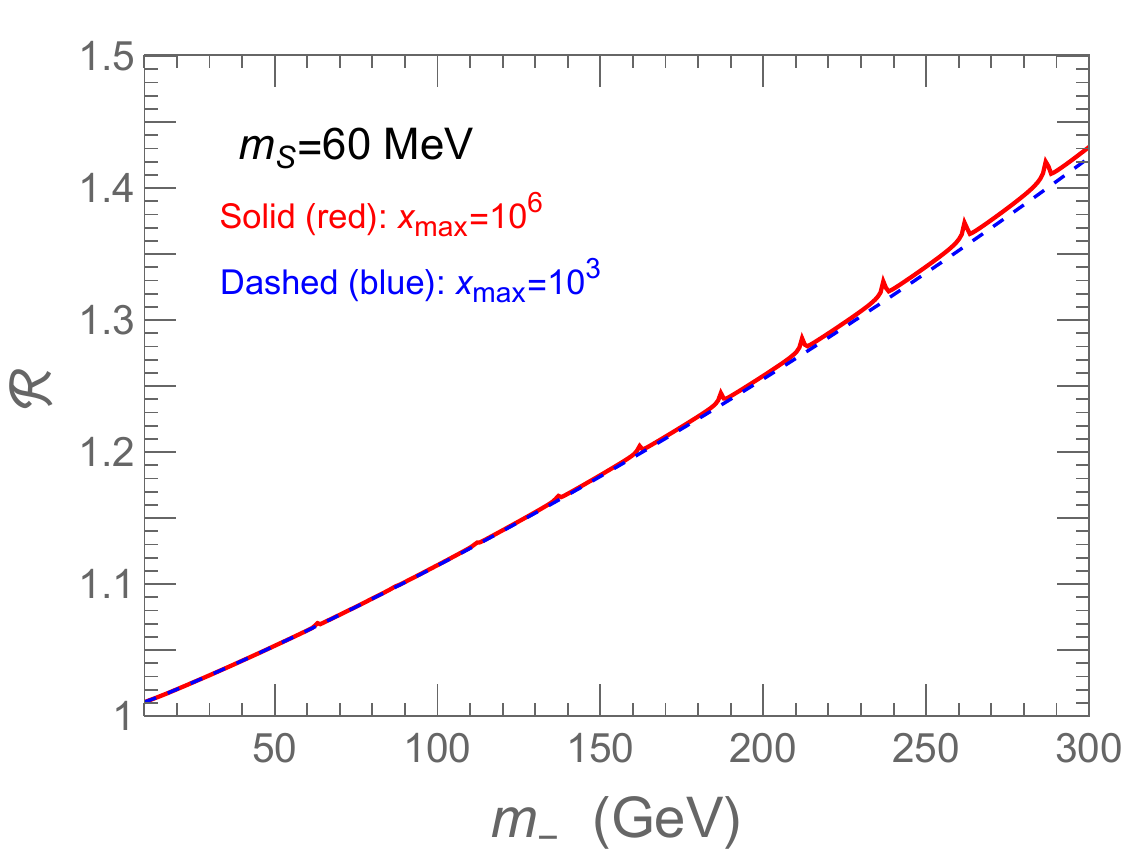}
\caption{The integrated Sommerfeld enhancement ratio $\mathcal{R}$ (Eq.~(\ref{eq:SEeffect})) as a function of the DM mass $m_-$, with $m_S = 60$ MeV. The perfect overlap of the $x_{\rm max} = 10^6$ (solid red) and $x_{\rm max} = 10^3$ (dashed blue) curves mathematically proves the absence of late-time reannihilation away from exact resonances. The baseline smoothly rises due to the Coulomb regime contribution, while narrow spikes isolate the zero-energy bound state resonances.
}
\label{fig:relic-ratio}
\end{center}
\end{figure}
In Fig.~\ref{fig:relic-ratio}, we plot $\mathcal{R}$ up to 300 GeV, amply covering the theoretically motivated region for addressing small-scale structure anomalies (see Sec.~\ref{sec:self-scatterings}). The perfect overlap of the $x_{\rm max} = 10^6$ and $10^3$ curves visually confirms our mathematical derivation: late-time reannihilation is strictly precluded in the generic parameter space. Crucially, the baseline enhancement inherently scales as $\mathcal{R} \propto \alpha_f$. For the dwarf-scale favored regime ($m_- \lesssim 100$ GeV), the overall Sommerfeld correction to the relic density remains minimal ($< 12\%$, i.e., $\mathcal{R} < 1.12$). Incorporating this integrated enhancement would strictly require a slight reduction of the coupling $f$ to maintain the exact relic density ($\alpha_{f, \rm true} \simeq \alpha_{f, \rm tree} / \mathcal{R}^{1/2}$). Because this required correction is negligible ($< 6\%$) for lighter masses, our tree-level-coupled self-interacting DM evaluations remain highly robust. For heavier masses ($m_- \gtrsim 300$ GeV), $\mathcal{R}$ grows moderately, dictating a smaller true $\alpha_f$. Consequently, the actual self-scattering cross sections at high masses are even further suppressed, rendering our subsequent predictions conservative overestimates that safely evade all massive cluster constraints. Finally, the distinct narrow spikes represent measure-zero zero-energy bound state resonances. Resolving small-scale anomalies only requires macroscopic proximity to these peaks (the ``shoulder''), inherently avoiding the exact pole and its associated reannihilation divergence.

\subsection{Indirect searches and constraints}\label{sec:sommerfeld-indirect}

\begin{figure}[t!]
\begin{center}
\includegraphics[width=0.41\textwidth]{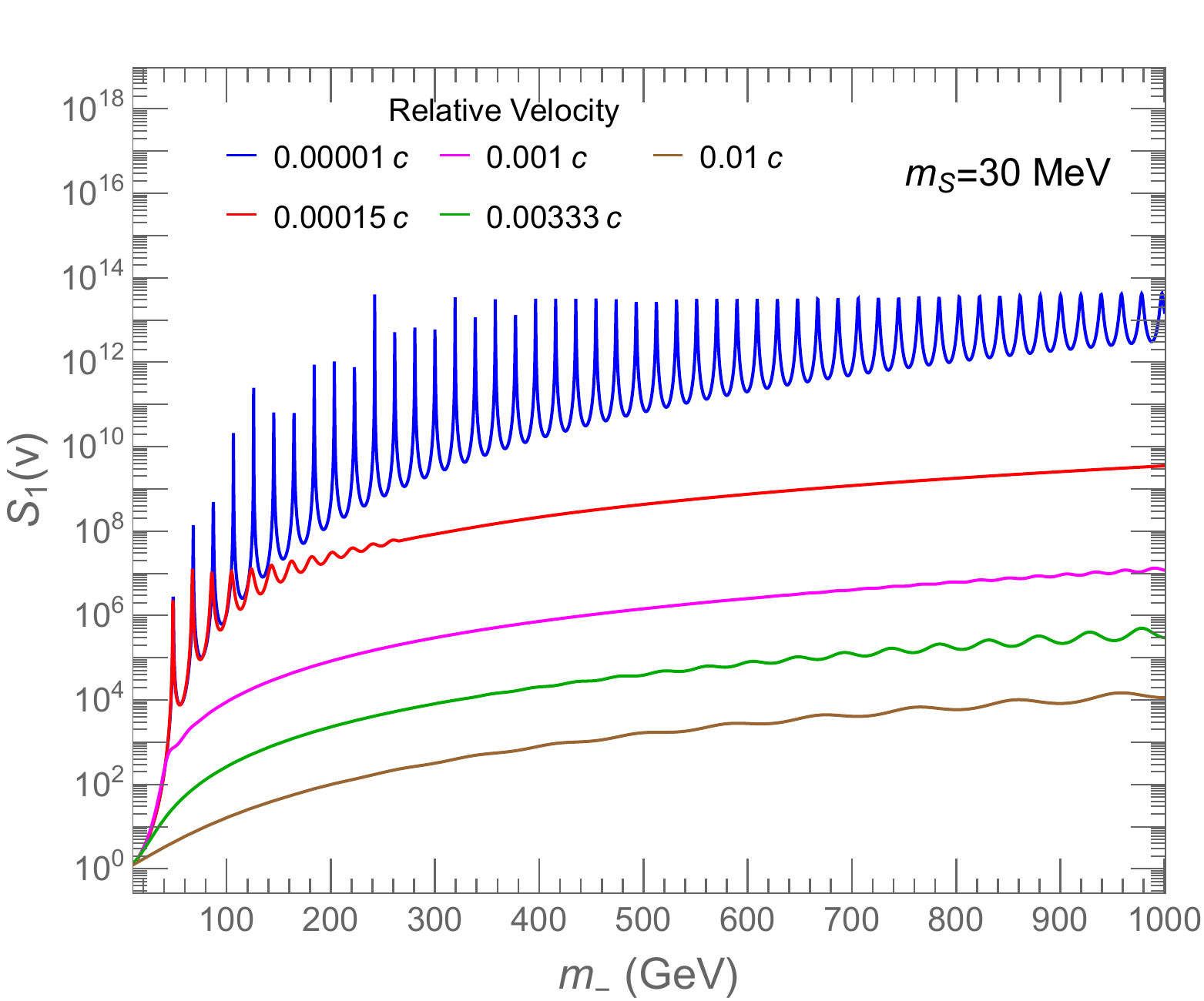}\hskip0.6cm
\includegraphics[width=0.41\textwidth]{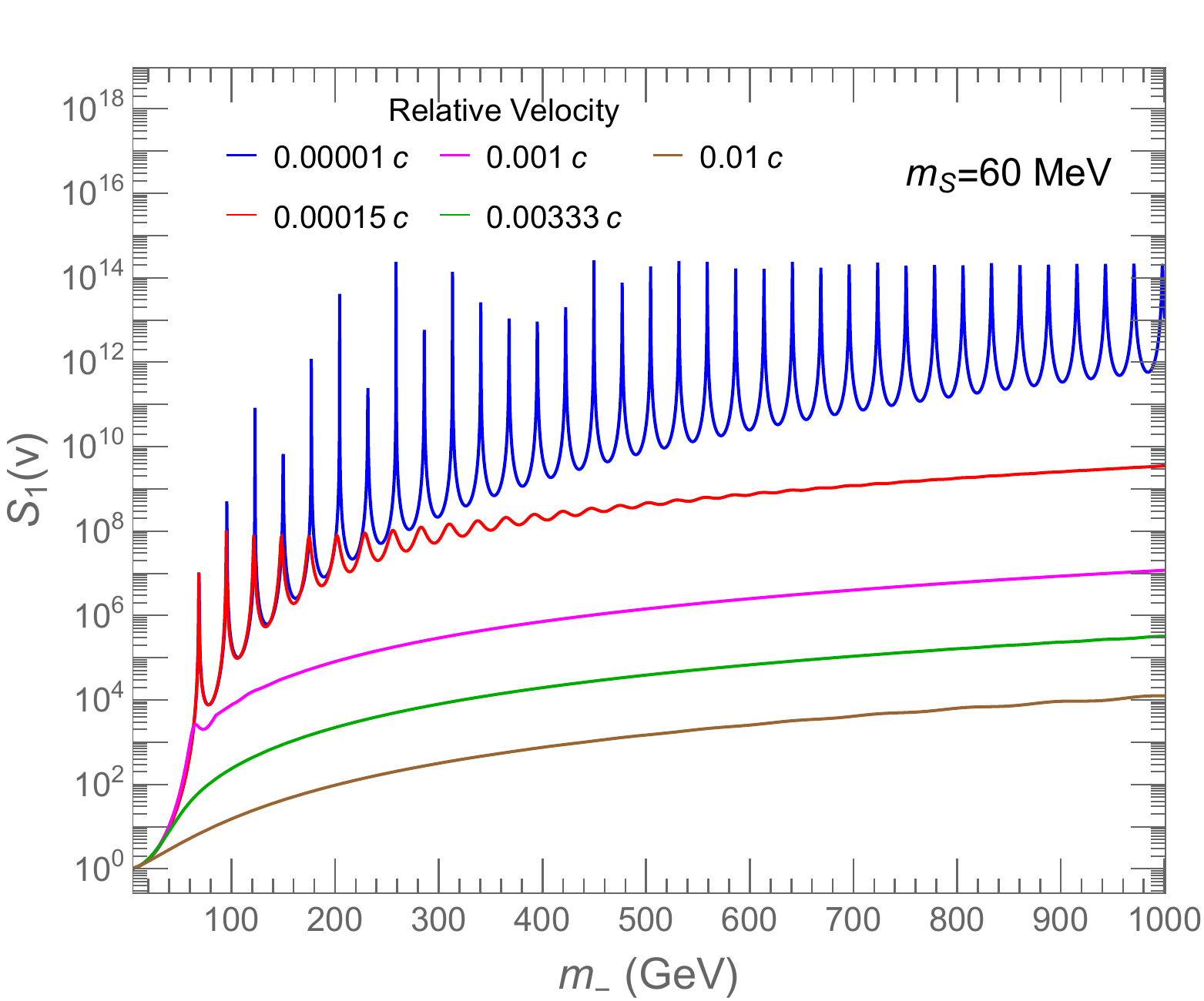}\hskip0.6cm\\
\includegraphics[width=0.41\textwidth]{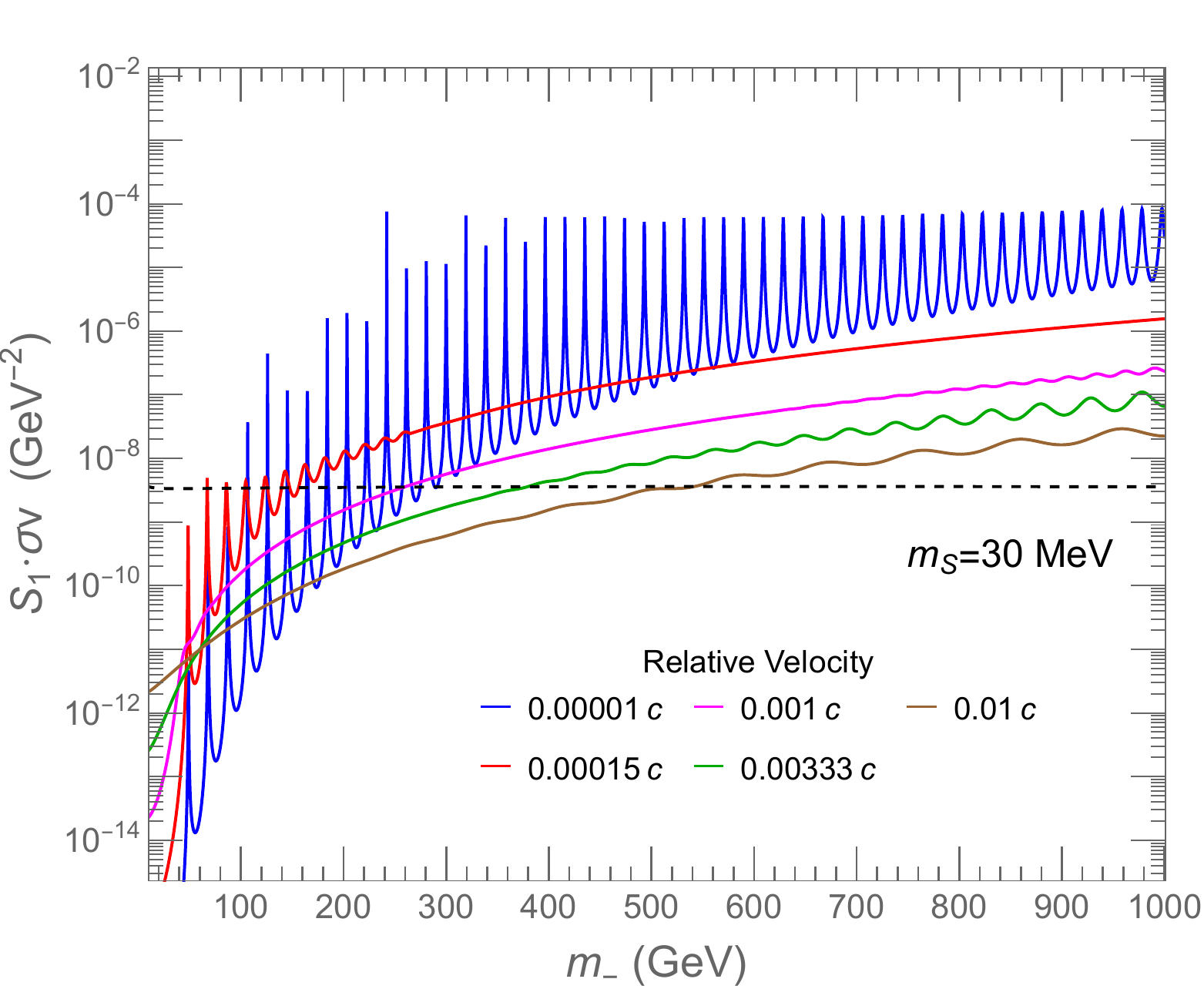}\hskip0.6cm
\includegraphics[width=0.41\textwidth]{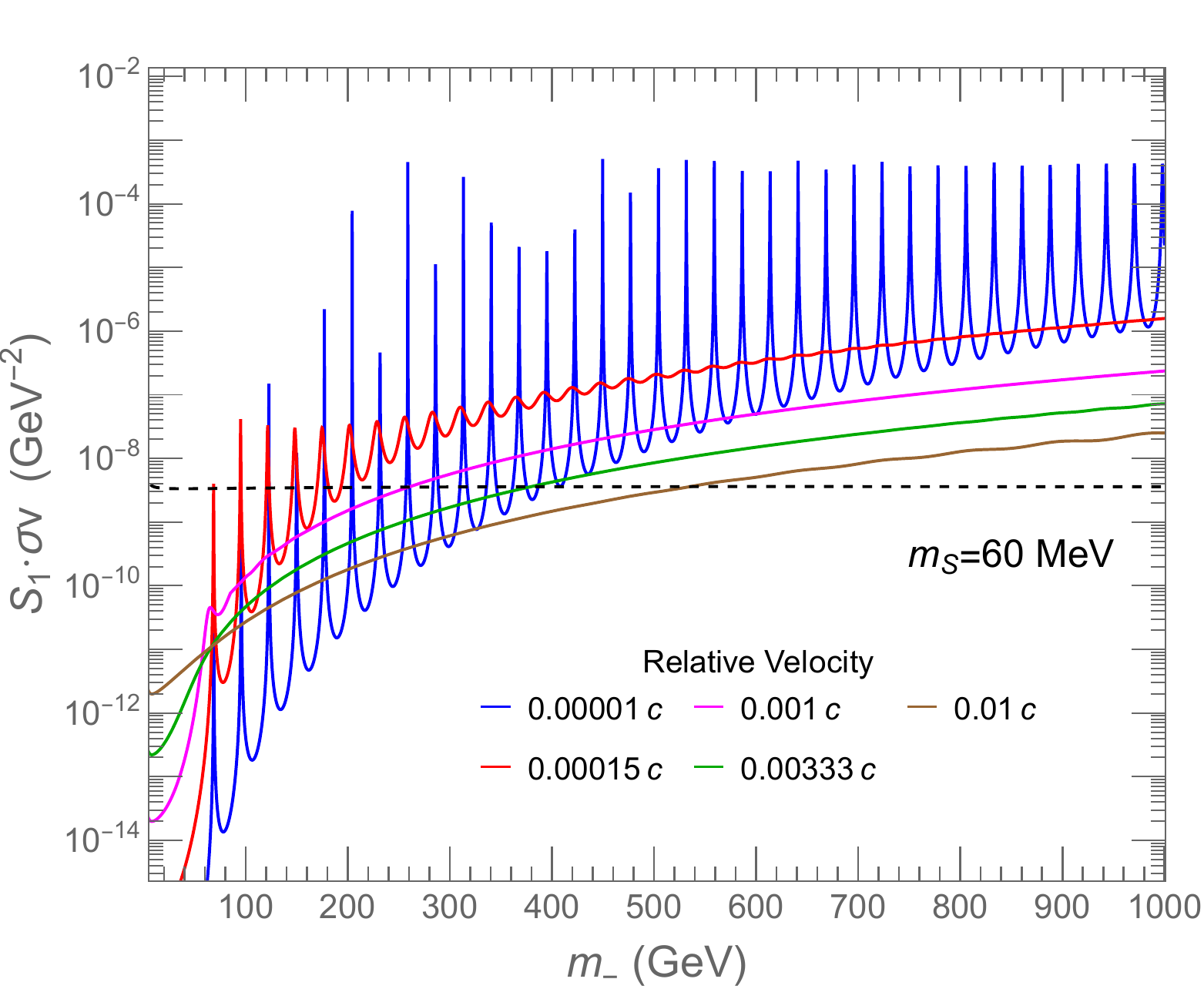}
\caption{ 
Top panels: The Sommerfeld enhancement factor as a function of $m_-$ for relative velocities $v/c = 0.00001, 0.00015, 0.001, 0.00333$, and $0.01$. The curves are ordered from top to bottom based on their peak values within the $m_- < 1000$~GeV range. Bottom panels: The corresponding annihilation cross sections for the same set of velocities. The black dashed line indicates the required canonical cross section $\langle \sigma v\rangle_{\chi_-\chi_-}$ at freeze-out for the thermal relic density. The mediator mass is $m_S = 30$~MeV (left) and 60~MeV (right).
}
\label{fig:s1-various-xs}
\end{center}
\end{figure}

Before dark matter annihilation, incident particles with low relative velocity exchange a ladder of light mediators (known as Sommerfeld enhancement effect), vastly boosting the annihilation rate. In Fig.~\ref{fig:s1-various-xs}, we display the $p$-wave Sommerfeld factor and the corresponding $\chi_-\chi_- \rightarrow SS, Z^\prime Z^\prime$ annihilation cross section for various non-relativistic relative velocities. Note that for a standard $s$-wave Majorana WIMP, the observed relic abundance typically requires $\langle \sigma v \rangle \sim 2 \times 10^{-26} \text{ cm}^3/\text{s}$. In this section, we evaluate the indirect detection signatures under the established premise that the present-day dark matter halo is composed entirely of the ground state $\chi_-$.

\begin{figure}[t!]
\begin{center}
\includegraphics[width=0.43\textwidth]{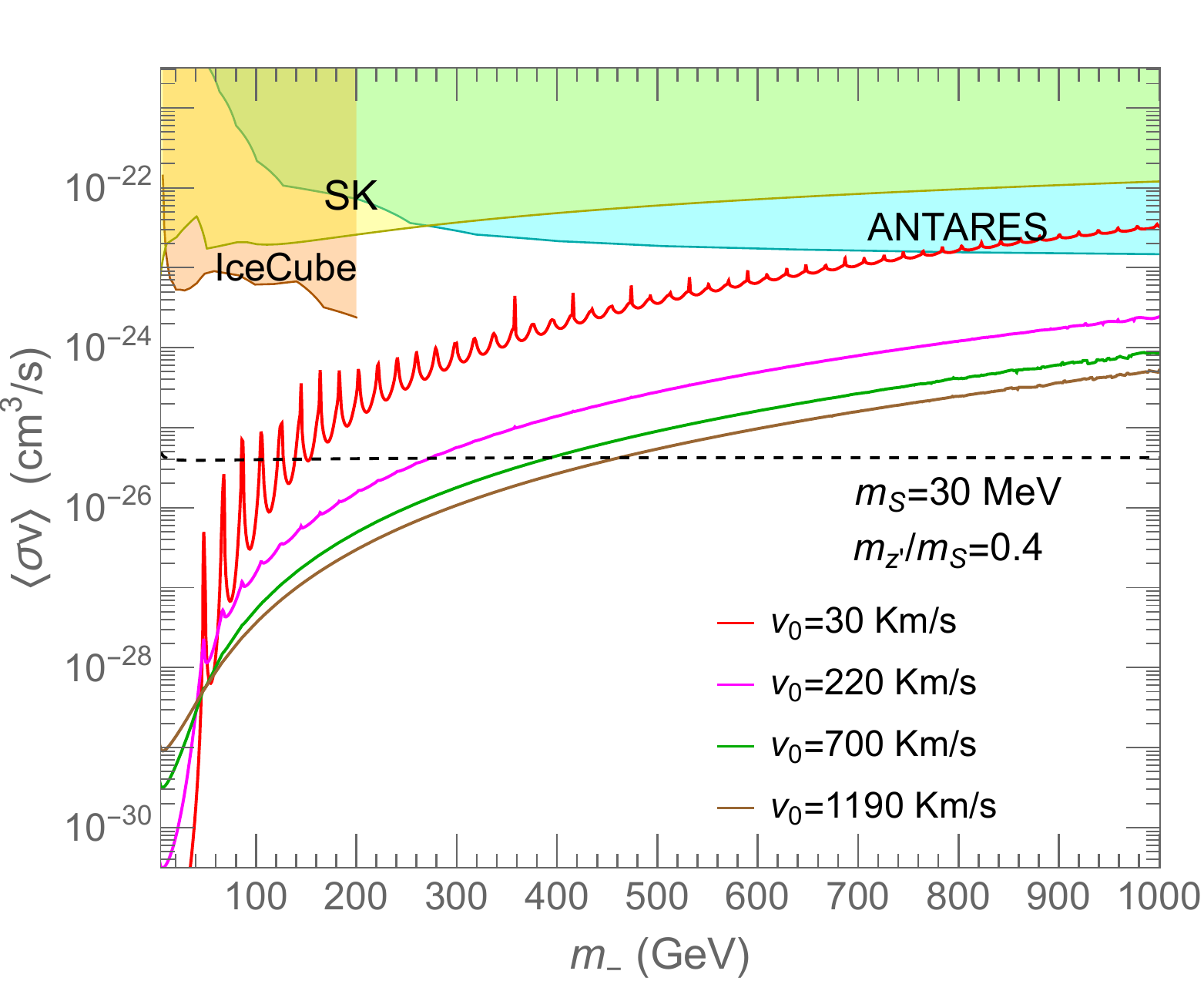}
\includegraphics[width=0.43\textwidth]{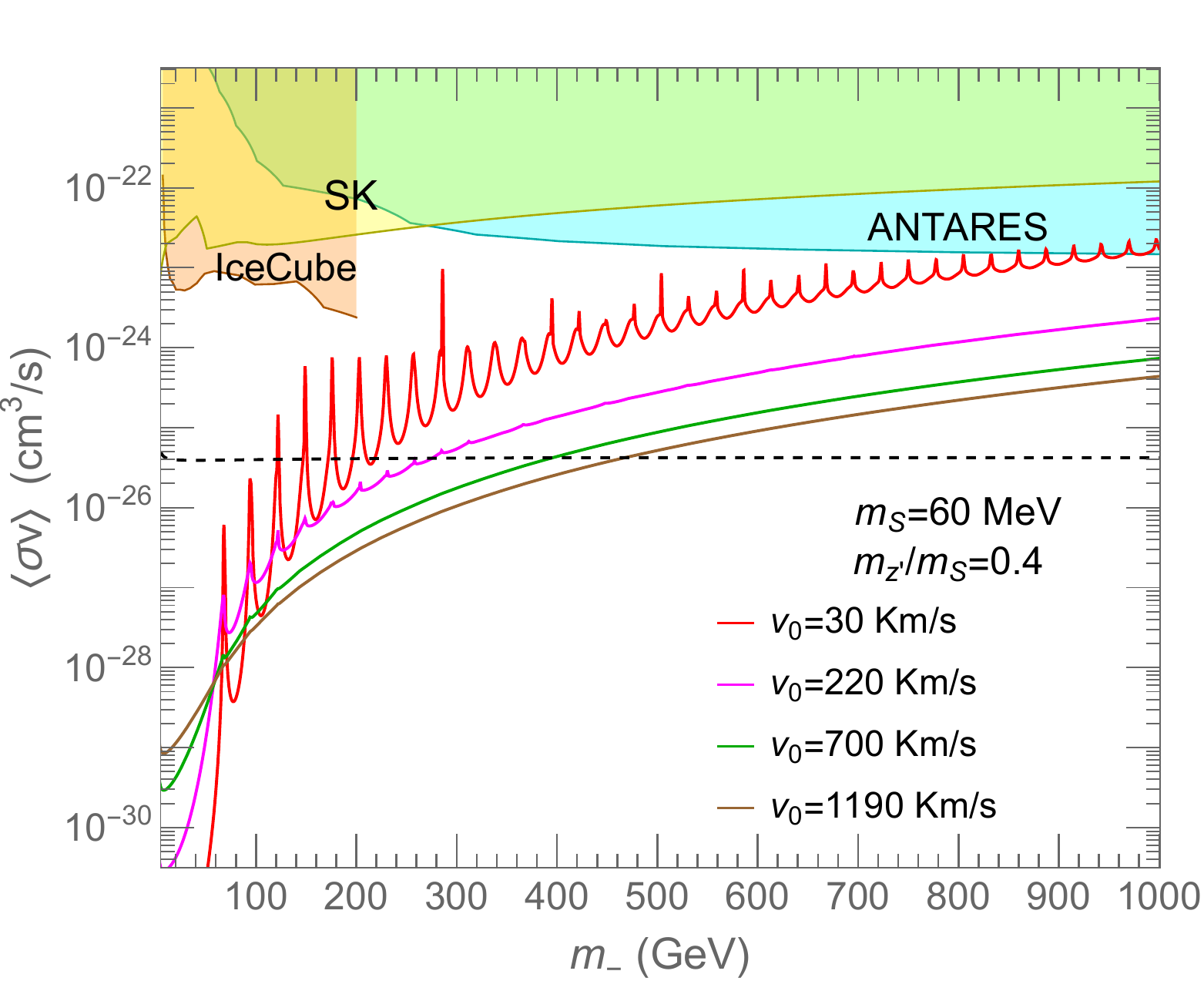}\hskip0.6cm
\caption{ Velocity-averaged annihilation cross sections for $\chi_- \chi_- \to S S, Z^\prime Z^\prime$ across different astrophysical scales. The curves represent most probable velocities of $v_0 = 30, 220, 700$, and $1190$ km/s (typical of dwarf galaxies, the local Milky Way, galaxy groups, and galaxy clusters), ordered from top to bottom at $m_- = 500$~GeV. Constraints from Super-Kamiokande (yellow), IceCube (orange), and ANTARES (cyan) apply only to the local Milky Way prediction (magenta curve, $v_0 = 220$ km/s). Benchmark inputs are $m_{Z^\prime}/m_S=0.4$ with $m_S=30$~MeV (left) and $60$~MeV (right). The black dashed line indicates the canonical cross section $\langle \sigma v\rangle_{\chi_-\chi_-}$ required for the thermal relic density.
}
\label{fig:s1-xs-ss}
\end{center}
\end{figure}

In comparison with the indirect detection constraints shown in Fig.~\ref{fig:s1-xs-ss}, we display the annihilation cross section averaged over the velocity distribution within each small-scale structure of the Universe. The benchmark parameters are $m_{Z^\prime}/m_S=0.4$ with $m_S=30$~MeV (left panel) and $60$~MeV (right panel).
Here, we adopt the most probable velocities of 30, 220, 700, and 1190 km/s, which correspond to the typical scales of the dwarf galaxies, local Milky Way, galaxy groups, and galaxy clusters, respectively.
The regions excluded by Super-Kamiokande (yellow) \cite{Super-Kamiokande:2020sgt, Frankiewicz, Super-Kamiokande:2015qek}, ANTARES (cyan) \cite{ANTARES:2015vis, Arguelles:2019ouk}, and IceCube (orange) \cite{IceCube:2025fcn} are compared only to the magenta curve ($v_0 = 220~$km/s), which represents the prediction for the local Milky Way.
These indirect searches for dark matter are constrained by the excess upper limit of high-energy muon neutrinos from the center of the Milky Way. 
Unlike Ref.~\cite{Kamada:2018zxi}, which assumes $p$-wave annihilations safely evade neutrino telescope bounds, our explicit inclusion of the Sommerfeld enhancement significantly boosts the late-time cross section, although it still safely satisfies current observational limits.
This model is consistent with existing constraints for indirect detection of dark matter mass, which ranges from several GeV to hundreds of GeV.
The details of these experiments, which indirectly search for dark matter, will be discussed in Appendix~\ref{app:SK}.

\section{Self-scatterings of dark matter and the small-scale structure problems}\label{sec:self-scatterings}

Self-interacting dark matter was introduced to address small-scale structure challenges that conflict with collisionless cold dark matter (CDM) predictions, particularly the core-cusp and too-big-to-fail problems \cite{Spergel:1999mh, Tulin:2017ara, Adhikari:2022sbh}. 

Dark matter self-interactions transfer heat toward the central regions of halos, flattening the core density profile. Lowering the central density of Milky Way subhalos addresses the "too big to fail" problem, while the resulting flattened density cores in low-mass galaxies naturally explain their observed rotation curves, resolving the core-cusp issue. Although the scarcity of satellite galaxies (the "missing satellites" problem) is now primarily attributed to baryonic feedback, self-interactions can still impact the subhalo mass function \cite{Tulin:2017ara}. More importantly, self-interacting DM provides a robust mechanism to explain the observed structural diversity of galaxies. Depending on the halo formation history, galaxies with the same maximum circular velocity can exhibit significant variations in their central densities, leading to the diverse internal rotation curves observed in similar-mass galaxies \cite{Tulin:2017ara, Kaplinghat:2015aga}.

In our model, a profound causal relationship naturally emerges between the early-Universe freeze-out and late-time halo dynamics. As established in Sec.~\ref{sec:boltz}, the $p$-wave dominant annihilation ($\chi_- \chi_- \to SS$) dictates that achieving the correct thermal relic abundance requires a remarkably large Yukawa coupling $f$. Note that utilizing the tree-level coupling $f$ yields a conservative overestimate for the scattering rates. As detailed in Sec.~\ref{subsec:sommerfeld-relic}, incorporating the late-time Sommerfeld enhancement effectively reduces the true $\alpha_f$ at higher masses, thereby further suppressing the actual cross sections and robustly ensuring our compliance with massive cluster constraints. Crucially, as the pure ground-state dark matter ($\chi_-$) becomes highly non-relativistic in the present-day Universe, this unsuppressed coupling, which is characterized by the dark fine-structure constant $\alpha_f = f^2/(4\pi)$, inevitably generates a strong attractive Yukawa potential $V(r) = - \frac{\alpha_f}{r} e^{-m_S r}$ for the elastic self-scattering process ($\chi_- \chi_- \to \chi_- \chi_-$) mediated by the light scalar $S$. Consequently, the Majorana nature of the dark matter not only dictates a strictly $p$-wave dominant annihilation during the freeze-out epoch but simultaneously guarantees the massive elastic self-interactions required to resolve small-scale anomalies.
 
Since the excited state is completely depleted ($r \to 0$), the dark matter halo consists purely of the ground state. The elastic scattering amplitude for the indistinguishable $\chi_-$ pairs can be expressed as a sum over partial waves,
\begin{align}
f(\theta) = \sum_{\ell=0}^{\infty} f_\ell (\theta) \equiv \frac{2}{m_- v} \sum_{\ell=0}^{\infty} (2\ell +1) e^{i\delta_\ell} P_\ell (\cos\theta) \sin\delta_\ell \,,
\label{eq:self-amp-dis}
\end{align}
where $\theta$ represents the scattering angle in the center-of-mass frame. Because the $\chi_-$-$\chi_-$ interaction involves two identical, unpolarized Majorana fermions, it includes a spin-singlet state (statistical weight 1/4) with a symmetric spatial wavefunction and a spin-triplet state (weight 3/4) with an antisymmetric spatial wavefunction. Considering $f_\ell (\pi-\theta) = (-1)^\ell f_\ell (\theta)$, the differential cross section is rigorously given by
\begin{align}
\left(\frac{d\sigma}{d\Omega} \right)_{--}= \frac{1}{2}\left( \left| \sum_{\rm even} f_\ell (\theta)\right|^2 + 3 \left| \sum_{\rm odd} f_\ell (\theta) \right|^2  \right)\,.
\label{eq:diff-xs-indis-1}
\end{align}
The phase shifts $\delta_\ell$ are extracted by matching the numerical solution of the radial Schrödinger equation to its asymptotic form \cite{Tulin:2013teo}. 

To evaluate the heat transfer inside halos, we calculate the transfer cross section $\sigma_T$, which accounts for the forward momentum loss. Because the differential cross section for identical particles remains unchanged under $\theta \leftrightarrow \pi -\theta$, the transfer cross section is defined as \cite{Sakurai:1994, Kahlhoefer:2013dca}:
\begin{equation}
\sigma_{\rm T} = \sigma_{\rm T} ^{--}= 4\pi \int_0^1 \left( \frac{d\sigma}{d\Omega}\right)_{--} (1-\cos\theta) d\cos\theta \,.
\label{xsT-indis}
\end{equation}
\begin{figure}[t!]
\begin{center}
\includegraphics[width=0.389\textwidth]{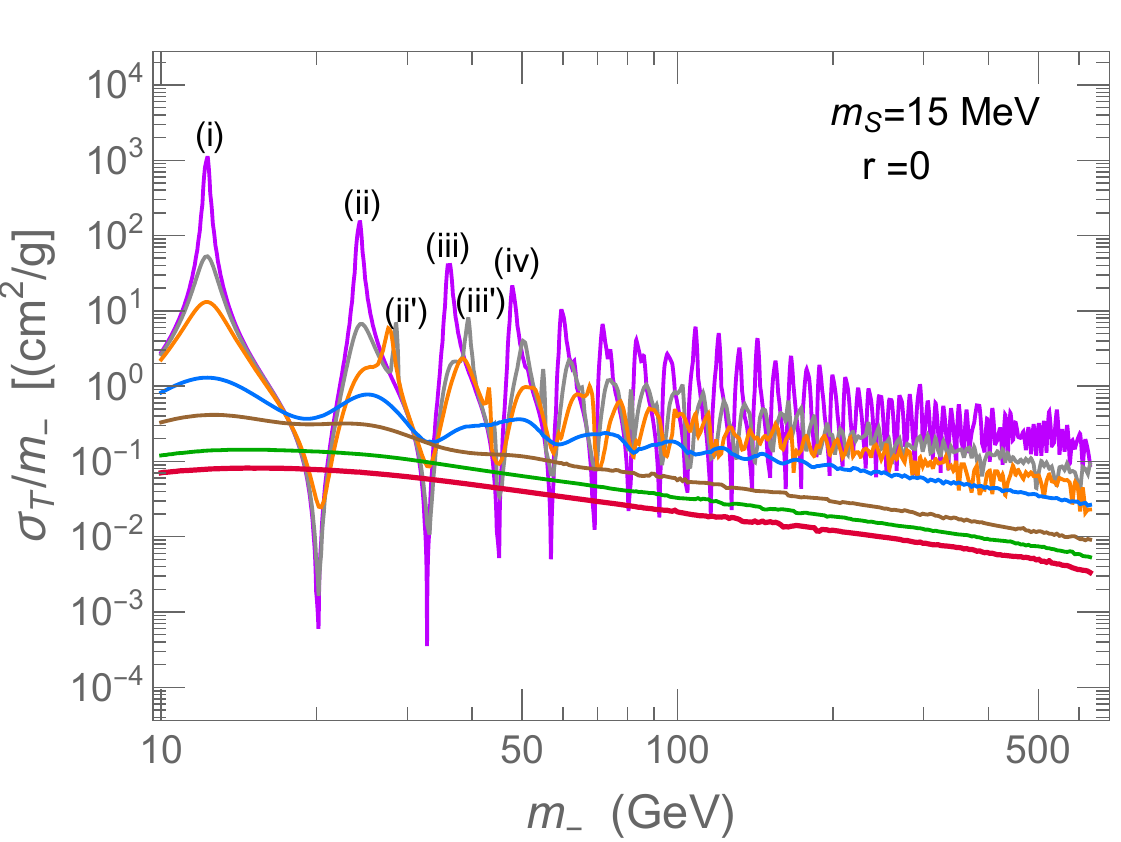}\hskip0.3cm
\includegraphics[width=0.383\textwidth]{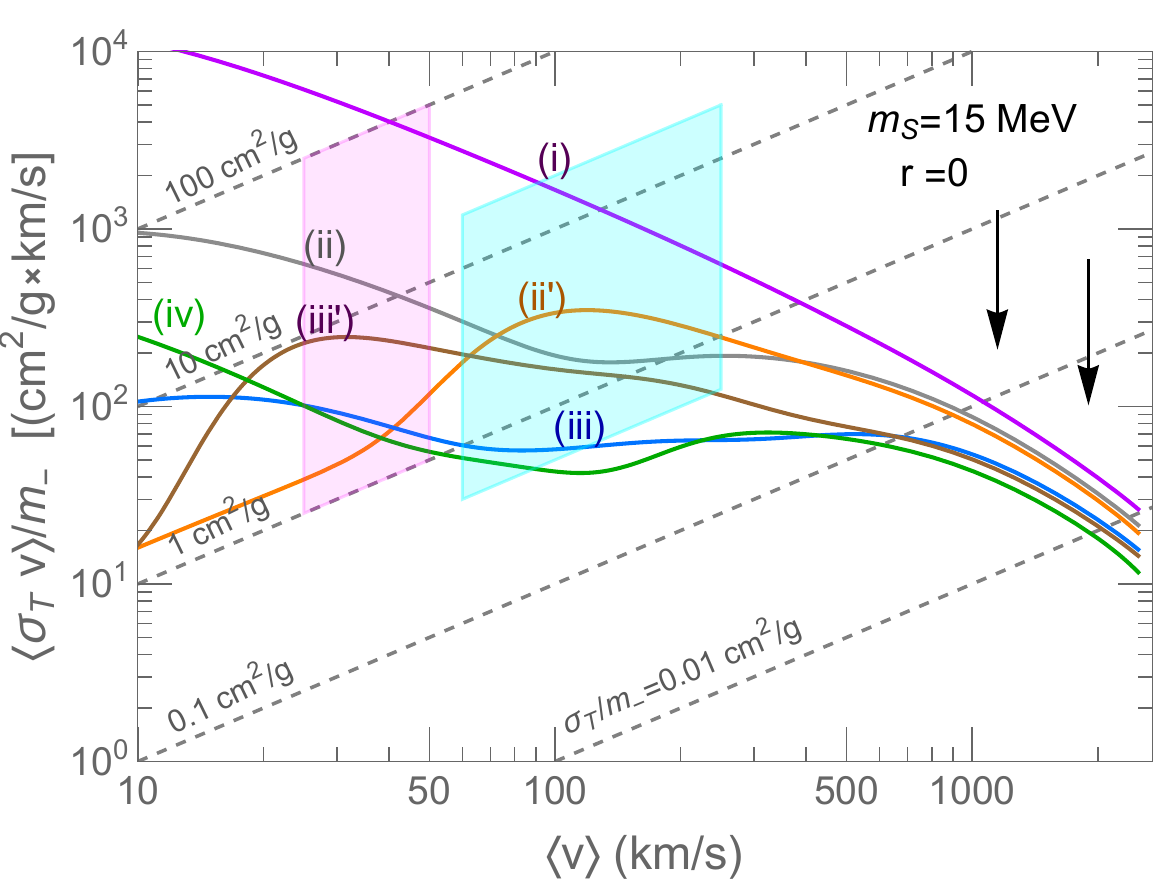} \\
\includegraphics[width=0.389\textwidth]{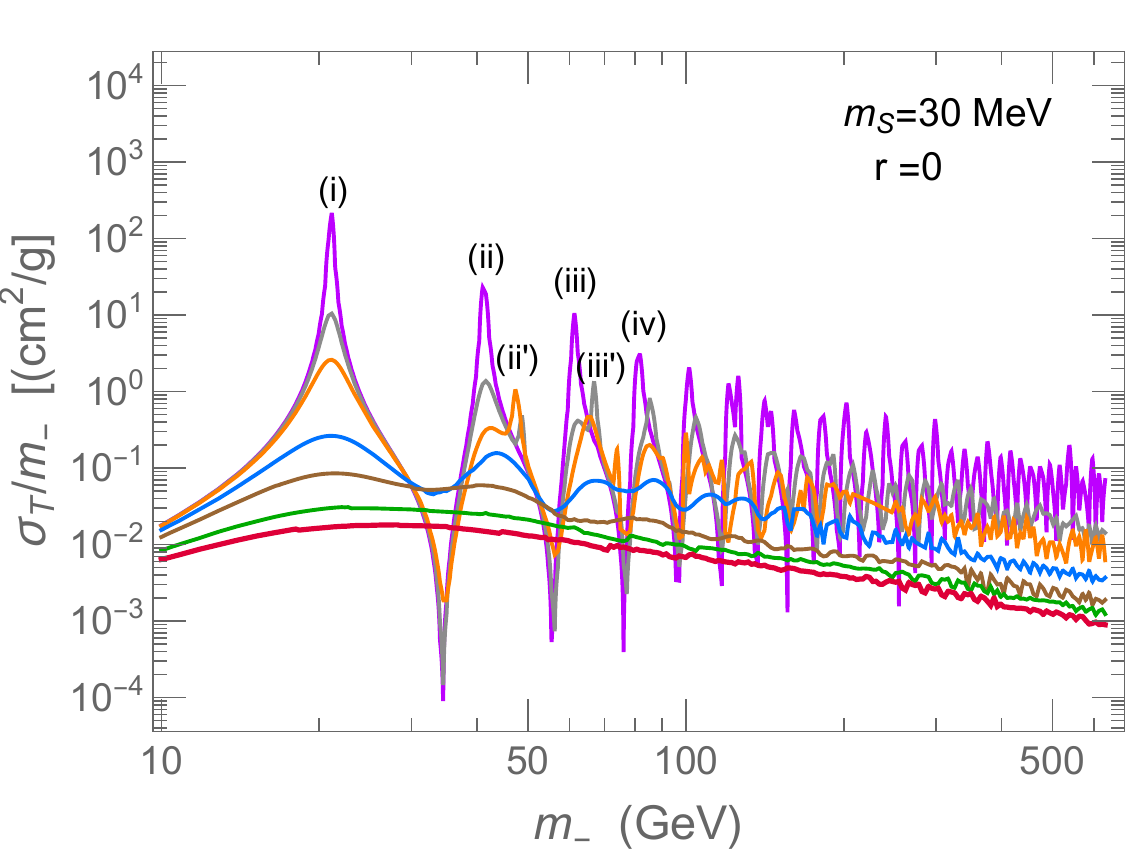}\hskip0.3cm
\includegraphics[width=0.383\textwidth]{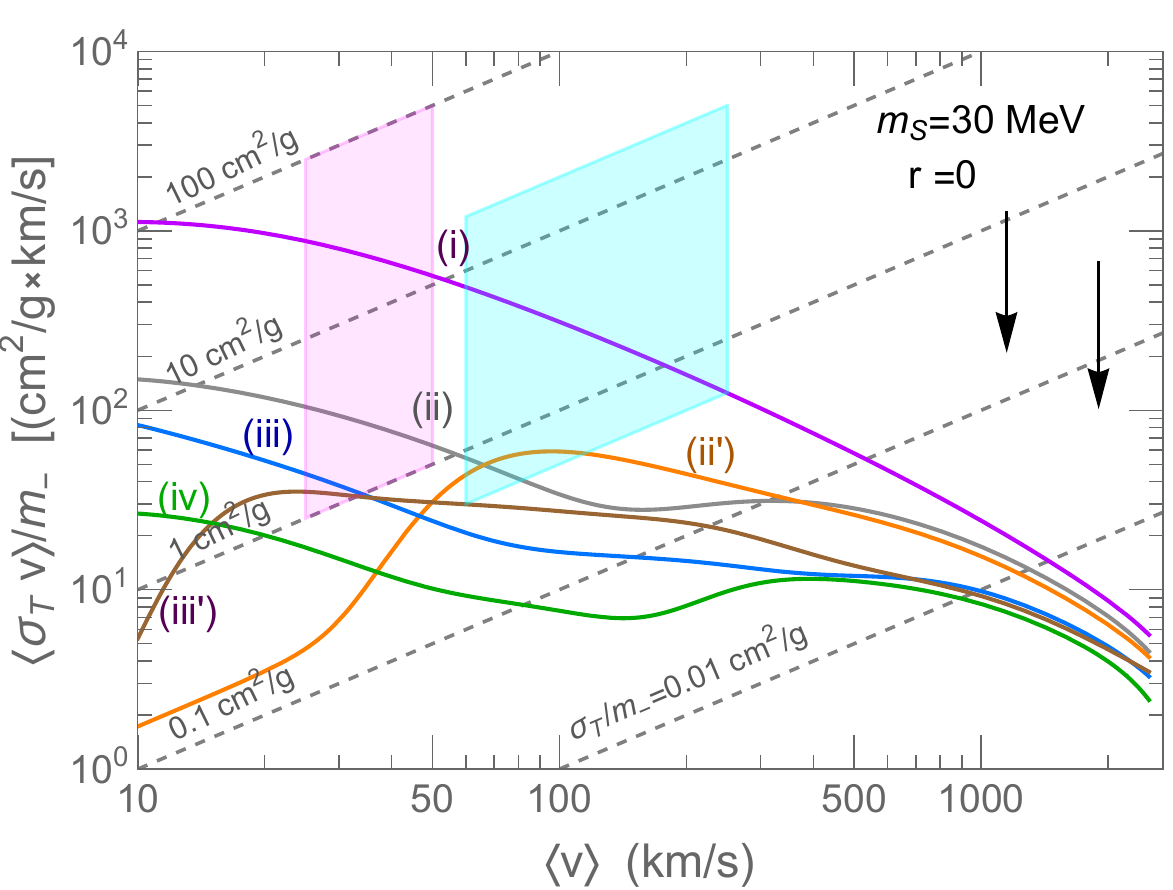} \\
\includegraphics[width=0.389\textwidth]{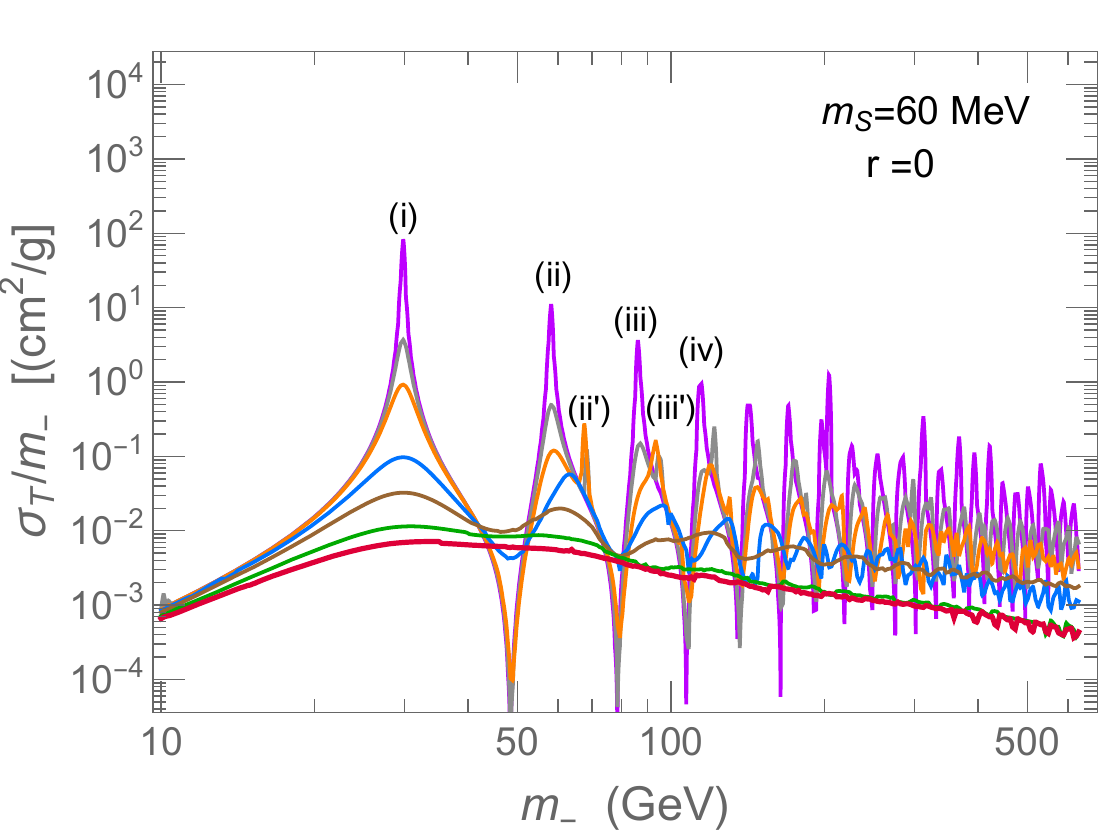}\hskip0.3cm
\includegraphics[width=0.383\textwidth]{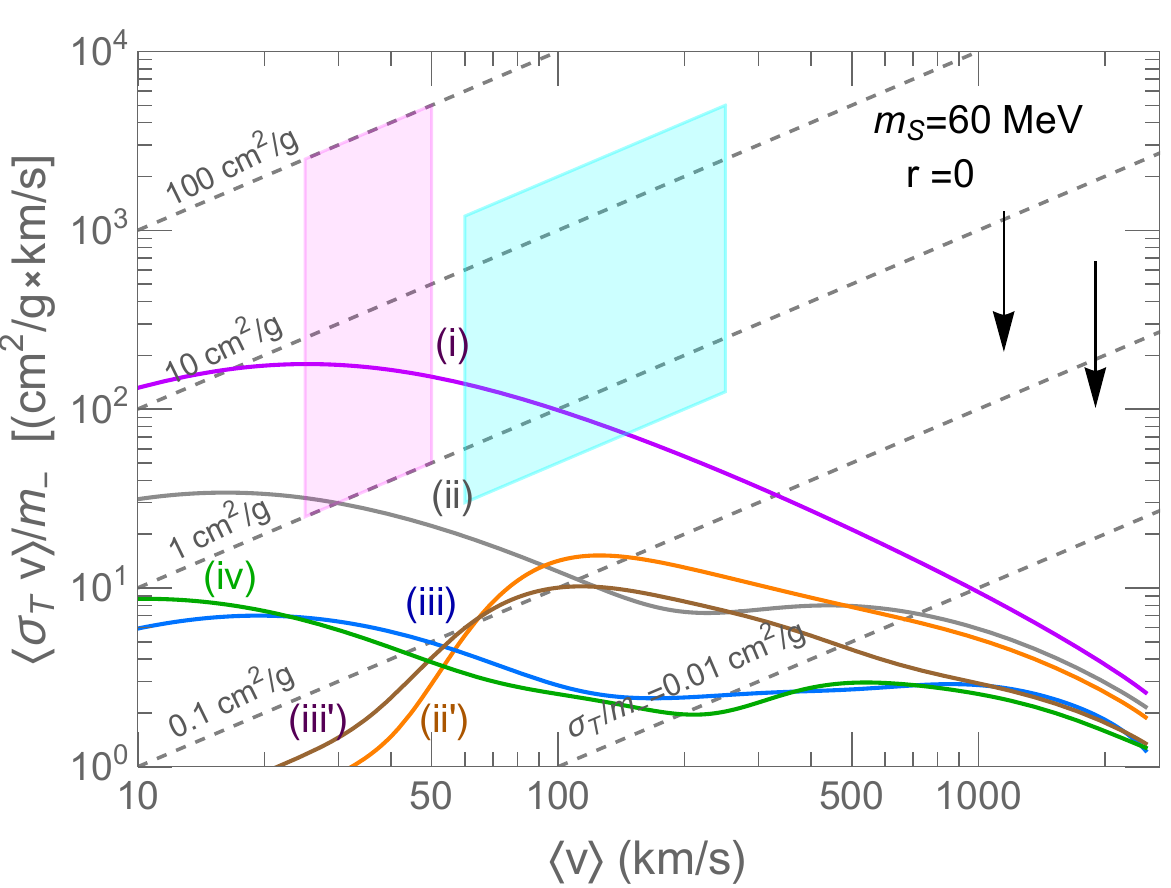} 
\caption{Left panels: The transfer cross section per unit mass $\sigma_T/m_-$ of the pure $\chi_-$ halo ($r=0$) as a function of $m_-$ for various mediator masses $m_S$. From top to bottom at $m_- \sim 15$~GeV, the curves denote $\langle v\rangle=$10 (purple), 50 (gray), 100 (orange), 300 (blue), 500 (brown), 800 (green), and 1000 (red) km/s. Right panels: The velocity-weighted cross section $\langle \sigma_T v \rangle / m_-$ vs. the average relative velocity $\langle v \rangle$. The solid curves, labeled (i)-(iv) and (ii')-(iii'), correspond to the primary and secondary resonant masses identified in the left panels. Diagonal dashed lines represent contours of constant $\sigma_T/m_-$. We compare the results with updated astrophysical constraints and theoretically favored regions: the stringent upper limits on galaxy groups ($\sigma_T/m_- \lesssim 1.1$ cm$^2$/g) and massive clusters ($\sigma_T/m_- \lesssim 0.35$ cm$^2$/g) indicated by downward arrows \cite{Sagunski:2020spe}; the parameter space favored by LSB spiral galaxies (central shaded box) \cite{Kaplinghat:2015aga}; and the broad favored region for dwarf galaxies encompassing both standard core formation \cite{Tulin:2017ara} and the onset of gravothermal collapse/diversity \cite{Correa:2020qam, Turner:2020vlf} (left shaded region).
}
\label{fig:self-scattering}
\end{center}
\end{figure}
A successful self-interacting DM model must accommodate the diverse structures of DM halos across vastly different astrophysical scales. To resolve small-scale controversies, the cross section is typically required to be around $\sigma_T/m_- \sim \mathcal{O}(1-10)$ cm$^2$/g at the scales of dwarf spheroidal galaxies \cite{Tulin:2017ara, Correa:2020qam, Turner:2020vlf} and low-surface-brightness (LSB) spirals  \cite{Kaplinghat:2015aga}. Conversely, at the scales of massive galaxy clusters and groups, the cross section must be significantly smaller---typically restricted to $\sigma_T/m_- \lesssim 0.1 - 0.5$ cm$^2$/g for clusters, and $\sigma_T/m_- \lesssim 1.1$ cm$^2$/g for groups---to remain consistent with various astronomical estimates, halo shape analyses, and cluster collision observations \cite{Tulin:2017ara, Adhikari:2022sbh, Kaplinghat:2015aga, Dave:2000ar, Elbert:2014bma, Ren:2018jpt, Harvey:2015hha, Bradac:2008eu, Randall:2008ppe, Sagunski:2020spe}. Our model naturally achieves this requisite velocity dependence through resonant Sommerfeld enhancement.

In Fig.~\ref{fig:self-scattering}, we present the velocity-weighted cross sections, $\langle \sigma_T v \rangle / m_-$, specifically evaluated for the pure ground-state DM scenario ($r=0$). To further investigate the kinetic properties of the model, we select specific dark matter mass values---corresponding to the resonance peaks identified in the left panels of Fig.~\ref{fig:self-scattering}---that best align with observational constraints for presentation. The resulting velocity-weighted cross sections, $\langle \sigma_T v \rangle / m_-$, are illustrated as a function of the average relative velocity $\langle v \rangle$ in the right panels of Fig.~\ref{fig:self-scattering}. To obtain these curves, we assume that the relative velocities of dark matter particles follow a Maxwell-Boltzmann distribution, as given by Eq.~(\ref{eq:dist-rel-v}), and calculate the statistical average of the velocity-weighted cross section. Under this distribution, the average collision velocity $\langle v \rangle$ on the horizontal axis is related to the most probable velocity $v_0$ by $\langle v \rangle = 2\sqrt{2/\pi} v_0$. As clearly illustrated in the figures, $\langle \sigma_T v \rangle / m_-$ exhibits a profound dependence on $\langle v \rangle$ due to the underlying resonant structures.

The motivation for plotting this specific rate stems from the dynamics of halo core formation. As pointed out in Ref.~\cite{Kaplinghat:2015aga}, the inner density $\rho$ of a galaxy halo is intrinsically linked to its age $t_{\rm age}$. Within the characteristic radius of a thermalized core, the condition $\langle \sigma_T v\rangle \rho / m_- \cdot t_{\rm age}\gtrsim 1$ holds, indicating that each dark matter particle has, on average, undergone at least one scattering event since formation. Here, the transfer cross section $\sigma_T$ is strictly used to appropriately account for the forward momentum loss. Because the characteristic DM velocities vary by orders of magnitude across different galactic systems---from dwarf galaxies to massive clusters---the efficacy of this scattering rate, and hence the DM self-interaction, must strongly depend on velocity.

A detailed analysis of the kinematic behavior reveals distinct features between the primary resonances (e.g., curves (i)-(iv)) and the secondary resonances (e.g., curves (ii') and (iii')). For the primary resonant masses, the cross section increases significantly as $\langle v \rangle$ decreases from cluster scales down to dwarf scales. Interestingly, for certain primary resonances (such as those in the $m_S = 30$ and $60$ MeV scenarios), the velocity-weighted cross section $\langle \sigma_T v \rangle / m_-$ reaches a local maximum around $\langle v \rangle \sim 20$--$25$ km/s and exhibits a slight decrease as $\langle v \rangle$ approaches $10$ km/s.
It is worth noting that while the rate $\langle \sigma_T v \rangle / m_-$ shows this non-monotonic behavior, the intrinsic cross section $\sigma_T/m_-$ remains monotonically increasing as $\langle v \rangle$ decreases, crossing the constant $\sigma/m$ contours.
In contrast, for secondary resonances, the cross section tends to saturate and approaches a constant value at $\langle v \rangle \to 10$ km/s.

In the dwarf galaxy regime ($\langle v \rangle \sim 30$--$50$ km/s), satisfying the canonical values for standard core formation \cite{Kaplinghat:2015aga, Tulin:2017ara} requires the model parameters to be tuned near these discrete resonance peaks.
The mediator mass, $m_S$, plays a crucial role in determining the density of these available peaks. For a lighter mediator ($m_S \sim 15$--$30$ MeV), the resonance peaks are more densely distributed along the dark matter mass spectrum. Consequently, within the range of $m_- \approx 20$--$75$ GeV, there are numerous distinct resonant values that successfully fall within the favored region (shaded areas). 
Notably, for lighter resonant dark matter ($10 \text{ GeV} \lesssim m_- \lesssim 40 \text{ GeV}$) coupled with a relatively light mediator ($m_S \lesssim 30$ MeV), the cross section can reach values as high as $\sigma_T/m_- \sim 10 \text{ cm}^2/\text{g}$. This enhancement aligns with the requirements for initiating gravothermal collapse, providing a natural explanation for the observed diversity in Milky Way satellites  \cite{Correa:2020qam, Turner:2020vlf}.

Conversely, for a heavier mediator ($m_S \sim 60$ MeV), the resonance peaks become much sparser. To satisfy the canonical core formation constraints in this scenario, the dark matter mass cannot take arbitrary values; it must be strictly fine-tuned to the specific vicinity of the sparse primary resonances, namely $m_- \approx 26.9$, $53.7$, or $79.4$ GeV, while simultaneously avoiding the stringent upper limits at galaxy group and cluster scales ($\langle v \rangle \gtrsim 1000$ km/s)  \cite{Sagunski:2020spe}.

 In summary, the desired velocity-dependent trend---large cross sections at dwarf scales and suppressed values at cluster scales---becomes more pronounced and easier to achieve without extreme fine-tuning when a much smaller mediator mass ($m_S$) is used.

\section{Discussions}\label{sec:discussions}

Below, we examine potential experimental signatures and observational consequences that may further probe and test the viability of this model.

\noindent{\underline{\bf $\mu N \to \mu N Z^\prime$}}

This model suggests that $Z^\prime$ bosons could be generated via the bremsstrahlung process $\mu N \rightarrow \mu N Z^\prime$ when a high-energy muon beam strikes a fixed target. For instance, the NA64$\mu$ experiment, employing a 160 GeV muon beam, projects that with $10^{11}$ muons on target (MOTs), its 90\% C.L. upper limit for the gauge coupling $g_{\mu\tau}$ could be reduced to below $3 \times 10^{-4}$ for $m_{Z^\prime} = 1 - 300$ MeV \cite{Sieber:2021fue}. Additionally, Fermilab’s $M^3$ (Muon Missing Momentum) experiment aims to investigate a similar parameter space in its Phase 1 projection \cite{Kahn:2018cqs}. Given the recent update on the SM prediction, which essentially resolves the longstanding $(g-2)_\mu$ anomaly, the allowed $L_\mu - L_\tau$ contribution is no longer restricted to a narrow favored band but is instead simply bounded by an upper limit. Consequently, the projected sensitivities of these future experiments will be crucial for thoroughly exploring this much broader viable parameter space.

\noindent{\underline{\bf The kinematic prohibition of late-time inelastic signatures}}

In conventional nearly degenerate dark matter models with microscopic mass splittings ($\delta m \lesssim \mathcal{O}(100 \text{ keV})$), inelastic transitions such as endothermic scattering with nucleons \cite{Cirelli:2013ufw, Graham:2010ca, Li:2022acp}, capture in neutron stars \cite{Potekhin:2013qqa, Goriely:2010bm, Zhang:2020wov, Garani:2018kkd, Feng:2015hja, Blennow:2015hzp, Lundberg:2004dn, Garani:2019fpa}, and upscattering at the Galactic Center yield prominent observational signatures. However, as established in Sec.~\ref{sec:mass-gap}, our $p$-wave dominated Majorana framework inherently necessitates a macroscopic mass splitting (i.e., $(\delta m / m_-) x_f \gg 1$, typically requiring $\delta m \gtrsim \mathcal{O}(1)$ GeV) to ensure a pure ground-state relic ($r \to 0$). Crucially, this large mass gap effectively shuts down the $s$-wave $\chi_- \chi_+ \to Z^\prime S$ coannihilation during freeze-out, which would otherwise severely restrict the parameter space.

As a natural consequence of this $p$-wave freeze-out condition, all late-time endothermic processes are kinematically forbidden. Because the mass gap $\delta m$ is vastly larger than the kinetic energy available in contemporary astrophysical environments, the minimum velocity threshold ($v_{\rm min}$) required to excite $\chi_-$ into $\chi_+$ far exceeds the local Galactic escape velocity. Consequently, spin-dependent inelastic scattering with target nuclei in direct detection experiments ($\chi_- p \rightarrow \chi_+ p$), inelastic capture by muons in neutron stars ($\chi_- \mu \rightarrow \chi_+ \mu$), and thermal upscattering in dense galactic centers ($\chi_- \chi_- \rightarrow \chi_+ \chi_+$) are completely closed. 

Ultimately, the model predicts an exceptionally clean observational footprint. By naturally shutting down the inelastic processes that typically constrain nearly degenerate models, our framework channels its predictive power strictly into the elastic scalar-mediated self-interactions (which resolve small-scale structure anomalies) and the pure gauge-driven $Z^\prime$ bremsstrahlung searches.

\section{Summary}\label{sec:summary}

We have investigated a $p$-wave dominant Majorana dark matter model based on a $U(1)_{L_\mu - L_\tau}$ gauge extension of the Standard Model. Following SSB, two mass-split Majorana dark matter particles, $\chi_-$ and $\chi_+$, are formed. The $U(1)_{L_\mu - L_\tau}$ gauge boson, $Z^\prime$, gains mass through symmetry breaking and interacts with muons, taus, and their corresponding neutrinos. While this interaction was historically proposed to explain the observed past deviation in the muon anomalous magnetic moment $(g-2)_\mu$, recent updates to the Standard Model prediction have essentially resolved this anomaly. Consequently, the $L_\mu - L_\tau$ contribution is now tightly constrained by new upper limits. Remarkably, our model's parameters naturally accommodate these stringent bounds while remaining fully consistent with a wide array of experimental, cosmological, and astronomical data.

The $Z^\prime$ gauge boson acts as a portal, interacting with particles in the dark sector. Due to SSB, the dark sector fermions split into the two Majorana states ($\chi_\pm$) with masses ranging from about 10 GeV to hundreds of GeV. A Dirac mass term introduces a macroscopic mass splitting ($\delta m$) between them, which is vital to ensure that the excited state $\chi_+$ is completely depleted during the early Universe. The scalar particle $S$, which drives the spontaneous symmetry breaking in the dark sector, has a mass around the MeV scale and is crucial for the thermal evolution, relic abundance, and self-interaction of dark matter. The main features and findings of this model are summarized below:

\begin{itemize}
\item{\bf Updated $(g-2)_\mu$ constraints:}
 We have examined the parameter space in light of the updated $(g-2)_\mu$ results \cite{Aliberti:2025beg}. The reduced $\Delta a_\mu$ value implies that the $L_\mu - L_\tau$ model remains a robust scenario, provided the gauge coupling $g_{\mu\tau}$ is sufficiently small. This smaller coupling comfortably satisfies the stringent bounds from neutrino trident production (CCFR) and collider searches (BABAR). Regarding other low-energy constraints (such as white dwarf cooling and Borexino), they impose restrictions on $g_{\mu\tau}$ and $m_{Z^\prime}$ via kinetic mixing. However, since these mixing effects may be influenced by primordial contributions from higher energy scales (Sec.~\ref{subsec:gauge}), the reliability of these limits in the $m_{Z^\prime} \sim 10$ MeV region (Fig.~\ref{fig:g-2}) is highly sensitive to the underlying UV physics. Future fixed-target experiments like NA64$\mu$ and ${\rm M}^3$ will be crucial to thoroughly investigate the much broader parameter space allowed by the new $(g-2)_\mu$ upper bounds (Sec.~\ref{sec:discussions}).

\item{\bf Thermal history and Hubble tension:} During the early thermal evolution, $S$ and $Z^\prime$ remain in equilibrium with $\nu_\mu$ and $\nu_\tau$. Because $Z^\prime$ can mix with photons at the one-loop level, its decay $Z^\prime \rightarrow e^+ e^-$ helps maintain thermal balance with the SM electromagnetic bath. About one second after the Big Bang, this decay process facilitates entropy transfer between the neutrino and electromagnetic sectors (Sec.~\ref{sec:neff}). This dynamic alters the effective number of relativistic species ($N_{\rm eff}$) (Fig.~\ref{fig:neff}) and offers a mechanism to alleviate the cosmological Hubble tension (Fig.~\ref{fig:g-2}).

\item{\bf Freeze-out dynamics:} The dark matter relic abundance is governed by strictly $p$-wave dominated annihilations of the ground state into scalars and gauge bosons ($\chi_- \chi_- \rightarrow SS, Z^\prime Z^\prime$), with the scalar channel playing the primary role. A relatively large, unsuppressed coupling constant $f$ between DM and $S$ is required to ensure the correct relic density (Sec.~\ref{sec:boltz}), which subsequently drives massive DM self-interactions in the late Universe. Our analysis demonstrates that the derived dark sector gauge coupling $g_\chi$ generally remains very small ($g_\chi \lesssim 10^{-3}$).

\item{\bf Constraints on the Higgs portal:}  
The scalar $S$ can couple to SM particles via a small mixing angle $\alpha$ with the SM Higgs boson. Since the spin-independent scattering cross section between dark matter and nucleons is highly sensitive to the scalar mediator mass (scaling roughly as $\alpha^2 / m_S^4$), direct detection experiments place strict joint constraints on the $(m_S, \alpha)$ parameter space. By exploring various experimental limits from both particle physics and direct detection (Secs.~\ref{sec:pp-expt} and \ref{sec:dd-expt}), we find that the recent results from the LZ experiment (LZ 2025) currently impose the most stringent bounds, requiring a severely suppressed mixing angle $\alpha$ particularly for lighter scalar masses (Fig.~\ref{fig:alpha-cons}).

\item{\bf Kinematic closure of inelastic channels and maximized self-interactions:} To achieve a pure ground-state relic ($r \to 0$) and safely neutralize the $s$-wave coannihilation during freeze-out, the model inherently necessitates a macroscopic mass splitting ($\delta m \gtrsim \mathcal{O}(1)$ GeV). This thermodynamical requirement allows the model to preserve a maximally large Yukawa coupling $f$ to drive massive self-interactions, while naturally acting as a strict kinematic barrier in the late Universe. Consequently, all endothermic inelastic processes are completely forbidden, channeling the model's predictive power entirely into purely elastic, scalar-mediated self-interactions and bremsstrahlung searches (Sec.~\ref{sec:discussions}).

\item{\bf Neutrino limits from Sommerfeld enhancement:} Although the late-time dark matter annihilations ($\chi_- \chi_- \rightarrow SS, Z^\prime Z^\prime$) are generally $p$-wave suppressed, the exchange of light $S$ particles induces a significant Sommerfeld enhancement in these channels (Sec.~\ref{sec:sommerfeld-indirect}). To derive the neutrino limits, we operate under the established result that the current dark matter relic is composed entirely of the ground state $\chi_-$. By comprehensively accounting for the cascade decays of the produced $S$ and $Z^\prime$ bosons from the $\chi_- \chi_- \rightarrow SS$ and $Z^\prime Z^\prime$ processes, we confront our theoretical neutrino spectra with the stringent upper limits from neutrino telescopes like Super-Kamiokande, ANTARES, and IceCube (Fig.~\ref{fig:s1-xs-ss}).

\item{\bf Self-interactions and small-scale structure:} In the DM mass range of $10 \sim 75$ GeV, with the scalar mass $m_S$ at the tens of MeV scale, the velocity-weighted transfer cross section $\langle \sigma_T v \rangle / m_-$ of the pure $\chi_-$ halo exhibits a profound dependence on the average collision velocity due to resonant enhancements (Fig.~\ref{fig:self-scattering}). This velocity-dependent self-interacting DM model elegantly explains the diversity of DM halos. At the scale of dwarf spheroidal and low surface brightness galaxies, the cross section reaches $\sigma_T / m_- \sim \mathcal{O}(1 - 10) \text{ cm}^2/\text{g}$, resolving the core-cusp problem and explaining halo diversity. Simultaneously, the cross section is naturally suppressed at larger scales, comfortably satisfying the stringent upper limits for galaxy groups ($\lesssim 1.1 \text{ cm}^2/\text{g}$) and massive clusters ($\lesssim 0.35 \text{ cm}^2/\text{g}$) (Sec.~\ref{sec:self-scatterings}).
\end{itemize}

\bigskip
\textbf{Acknowledgments.}
 This work was partly supported by the National Center for Theoretical Sciences and the National Science and Technology Council of Taiwan under grant number NSTC 113-2112-M-033-005.


\appendix

\section{Derivation of the mass-split Majorana eigenstates after SSB}\label{app:pseudo-M}

After SSB, the mass term of the dark fermions is given by 
\begin{align}
{\cal L}_m= {\cal L}_m^D +{\cal L}_m^M \,,
\end{align}
where the Dirac mass part is 
\begin{align}
-{\cal L}_m^D&=  \frac{m_D}{2} (\bar\chi \chi + \overline{\chi^c} \chi^c) \nonumber\\
                      &= \frac{m_D}{2} ( \overline{\chi_R} \chi_L + \overline{\chi_L} \chi_R + \overline{\chi^c_R} \chi^c_L + \overline{\chi^c_L} \chi^c_R  ) \,,
\end{align}
and the Majorana mass part with $f v_S \equiv m_M$ is 
\begin{align}                    
-{\cal L}_m^M&=  \frac{m_M}{2}  (\overline{\chi^c} \chi + \overline{\chi} \chi^c )
=  \frac{m_M}{2}( \overline{\chi^c_R} \chi_L + \overline{\chi_L} \chi^c_R ) +  \frac{m_M}{2}(\overline{\chi^c_L} \chi_R + \overline{\chi_R} \chi^c_L  ) \nonumber\\
                      &= \frac{m_M}{2} \left(\overline{\chi_L} + \overline{\chi^c_R} \right)  \left(\chi_L  + \chi^c_R \right) 
                           + \frac{m_M}{2} \left(\overline{\chi_R} + \overline{\chi^c_L} \right)  \left(\chi_R  + \chi^c_L \right) \,.
\end{align}
In the absence of $m_D$, the dark fermion can be decomposed into two exactly degenerate Majorana states, $\chi_1 = \chi_L + \chi_R^c$ and $\chi_2 = \chi_R + \chi_L^c$, which can be defined up to an arbitrary rotation. In the basis $ \mathbf{X} = (\chi_1, \chi_2)^T$, the mass Lagrangian can be recast into a $4 \times 4$ matrix form:\footnote{Alternatively, we can write the mass Lagrangian as a $2\times 2$ matrix,
\begin{equation}
-{\cal L}_m=
\frac{1}{2}
\left(\begin{array}{cc}
\overline{\chi^c_R} & \overline{\chi_R}
\end{array}\right)
 \left(\begin{array}{cc}
 m_M  &  m_D  \\
 m_D  &   m_M 
 \end{array}\right) 
 \left(\begin{array}{c}
 \chi_L \\ \chi^c_L
 \end{array}\right)
 +\text{h.c.} ,
\end{equation}
 to find the Majorana mass eigenstates.
}
\begin{equation}
-{\cal L}_m=
\frac{1}{2}
\begin{tikzpicture}[baseline=0cm,mymatrixenv]
 \matrix [mymatrix,inner sep=4pt,row sep=1em] (m)  
    {
   \overline{\chi_L} & 
  \overline{\chi^c_R}  &
  \overline{\chi^c_L}  &
   \overline{\chi_R} \\
    };
   \path (m-1-1.east) -- (m-1-4.west) coordinate[midway] (X);
  \draw [cheating dash=on 2pt off 2pt,blue]
       (X |- m.north) --   (X |- m.south);
   \mymatrixbracebottom{1}{4}{$\mathbf{\overline{X}}$}
 \end{tikzpicture}
\begin{tikzpicture}[baseline=0cm,mymatrixenv]
 \matrix [mymatrix,inner sep=4pt,row sep=1em] (m)  
    {
   0 & 1         &  0 &  0  \\ 
   1        & 0    &  0       &  0   \\
    0 &  0         & 0  &  1  \\
    0       & 0    & 1       &  0  \\
    };
    \path (m-2-1.east) -- (m-2-4.west) coordinate[midway] (X)
    (m-1-2.south) -- (m-4-2.north) coordinate[midway] (Y);
    \draw [cheating dash=on 2pt off 2pt,blue]
         (X |- m.north) --   (X |- m.south);
    \draw [cheating dash=on 2pt off 2pt,blue]
         (Y -| m.west) -- (Y -| m.east);
 \end{tikzpicture}
 \begin{tikzpicture}[baseline=0cm,mymatrixenv]
\matrix [mymatrix,inner sep=4pt,row sep=1em] (m)  
    {
    m_M & 0          &  m_D &  0  \\ 
    0       & m_M    &  0       &  m_D    \\
    m_D &  0         & m_M  &  0  \\
    0       & m_D    & 0        &  m_M  \\
    };
    \path (m-2-1.east) -- (m-2-4.west) coordinate[midway] (X)
    (m-1-2.south) -- (m-4-2.north) coordinate[midway] (Y);
    \draw [cheating dash=on 4pt off 2pt,blue]
         (X |- m.north) --   (X |- m.south);
    \draw [cheating dash=on 2pt off 2pt,blue]
         (Y -| m.west) -- (Y -| m.east);
   \mymatrixbracebottom{1}{4}{$\mathbf{M}$}
 \end{tikzpicture}
\begin{tikzpicture}[baseline=0cm,mymatrixenv]
 \matrix [mymatrix,inner sep=4pt,row sep=1em] (m)  
    {
   \chi_L  \\ 
  \chi^c_R  \\
  \chi^c_L  \\
   \chi_R \\
    };
    \path 
    (m-1-2.south) -- (m-4-2.north) coordinate[midway] (Y);
    \draw [cheating dash=on 2pt off 2pt,blue]
         (Y -| m.west) -- (Y -| m.east);
    \mymatrixbracebottom{1}{1}{$\mathbf{X}$}
 \end{tikzpicture}
 .
\end{equation}
We diagonalize the mass matrix {$\mathbf{M}$} by a unitary transformation, $\mathbf{U_D M U_D^\dag = D} $, where 
$\mathbf{D}={\rm diag} (m_M -m_D, m_M -m_D, m_M +m_D, m_M +m_D)$. 
The physical eigenstates can be read from
\begin{equation}
\mathbf{X_D} = \mathbf{U_D X}=  \frac{1}{\sqrt{2}}
\begin{tikzpicture}[baseline=0cm,mymatrixenv]
\matrix [mymatrix,inner sep=4pt,row sep=1em] (m)  
    {
    0        & -1   &  0    &  1  \\ 
   -1        &  0   &  1    &  0   \\
    0        &  1   &  0    &  1  \\
    1        &  0   &  1    &  0  \\
    };
 \end{tikzpicture}
 \begin{tikzpicture}[baseline=0cm,mymatrixenv]
 \matrix [mymatrix,inner sep=4pt,row sep=1em] (m)  
    {
   \chi_L  \\ 
  \chi^c_R  \\
  \chi^c_L  \\
   \chi_R \\
    };
 \end{tikzpicture}
 =   \frac{1}{\sqrt{2}}
 \begin{tikzpicture}[baseline=0cm,mymatrixenv]
 \matrix [mymatrix,inner sep=4pt,row sep=1em] (m)  
    {
   \chi_R - \chi^c_R  \\ 
  -\chi_L + \chi^c_L  \\
   \chi_R + \chi^c_R  \\
   \chi_L + \chi^c_L \\
    };
    \path 
    (m-1-2.south) -- (m-4-2.north) coordinate[midway] (Y);
    \draw [cheating dash=on 2pt off 2pt,blue]
         (Y -| m.west) -- (Y -| m.east);
 \end{tikzpicture}
 \,.
\end{equation}
The first two elements of $\mathbf{X_D}$ sum to $\chi_-$, and the last two elements sum to $\chi_+$, given  by
\begin{align}
\chi_- = \frac{1}{\sqrt{2}} (-\chi_L + \chi_R - \chi^c_R + \chi^c_L) 
\hskip0.5cm {\rm and}   
\hskip0.5cm
\chi_+ &= \frac{1}{\sqrt{2}} ( \chi_L + \chi_R + \chi^c_R + \chi^c_L)
  \,,
\end{align}
with masses $m_M-m_D$ and $m_M+m_D$, respectively.

\section{Diagonalization of the gauge boson sector}\label{app:kinetic-to-mass}

 Considering the possible kinetic mixings among the neutral gauge fields at low energy scale $\mu \lesssim m_\mu$, the neutral gauge boson sector of Lagrangian is given by
\begin{align}
{\cal L}_{\rm gauge} =& - \frac{1}{4} \hat{Z}_{\mu\nu} \hat{Z}^{\mu\nu}  - \frac{1}{4} \hat{F}_{\mu\nu} \hat{F}^{\mu\nu} - \frac{1}{4} \hat{Z}^\prime_{\mu\nu} \hat{Z}^{\prime \mu\nu} 
+ \frac{1}{2} m_Z^2 \hat{Z}_{\mu} \hat{Z}^{\mu} +  \frac{1}{2} m_{Z^\prime}^2 \hat{Z}'_{\mu} \hat{Z}'^{\mu} \nonumber\\
 &  - \frac{1}{2} \epsilon_A \hat{Z}^\prime_{\mu\nu} \hat{F}^{\mu\nu}  - \frac{1}{2} \epsilon_Z \hat{Z}^\prime_{\mu\nu} \hat{Z}^{\mu\nu}  \,,
\label{app:gaugeL}
\end{align}
where hatted fields denote interaction eigenstates. To eliminate the kinetic mixing, we transform to a canonical basis ${\bf \tilde V}_\mu= (\tilde Z_\mu,\,  \tilde A_\mu,\,  \tilde Z^\prime_\mu)^T$ via ${\bf \tilde V}_\mu= {\bf \zeta}\, {\bf \hat V}_\mu$, where
 ${\bf \hat V}_\mu= (\hat Z_\mu,\,  \hat A_\mu,\,  \hat Z^\prime_\mu)^T$, and
  \begin{align}
{\bf \zeta}=
 \left(\begin{array}{ccc}
 1 \  \ & \ \ 0 \   &   \epsilon_Z \\
 0 \ \ &\ \   1  & \epsilon_A \\
 0 \ \ &\ \  0  &1-\frac{\epsilon_Z^2}{2} -\frac{\epsilon_A^2}{2} 
\end{array}\right) 
+ {\cal O} (\epsilon_{A,Z}^3) \,.
 \end{align}
In terms of ${\bf \tilde V}_\mu$, the mass term in Eq.~(\ref{eq:gaugeL}) can be written as
 ${\cal L}_{\rm gauge}^{\rm mass} =(1/2) {\bf \tilde V}_\mu^T  (\zeta^{-1})^T {\bf M}\zeta^{-1} {\bf \tilde V}^\mu $, where ${\bf M} ={\rm diag}(m_Z, 0, m_{Z^\prime})$.
We can then diagonoalize the mass matrix to be $ \xi^T( (\zeta^{-1})^T {\bf M}\zeta^{-1}  )\xi  =   {\bf M}_{ph}$, where
  \begin{align}
{\bf \xi}=
 \left(\begin{array}{ccc}
 1-\frac{\epsilon_Z^2 m_Z^4}{2 (m_Z^2 -m_{Z^\prime}^2 )^2} \  \ & \ \ 0 \ \ \  &   \frac{\epsilon_Z m_Z^2}{ m_Z^2 -m_{Z^\prime}^ 2 } \\
 0 \ \ &\ \   1  \ \ \ &  0  \\
-\frac{\epsilon_Z m_Z^2}{ m_Z^2 -m_{Z^\prime}^2 } \ \ &\ \  0  \ \ \  & 1-\frac{\epsilon_Z^2 m_Z^4}{2 (m_Z^2 -m_{Z^\prime}^2 )^2}
\end{array}\right) 
+ {\cal O} (\epsilon_{A,Z}^3) \,,
 \end{align}
and diagonalized mass matrix is
  \begin{align}
  {\bf M}_{ph}=
 \left(\begin{array}{ccc}
 m_Z^2 + \frac{\epsilon_Z^2 m_Z^4}{m_Z^2 -m_{Z^\prime}^ 2 } \  \ & \ \ 0 \ \ \  &  0  \\
 0 \ \ &\ \   0  \ \ \ &  0  \\
 0 \ \ &\ \  0  \ \ \  & m_{Z^\prime}^2 (1+\epsilon_A^2) -\frac{\epsilon_Z^2 m_Z^4}{ m_Z^2 -m_{Z^\prime}^2 }
\end{array}\right) 
+ {\cal O} (\epsilon_{A,Z}^3) \,.
 \end{align}
 Finally, the physical mass eigenstates relate to the original basis via ${\bf V}_\mu  =(Z_\mu, \, A_\mu, \, Z^\prime_\mu)^T = {\bf U}\, {\bf \hat V}_\mu$, with the transformation matrix given by 
 \begin{align}
{\bf U}= \xi^{-1} \zeta =
 \left(\begin{array}{ccc}
 1-\frac{\epsilon_Z^2 m_Z^4}{2 (m_Z^2 -m_{Z^\prime}^2 )^2}  \  \ & \ \ 0 \ \  &  -\frac{\epsilon_Z m_{Z^\prime}^2}{m_Z^2 -m_{Z^\prime}^2} \\
 0 \ \ &\ \   1 \ \  & \epsilon_A \\
 \frac{\epsilon_Z m_{Z^\prime}^2}{m_Z^2 -m_{Z^\prime}^2}  \ \ &\ \  0 \ \ & 1 - \frac{\epsilon_A^2}{2} -\frac{\epsilon_Z^2 m_{Z^\prime}^4}{2 (m_Z^2 -m_{Z^\prime}^2 )^2}
\end{array}\right) + {\cal O} (\epsilon_{A,Z}^3)  \,.
 \end{align}
 
The interaction Lagrangian in Eq.~(\ref{eq:current-coupling}) is written in terms of the interaction eigenstates ${\bf \hat V}_{\mu}$, which are related to the physical mass eigenstates $V_\mu$ by ${\bf \hat{V}_\mu = U^{-1} V_\mu}$. To the leading order in $\epsilon_{A,Z}$, the inverse matrix ${\bf U^{-1}}$ is given by
\begin{equation}
{\bf U^{-1}} \simeq
\begin{pmatrix}
1 & 0 & \frac{\epsilon_Z m_Z^2}{m_Z^2 - m_{Z^\prime}^2} \\
0 & 1 & -\epsilon_A \\
-\frac{\epsilon_Z m_{Z^\prime}^2}{m_Z^2 - m_{Z^\prime}^2} & 0 & 1
\end{pmatrix}.
\label{eq:U_inverse}
\end{equation}


\section{Explicit formulas for the muon anomalous magnetic moment}\label{app:gminus2}

From Eq.~(\ref{eq:amu-L}), the Lagrangian for the muon current coupling with the gauge bosons, relevant to the contributions to $\Delta a_\mu$, is given by
\begin{align}
{\cal L} \supset  & - (g_{\mu\tau}  + e \, \epsilon_A + \epsilon_Z \, r_{ZZ^\prime}\,  g_V \, g_Z) \bar\mu \gamma^\mu \mu \, Z^\prime_\mu
 - \epsilon_Z \, r_{ZZ^\prime}\,  g_A \, g_Z \bar\mu \gamma^\mu\gamma_5  \mu \, Z^\prime_\mu \nonumber \\
 & -\big(g_Z \, g_V - \epsilon_Z \, (1+r_{Z Z^\prime} ) g_{\mu\tau} \big) \,  \bar\mu \gamma^\mu \mu \,  Z_\mu 
 - g_Z \, g_A  \bar\mu \gamma^\mu \gamma_5 \mu \,  Z_\mu   \,,
\end{align}
where $g_V=-1/4 + \sin^2\theta_W$, $g_A=1/4$, and $r_{Z Z^\prime} = m_{Z^\prime}^2 /(m_Z^2 -m_{Z^\prime}^2)$.
Using the generic formulas for neutral vector bosons coupling to muons \cite{Jegerlehner:2009ry}, the total contribution to the anomaly in this model is evaluated as
\begin{align}
\Delta a_\mu  & =  \frac{ (g_{\mu\tau}  + e \, \epsilon_A + \epsilon_Z \, r_{Z Z^\prime}\,  g_V \, g_Z)^2 }{ 4 \pi^2} \frac{m_\mu^2}{m_{Z^\prime}^2} I_V(m_{Z^\prime}) 
                       + \frac{(\epsilon_Z \, r_{Z Z^\prime}\,  g_A \, g_Z)^2 }{ 4 \pi^2} \frac{m_\mu^2}{m_{Z^\prime}^2} I_A(m_{Z^\prime})  \nonumber\\
                      & \ \ + \frac{ \big( g_Z\, g_V - \epsilon_Z \, (1+ r_{Z Z^\prime} ) \,  g_{\mu\tau}  \big)^2  }{ 4 \pi^2} \frac{m_\mu^2}{m_{Z}^2} I_V(m_{Z}) 
                       + \frac{ ( g_Z \, g_A )^2 }{ 4 \pi^2} \frac{m_\mu^2}{m_{Z}^2} I_A(m_{Z}) \,,
\end{align}
with
\begin{align}
I_V(m_0)  &= \int_0^1 dx \frac{m_0^2 \, x^2 (1-x)}{m_0^2 (1-x) + m_\mu^2 \, x^2}  \,,  \\
I_A (m_0)  &= \int_0^1 dx \frac{m_0^2 \, x (1-x) (x-4) -2 x^3 m_\mu^2 }{m_0^2 (1-x) + m_\mu^2 \, x^2} \,.
\end{align}

Using the relation $\epsilon_Z = -\epsilon_A \tan\theta_W$ and considering the limit $r_{Z Z^\prime} \ll 1$ (i.e., $m_{Z^\prime}^2 \ll m_Z^2 $), the dominant contribution comes from the $\epsilon_A$ term while the $\epsilon_Z$ contribution is suppressed. This leads to the simplified expression as
\begin{align}
\Delta a_\mu  & \simeq \frac{ (g_{\mu\tau} + e \, \epsilon_A )^2 }{ 4 \pi^2} \int_0^1 dx \frac{m_{\mu}^2 \, x^2 (1-x)}{m_{Z^\prime}^2 (1-x) + m_\mu^2 \, x^2} \,.
\end{align}

\section{Decay widths}\label{app:widths}

\allowdisplaybreaks
\begin{align}
\Gamma_{S \to Z^\prime Z^\prime} &=\frac{ (g_{\chi}  \cos\alpha)^2 }{8 \pi} \frac{ m_S^3}{m_{Z^\prime}^2} 
\left(1 -\frac{4m_{Z^\prime}^2}{m_S^2} \right)^{1/2} \left(1 -\frac{4m_{Z^\prime}^2}{m_S^2} + \frac{12 m_{Z^\prime}^4}{m_S^4}\right) \,,
\label{eq:width-SZpZp} 
\\
%
\Gamma_{Z^\prime\to f \bar{f}} &=   \left\{ \begin{array}{lr} 
\frac{g_{\mu\tau}^2}{12\pi} m_{Z^\prime}\Big(  1+ \frac{2 m_f^2}{m_{Z^\prime}^2}\Big) \Big(  1- \frac{4 m_f^2}{m_{Z^\prime}^2}  \Big)^{1/2} \theta(m_{Z^\prime} - 2m_f)
        \,,  & \text{for}\  f=\mu \  {\rm or\ } \tau\,,
        \\ 
        \\
   \frac{g_{\mu\tau}^2}{24\pi} m_{Z^\prime}  \,, \  & \text{for}\  f=\nu_\mu\  {\rm or\ }  \nu_\tau\,,
    \end{array}
    \right.  \\
   \nonumber \\
\Gamma_{Z^\prime\to e^+ e^-} &= \frac{\alpha \epsilon_A^2 m_{Z^\prime}}{3} \left(1+ \frac{2 m_e^2}{m_{Z^\prime}^2} \right)  \left(1- \frac{4 m_e^2}{m_{Z^\prime}^2} \right)^{1/2} \,,
\label{eq:Zp-partial-width-f} 
\\
%
\Gamma_{H\to \chi_\pm \chi_\pm} &= \frac{f^2 \sin^2\alpha}{16\pi m_H^2} (m_H^2 - 4 m_\pm^2)^{3/2}\, \theta(m_H-2m_\pm)\,,
\label{eq:width-Hchichi} 
\\
%
\Gamma_{H\to SS} &= \frac{1}{128 \pi m_H^2 v_S^2 v_H^2} (m_H^2 + 2 m_S^2)^2 (m_H-4 m_S^2)^{1/2}
                                  (v_H \cos\alpha +v_S\sin\alpha)^2 \sin^2 2\alpha \,,
\label{eq:width-HSS}                                   
\\
%
\Gamma_{H \to Z^\prime Z^\prime} &=\frac{ (g_\chi \sin\alpha)^2 }{8 \pi} \frac{ m_H^3}{m_{Z^\prime}^2} 
\left(1 -\frac{4m_{Z^\prime}^2}{m_H^2} \right)^{1/2}  \left(1 -\frac{4m_{Z^\prime}^2}{m_H^2} + \frac{12 m_{Z^\prime}^4}{m_H^4}\right) \,.
\end{align}

\section{Cross sections relevant to the DM relic abundance}\label{appsec:xs}

The thermally averaged annihilation cross section for the process $i \,j \to k\, \ell $ at temperature $T$ is represented by
\cite{Edsjo:1997bg}
\begin{align}
\langle \sigma v  \rangle_{ij\to k\ell} =& \frac{g_i g_j T}{8 \pi^4 \,  n_i^{\rm eq} (T) \,  n_j^{\rm eq}(T)}
 \int_{ (m_i +m_j)^2 }^{\infty} 
\sigma_{ij\to k\ell}  \,  p_{ij}^2 \,  \sqrt{s}  K_1 \left( \frac{\sqrt{s}}{T} \right) ds, 
\label{app:thermal-average}
\end{align}
where $K_{n}$ is the modified Bessel function of the second kind, and the Mandelstam variable $s$ is the square of the center-of-mass energy. The momentum of particle $i$ (or $j$) in the center-of-mass frame is
\begin{align}
p_{ij} & =\frac{ [s-  (m_i+m_j)^2]^{1/2}  [s-  (m_i - m_j)^2]^{1/2} }{2 \sqrt{s}},
\end{align}
and the equilibrium number density of $i$ at temperature $T$ is given by
\begin{equation}
n_{i}^{\rm eq} (T) =\frac{T}{2\pi^2} g_i m_i^2  K_2 \Big( \frac{m_i}{T} \Big),
\end{equation}
with $g_i$ being its internal degrees of freedom.

Expanding the cross section in powers of the relative velocity $v$, $(\sigma v)_{ij\rightarrow k \ell} = a + b v^2 + c v^4 + \dots$, the thermal average in the $m_i \approx m_j$ limit yields \cite{Gondolo:1990dk}:
\begin{align}
\langle \sigma v  \rangle_{ij\to k\ell} =& a + 6 b \, \frac{T}{m_i} + ( 18 b + 60 c ) \Big( \frac{T}{m_i} \Big)^2 + \cdots \,.
 \label{app:thermal-average-NR}
\end{align}

\subsection{\texorpdfstring{$\chi_\mp \chi_\mp \to S \, S$}{}}\label{appsub:xs-chichi2ss}

The cross section for $\chi_\mp \chi_\mp \rightarrow SS$, proceeding via $t$- and $u$-channel $\chi_\pm$ exchanges and the $s$-channel $S$ contribution, is given by 
{
\begin{align}
\sigma_{\chi_\mp \chi_\mp \to S \, S}& =
 \frac{f^2 c_\alpha^2\sqrt{s-4 m_S^2} }{32 \pi  s m_Z^2 \sqrt{s-4 m_{\mp}^2} \left(\Gamma_S^2 m_S^2+ (s-m_S^2 )^2\right)}
   \nonumber\\
   & \times
 \Bigg[ -\frac{f^2 c_\alpha^2 m_Z^2 \left( ( \Gamma_S^2-2 s ) m_S^2+m_S^4+s^2\right)
   \left(2 m_{\mp}^2 \left(s-8 m_S^2\right)+16 m_{\mp}^4+3 m_S^4\right)}{m_{\mp}^2 \left(s-4
   m_S^2\right)+m_S^4}  \nonumber\\
   & \ \ \
    +  \frac{f c_\alpha m_Z}{\sqrt{s-4 m_{\mp}^2} \sqrt{s-4 m_S^2} \left( s - 2 m_S^2 \right)}
     \left(
    \log \frac{ s+ \sqrt{s-4 m_{\mp}^2} \sqrt{s-4 m_S^2}- 2 m_S^2 }{  s- \sqrt{s-4 m_{\mp}^2} \sqrt{s-4 m_S^2} - 2 m_S^2 } \right) 
 \nonumber\\
 & \ \ \ \  \times
   \bigg(f c_\alpha m_Z
   \left(\Gamma_S^2 m_S^2+ ( s-m_S^2 )^2\right) \left( s^2 -4 m_S^2 (4 m_{\mp}^2+s )+16 s m_{\mp}^2-32 m_{\mp}^4+6 m_S^4 \right) \nonumber\\
  & \ \ \ \ \
   -12 g_\chi c_\alpha^3 m_{\mp} m_S^2 ( s^2 -3 s m_S^2+2 m_S^4) ( s-8 m_{\mp}^2+2 m_S^2) \bigg)
   \nonumber\\
 & 
  - 24 f  g_\chi c_\alpha^4 m_{\mp} m_S^2 m_Z  (s- m_S^2)+18 g_\chi^2 c_\alpha^6 m_S^4 (s-4  m_{\mp}^2 ) \Bigg] \,,
 \end{align}
 }
 where $c_\alpha \equiv \cos\alpha$, and $s$ is the Mandelstam invariant mass squared of two incident particles. Here, we have used the triple coupling to calculate the $s$-channel amplitude,
 \begin{equation}
 {\cal L}\supset \frac{1}{6} g_{SSS} SSS + \cdots\,, \ \text{with} \ g_{SSS} \simeq -\frac{6g_\chi c_\alpha^3 m_S^2}{m_{Z^\prime}} \,.
 \label{app:gSSS}
 \end{equation}
 Because the triple scalar coupling $g_{SSS}$ is explicitly proportional to $m_S^2$, the $s$-channel $S$-exchange amplitude is parametrically suppressed compared to the dominant $t/u$-channel $\chi$-exchanges. Consequently, its contribution is safely absorbed into the $\mathcal{O}(m_S^2/m_\mp^2)$ mass ratio corrections.
 Considering two annihilating particles in the non-relativistic limit, the cross section is approximated as
 \begin{align}
(\sigma v)_{\chi_\mp \chi_\mp \to SS}  \simeq 
\frac{3 f^4}{ 128 \pi m_\mp^2} \left[ v^2  - \frac{13 }{15} v^{4} +  {\cal O} \bigg(\frac{m_S^2}{m_\mp^2}, \frac{m_{Z'}^2}{m_\mp^2}, v^6 \bigg) \right] \,.
\end{align}
The $s$-wave contribution is heavily suppressed, rendering the annihilation inherently $p$-wave dominant, with a leading-order velocity dependence proportional to $v^2$.

\subsection{\texorpdfstring{$\chi_\mp \chi_\mp \to  Z^\prime \, Z^\prime$}{}}\label{appsubsec:xs-chimpchimp2zpzp}

The cross-section is given by
{\footnotesize 
  \begin{align} 
& \sigma_{\chi_\mp \chi_\mp \to  Z^\prime  Z^\prime} 
 = \frac{g_{\chi}^2 \sqrt{s-4 m_{Z^\prime}^2}  \big(s - 2 (m_{\mp}^2-m_{\pm}^2+m_{Z^\prime}^2 ) \big){}^2  }{32 \pi m_{Z^\prime}^4 \big( ( s-m_S^2 )^2 + \Gamma_S^2 m_S^2 \big)
  s  \sqrt{s- 4 m_{\mp}^2}  \big(2 (m_{\mp}^2-m_{\pm}^2+m_{Z^\prime}^2 )-s \big){}^3}\nonumber\\
    &
\times \Bigg\{
      \frac{ \big( s -2 (m_{\mp}^2-m_{\pm}^2+m_{Z^\prime}^2 ) \big)}{m_{\mp}^4-2 (m_{\pm}^2+m_{Z^\prime}^2 ) m_{\mp}^2+s
   m_{\pm}^2+ (m_{\pm}^2-m_{Z^\prime}^2 ){}^2}
   \Big[ g_{\chi}^2 \Gamma_S^2 m_S^2   
  \big[ 3
   m_{\mp}^8-12 m_{\pm}^2 m_{\mp}^6+6 m_{Z^\prime}^2 m_{\mp}^6
   \nonumber\\
   & +28 m_{\pm} m_{Z^\prime}^2 m_{\mp}^5 +18 m_{\pm}^4 m_{\mp}^4-13 m_{Z^\prime}^4 m_{\mp}^4-6 m_{\pm}^2 m_{Z^\prime}^2
   m_{\mp}^4-20 m_{\pm} m_{Z^\prime}^4 m_{\mp}^3-56 m_{\pm}^3 m_{Z^\prime}^2 m_{\mp}^3
   \nonumber\\
   & -12 m_{\pm}^6 m_{\mp}^2 -4 m_{Z^\prime}^6 m_{\mp}^2+2 m_{\pm}^2 m_{Z^\prime}^4 m_{\mp}^2-6
   m_{\pm}^4 m_{Z^\prime}^2 m_{\mp}^2-8 m_{\pm} m_{Z^\prime}^6 m_{\mp}-20 m_{\pm}^3 m_{Z^\prime}^4   m_{\mp}
   \nonumber\\
   & +28 m_{\pm}^5 m_{Z^\prime}^2 m_{\mp} +3 m_{\pm}^8+8 m_{Z^\prime}^8-4 m_{\pm}^2   m_{Z^\prime}^6-13 m_{\pm}^4 m_{Z^\prime}^4-2 s^2 m_{\pm}^2   (m_{\mp}+m_{\pm} ){}^2+6 m_{\pm}^6 m_{Z^\prime}^2
   \nonumber\\
   & -2 s   \big(  (m_{\mp}^2+2 m_{\pm} m_{\mp}-m_{\pm}^2 ) m_{Z^\prime}^4-2   (m_{\mp}+m_{\pm} ){}^2  (m_{\mp}^2+3 m_{\pm}^2 ) m_{Z^\prime}^2 
   +m_{\mp}  (m_{\mp}+2  m_{\pm} )  (m_{\mp}^2-m_{\pm}^2 ){}^2 \big) 
    \big] 
    \nonumber\\
    &  - g_{\chi}^2 (m_S^2-s )^2  \big[ -3 m_{\mp}^8+12 m_{\pm}^2 m_{\mp}^6-6 m_{Z^\prime}^2 m_{\mp}^6  
     -28 m_{\pm} m_{Z^\prime}^2 m_{\mp}^5-18 m_{\pm}^4 m_{\mp}^4+13 m_{Z^\prime}^4 m_{\mp}^4
   \nonumber\\
   & +6 m_{\pm}^2 m_{Z^\prime}^2 m_{\mp}^4+20 m_{\pm} m_{Z^\prime}^4
   m_{\mp}^3+56 m_{\pm}^3 m_{Z^\prime}^2 m_{\mp}^3 +12 m_{\pm}^6 m_{\mp}^2+4 m_{Z^\prime}^6
   m_{\mp}^2-2 m_{\pm}^2 m_{Z^\prime}^4 m_{\mp}^2
   \nonumber\\
   & +6 m_{\pm}^4 m_{Z^\prime}^2 m_{\mp}^2 
   +8 m_{\pm} m_{Z^\prime}^6 m_{\mp}+20 m_{\pm}^3 m_{Z^\prime}^4 m_{\mp}
   -28 m_{\pm}^5 m_{Z^\prime}^2  m_{\mp}-3 m_{\pm}^8-8 m_{Z^\prime}^8+4 m_{\pm}^2 m_{Z^\prime}^6
   \nonumber\\
   & +13 m_{\pm}^4 m_{Z^\prime}^4+2 s^2 m_{\pm}^2  (m_{\mp}+m_{\pm} ){}^2
    -6 m_{\pm}^6 m_{Z^\prime}^2 + 2 s  \big( (m_{\mp}^2+2 m_{\pm} m_{\mp}-m_{\pm}^2 ) m_{Z^\prime}^4
   \nonumber\\
   & -2  (m_{\mp}+m_{\pm} ){}^2  (m_{\mp}^2+3 m_{\pm}^2 )  m_{Z^\prime}^2
   +m_{\mp}  (m_{\mp}+2 m_{\pm} )  (m_{\mp}^2-m_{\pm}^2 ){}^2 \big) \big] 
   \nonumber\\
   & +4   g_{\chi} f m_{Z^\prime} (m_S^2-s ) \big( -  (m_{\mp}+m_{\pm} ) s^2
    +2 (m_{\mp}+m_{\pm} )  (m_{\mp}^2-m_{\pm} m_{\mp}+m_{Z^\prime}^2 ) s
  \nonumber\\
  &  +4 m_{\mp} m_{Z^\prime}^2  (m_{\mp}^2-m_{\pm}^2-2 m_{Z^\prime}^2 ) \big) 
   \big(m_{\mp}^4-2  (m_{\pm}^2+m_{Z^\prime}^2 ) m_{\mp}^2
    +s   m_{\pm}^2+ (m_{\pm}^2-m_{Z^\prime}^2 ){}^2 \big) 
    \nonumber\\
    & +2 f^2  (4 m_{\mp}^2-s ) m_{Z^\prime}^2  (12 m_{Z^\prime}^4-4 s m_{Z^\prime}^2+s^2 )  
   \big(m_{\mp}^4-2 (m_{\pm}^2+m_{Z^\prime}^2 ) m_{\mp}^2
   +s m_{\pm}^2+ (m_{\pm}^2-m_{Z^\prime}^2 ){}^2 \big) \Big]
   \nonumber\\
   & -   \frac{g_{\chi} }{\sqrt{s-4 m_{\mp}^2}
   \sqrt{s-4 m_{Z^\prime}^2}} 
   \Bigg[  \log  \Bigg(\frac{ 2 m_{\mp}^2-2 m_{\pm}^2+2 m_{Z^\prime}^2-s-\sqrt{s-4 m_{\mp}^2} \sqrt{s-4 m_{Z^\prime}^2} }
   { 2  m_{\mp}^2-2 m_{\pm}^2+2 m_{Z^\prime}^2-s+\sqrt{s-4 m_{\mp}^2} \sqrt{s-4 m_{Z^\prime}^2} }\Bigg) 
   \nonumber\\
   &\times \bigg[ \Gamma_S^2 g_{\chi}   \big[ 16 m_{Z^\prime}^8-8  (5 m_{\mp}^2+2 m_{\pm} m_{\mp}+m_{\pm}^2 ) m_{Z^\prime}^6-2  (m_{\mp}-m_{\pm} )  
   (5 m_{\mp}^3-23 m_{\pm} m_{\mp}^2
   \nonumber\\
   & -33 m_{\pm}^2 m_{\mp}-13 m_{\pm}^3 )
   m_{Z^\prime}^4+4  (m_{\mp}^2-m_{\pm}^2 ){}^2  (7 m_{\mp}^2+14 m_{\pm}
   m_{\mp}+3 m_{\pm}^2 ) m_{Z^\prime}^2+6  (m_{\mp}^2-m_{\pm}^2 ){}^4
   \nonumber\\
   & +s^2   (4 m_{Z^\prime}^4+4  (m_{\mp}+m_{\pm} ){}^2 m_{Z^\prime}^2+ (m_{\mp}-3
   m_{\pm} )  (m_{\mp}+m_{\pm} ){}^3 )
   -4 s  \big(4 m_{\mp} (m_{\mp}+2 m_{\pm} ) m_{Z^\prime}^4
    \nonumber\\
    & + (3 m_{\mp}-5 m_{\pm} )
    (m_{\mp}+m_{\pm} ){}^3 m_{Z^\prime}^2+2 m_{\mp}
    (m_{\mp}-m_{\pm} ){}^2  (m_{\mp}+m_{\pm} ){}^3 \big) \big]
   m_S^2
   \nonumber\\
   & + (m_S^2-s )  \Big[ 4 f m_{Z^\prime}  \big(2   (m_{\mp}^2-m_{\pm}^2+m_{Z^\prime}^2 )-s \big)  
    \big[ s^2 m_{\pm}  (m_{\mp}+m_{\pm} ){}^2
    \nonumber\\
    & + 4 m_{\mp}  (m_{\mp}+m_{\pm}-m_{Z^\prime} )
   m_{Z^\prime}^2  (m_{\mp}+m_{\pm}+m_{Z^\prime} )
    \big( (m_{\mp}-m_{\pm} ){}^2+2 m_{Z^\prime}^2 \big) + 2 s  
    \big(2 (m_{\mp}-m_{\pm} ) m_{Z^\prime}^4
    \nonumber\\
    & - (m_{\mp}+m_{\pm} )  (3  m_{\mp}^2+m_{\pm}^2 ) m_{Z^\prime}^2+m_{\mp} (m_{\mp}^2-m_{\pm}^2 ){}^2 
   \big) 
   \big]  + (m_S^2-s ) g_{\chi}
   \big[ 16 m_{Z^\prime}^8-8  (5 m_{\mp}^2
    \nonumber\\
    & +2 m_{\pm} m_{\mp}+m_{\pm}^2 ) m_{Z^\prime}^6-2
    (m_{\mp}-m_{\pm} )  (5 m_{\mp}^3-23 m_{\pm} m_{\mp}^2-33
   m_{\pm}^2 m_{\mp}-13 m_{\pm}^3 ) m_{Z^\prime}^4
   \nonumber\\
   &  +4   (m_{\mp}^2-m_{\pm}^2 ){}^2  (7 m_{\mp}^2+14 m_{\pm} m_{\mp}+3
   m_{\pm}^2 ) m_{Z^\prime}^2+6  (m_{\mp}^2-m_{\pm}^2 ){}^4
   +s^2  \big(4 m_{Z^\prime}^4+4  (m_{\mp}+m_{\pm} ){}^2 m_{Z^\prime}^2
    \nonumber\\
    &  + (m_{\mp}-3 m_{\pm} ) (m_{\mp}+m_{\pm} ){}^3 \big)
    -4 s  \big(4 m_{\mp}  (m_{\mp}+2 m_{\pm} ) m_{Z^\prime}^4+ (3 m_{\mp}-5 m_{\pm} ) (m_{\mp}+m_{\pm} ){}^3 m_{Z^\prime}^2
    \nonumber\\
    &  +2 m_{\mp}    (m_{\mp}-m_{\pm} ){}^2  (m_{\mp}+m_{\pm} ){}^3 \big)  \big]   \Big]   \bigg]
    \Bigg]
   \Bigg\} 
   \,. \label{app:zpzp}
  \end{align} 
  } 
  
 For two incident particles in the non-relativistic limit, the annihilation cross section, performed the Laurent series expansion around $m_{Z^\prime}=0$, is approximated as (for $(\delta{m}/m_-)^2 \ll 1$)
\begin{align}
&(\sigma v)_{\chi_- \chi_- \to Z^\prime Z^\prime}  \nonumber\\
& \simeq 
\frac{g_\chi^4}{16 \pi m_-^2} \Bigg\{ \left(1- 2\frac{\delta{m}}{m_-}\right) + \frac{1}{2} \bigg(\frac{ f m_-}{ g_\chi m_{Z^\prime}}\bigg)^2  
   \left[ 1 - \frac{8}{3} \frac{g_\chi m_-}{f m_{Z^\prime}}  \left(1-\frac{\delta{m}}{ 4m_-} \right)
   + \frac{8}{3} \Big( \frac{g_\chi m_-}{f m_{Z^\prime}} \Big)^2  \left(1-\frac{\delta{m}}{ m_-} \right) \right] v^2
\nonumber\\
& -  \frac{1}{2} \bigg(\frac{ f m_-}{ g_\chi m_{Z^\prime}}\bigg)^2  
   \left[ \frac{1}{2} - \frac{6}{5} \frac{g_\chi m_-}{f m_{Z^\prime}}   \left(1-\frac{\delta{m}}{2 m_-} \right)  
   + \frac{6}{5} \Big( \frac{g_\chi m_-}{f m_{Z^\prime}}   \Big)^2 \left(1-\frac{17 \delta{ m}}{9 m_-} \right)  \right] v^4
  \Bigg\} \,,
\end{align}
where the contributions proportional to $v^0$ and $v^4$ are much smaller than the $v^2$ term. Compared with the $s$-wave, the $p$-wave term is kinematically enhanced by a factor of $(fm_-/(g_\chi m_{Z^\prime}))^2$.

\subsection{\texorpdfstring{$\chi_- \chi_+ \to \ell \bar\ell, \nu_\ell \bar\nu_\ell$}{}}\label{appsubsec:xs-chimchip2ll}

\begin{align}
 & \sigma_{\chi_- \chi_+ \to \ell \bar\ell}  =
 \frac{g_{\mu\tau}^2 g_\chi^2  \sqrt{s- (m_+ + m_- )^2}  \left( 2s+ (m_+ - m_- )^2 \right) \sqrt{s-4 m_\ell^2}  (2 m_\ell^2+s )}{24 \pi  s^{3/2} \sqrt{s- (m_+ - m_- )^2} 
   \left( (s -m_{Z^\prime})^2+\Gamma ^2 m_{Z^\prime}^2 \right)} \,, \nonumber\\
 & \sigma_{\chi_- \chi_+ \to \nu_\ell \bar\nu_\ell}  =
 \frac{g_{\mu\tau}^2 g_\chi^2 \sqrt{s-  (m_+ + m_- )^2}
   \left( 2s + (m_+ - m_- )^2 \right)}{48 \pi  \sqrt{s- (m_+ - m_- )^2} \left( (s -m_{Z^\prime})^2+\Gamma ^2 m_{Z^\prime}^2 \right)} \,,
 \end{align}
where $\ell =\mu$ or $\tau$, 
$\Gamma_{Z^\prime}$  is the $Z^\prime$ width, which in general reads
$\Gamma_{Z^\prime} = \sum_f \Gamma_{Z^\prime \to f\bar f} +  \Gamma_{Z^\prime \to \chi_- \chi_+} +  \Gamma_{Z^\prime \to e\bar{e}} $ with the sums over $f=\mu, \tau, \nu_{\mu}$ and $\nu_{\tau}$, given by
\begin{align}
\Gamma_{Z^\prime} = 
 &\frac{g_{\mu\tau}^2 m_{Z^\prime}}{12\pi}
    \bigg[
      1 + \sum_{i = \mu,\tau} (1+ 2 \kappa_i)(1-4 \kappa_i)^{1/2} \, \theta(1-2 \sqrt{\kappa_i} )
    \bigg] \nonumber\\
 & +  \frac{g_{\mu\tau}^2 m_{Z^\prime}}{12\pi}
    \left[ 1 -(\kappa_+ + \kappa_-)^2 \right]^{3/2}
     \left[ 1 -(\kappa_+ - \kappa_-)^2 \right]^{1/2}
      \left[ 1 + \frac{(\kappa_+ - \kappa_-)^2}{2} \right] \theta(1- \sqrt{\kappa_+} - \sqrt{\kappa_-} ) \nonumber\\
   &+  \frac{e^2 \epsilon_A^2 m_{Z^\prime}}{12\pi}
    \bigg[
      (1+ 2 \kappa_e)(1-4 \kappa_e)^{1/2} \, \theta(1-2 \sqrt{\kappa_e} )
    \bigg] 
    \,,
\end{align}
with $\kappa_i \equiv m_i^2/ m_{Z^\prime}^2$. In the present study, $Z^\prime \to \chi_- \chi_+$ is kinetically blocked because $m_{Z^\prime} < m_+ +m_-$.
Considering two annihilating particles in the non-relativistic limit, the cross section is given by
\begin{align}
 & (\sigma v)_{\chi_- \chi_+ \to \ell \bar\ell}  =
 \frac{g_{\mu\tau}^2 g_\chi^2 }{96 \pi m_-^2} v^2   
 \left[  1 + \frac{\delta m}{m_-} +  {\cal O}\left(\frac{\delta m^2}{m_-^2}, \frac{m_{Z^\prime}^2}{m_-^2}, \frac{m_\ell^2}{m_-^2}\right) \right]
  \,, \nonumber\\
 & (\sigma v)_{\chi_- \chi_+ \to \nu_\ell \bar\nu_\ell}  =
\frac{g_{\mu\tau}^2 g_\chi^2 }{192 \pi m_-^2} v^2   
 \left[  1 + \frac{\delta m}{m_-} +  {\cal O}\left(\frac{\delta m^2}{m_-^2}, \frac{m_{Z^\prime}^2}{m_-^2} \right) \right]
 \,,
 \end{align}
where $m_+ = m_- + \delta m$.

\subsection{\texorpdfstring{$\chi_- \chi_+ \to Z^\prime S$}{}}\label{appsubsec:xs-chimchip2zps}

The annihilation cross section for the coannihilation channel $\chi_- \chi_+ \to Z^\prime S$ is given by
\begin{align}
&\sigma_{\chi_- \chi_+ \to Z^\prime S}  \simeq
 \frac{g_\chi^4}{16 \pi  m_-^3 m_{Z'}^4 v}
  \left(16 \delta{m} m_-^4-3 \delta{m}
   m_{Z'}^4+16 m_-^5-12 m_-^3 m_{Z'}^2+3 m_-
   m_{Z'}^4\right)
 \nonumber\\
&\ \ + \frac{g_\chi^2 v}{384 \pi  m_-^4 m_{Z'}^4}
 \bigg[ m_-  \Big(16 c_\theta^2 f^2 m_- m_{Z'}^2 \left(m_-^2+2
   m_{Z'}^2\right)\nonumber\\
&\ \ -8 c_\theta f g_\chi m_{Z'} \left(8
   m_-^4-2 m_-^2 m_{Z'}^2+5 m_{Z'}^4\right)+g_\chi^2
   m_- \left(48 m_-^4+36 m_-^2 m_{Z'}^2-23
   m_{Z'}^4\right) \Big)
  \nonumber\\
&\ \ + \delta{m} \Big(-32 c_\theta^2 f^2  m_- m_{Z'}^4+4 c_\theta f g_\chi m_{Z'} \left(-8 m_-^4-2 m_-^2 m_{Z'}^2+15 m_{Z'}^4\right) 
 \nonumber\\
 &\ \ + g_\chi^2 m_- \left(48 m_-^4+23 m_{Z'}^4\right) \Big)   
  \bigg] \,,
 \end{align}
where we have accounted for two annihilating particles in the non-relativistic limit, neglected $m_S$, and used $m_+ = m_- + \delta m$ in the expansion. Using the relation $g_\chi = f m_{Z'}/(m_- + m_+)$, the cross section can be further approximated as
\begin{align}
( \sigma v )_{\chi_- \chi_+ \to Z^\prime S}  = \frac{f^4}{ 16 \pi m_-^2}  \left(1 - \frac{\delta{m}}{m_-} \right)
\left(1 + \frac{v^2 (4 c_\theta^2 - 8c_\theta +3 )}{24} \right)
 + {\cal O}(m_{Z'}, m_S, \delta m^2, v^4)
\,,
\end{align}
As explicitly shown above, the $s$-wave contribution does not vanish in the degenerate limit ($\delta m \to 0$), yielding a substantial cross section proportional to $f^4 / m_-^2$. If $\chi_+$ were abundant, this powerful $s$-wave channel would strictly dominate the freeze-out process, severely restricting $f$ and spoiling the required large self-interactions. However, as detailed in Sec.~\ref{sec:boltz}, the mass splitting in our framework exponentially suppresses the primordial $\chi_+$ abundance. This thermodynamical depletion safely neutralizes the $s$-wave coannihilation, naturally ensuring robust $p$-wave dominance for the $\chi_-$ ground state.

\section{\texorpdfstring{Kinetic decoupling of $\chi_\pm$ and $S$}{}}\label{app:chiS-el}

\begin{figure}[th!]
\begin{center}
\includegraphics[width=0.76\textwidth]{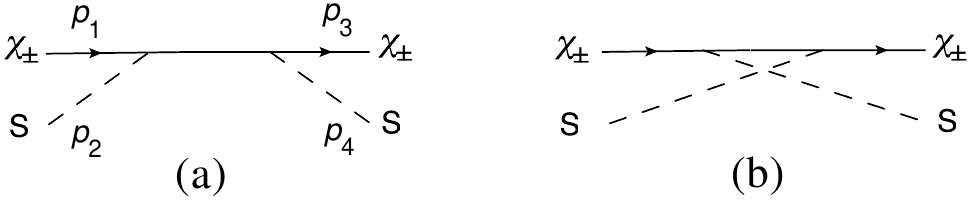}
\caption{Feynman diagrams for the $\chi_\pm \, S \to \chi_\pm \, S$ elastic scattering process, where the corresponding momenta $p_i$ with $i=1,2,3,4$ are shown.}
\label{fig:maj_el}
\end{center}
\end{figure}

Since the $\chi_+$ population is exponentially suppressed well before the epoch of kinetic decoupling, the kinetic equilibrium of the dark sector is overwhelmingly maintained by the ground-state scattering $\chi_- S \leftrightarrow \chi_- S$. For generality in the mathematical derivation below, we use the notation $\chi_\pm$, noting that the physical dynamics are entirely governed by $\chi_-$ in our relevant parameter space.

Following the chemical freeze-out of dark matter, the mediator particles $S$ and $Z^\prime$ remain in thermal equilibrium with the SM bath through the interactions $S \leftrightarrow Z^\prime Z^\prime$ and $Z^\prime \leftrightarrow \nu_\mu \bar\nu_\mu, \nu_\tau \bar\nu_\tau$. As illustrated in Fig.~\ref{fig:temp-ratio}, the interaction $Z^\prime \leftrightarrow e^+ e^-$ facilitates the thermal connection between the $(\gamma, e^\pm$) and ($\nu_\mu, \nu_\tau, S, Z^\prime$) baths, ensuring that $T =T_{\nu_\mu}$ is maintained until the temperature roughly drops below $m_e$. As the universe further cools, the $S$ particles, which are exponentially suppressed by Boltzmann factors with a vanishing chemical potential, strictly follow the temperature $T_{\nu_\mu}$.

Through the aforementioned $\chi_- S \leftrightarrow \chi_- S$ elastic scattering, dark matter remains in kinetic equilibrium with the SM-linked bath long after its annihilation ceases. This extended equilibrium occurs because the elastic scattering rate is proportional to the number density of $S$, which remains highly relativistic and avoids Boltzmann suppression until the temperature drops well below $T_{\rm fo}$. In the following analysis, we demonstrate that dark matter with masses ranging from a few GeV to hundreds of GeV kinetically decouples from $S$ at temperatures between 210 and 250 MeV. Therefore, when studying the kinetic decoupling of dark matter, the $S$ particles can be safely treated as an integral component of the thermal bath, sharing the same temperature as the photons.

Fig.~\ref{fig:maj_el} depicts the diagrams for elastic scattering $\chi_\pm \ S \to \chi_\pm \ S$, which are relevant for the kinetic energy transfer between $\chi_\pm$ and $S$. The $u$-channel diagram (on the right) is derived from the $s$-channel diagram (on the left) through crossing symmetry. Driven by $g_{SSS} \propto m_S^2$ (Eq.~(\ref{app:gSSS})), the $t$-channel $S$-exchange amplitude is relatively suppressed compared with the dominant $s/u$-channels and thus safely negligible. For nonrelativistic dark matter, its temperature evolution follows the equation
\begin{align}
      \frac{d T_{\rm DM}}{dt} + & 2  H  T_{\rm DM} 
      =  
         -  \frac{1}{n} \sum_{i=+,-} 2 n_i\  \gamma_i  \,  (T_{\rm DM} -T)  +\dots         \,,
       \label{app:boltz-t-Tdm}
\end{align}
where $n=n_+ + n_-$, and $2\gamma_\pm$ is the average rate of kinetic energy transfer from $S$ to $\chi_\pm$ at temperature $T$. 

The $\gamma_i$ values can be obtained by reformulating the thermally averaged collision term for elastic scattering $\chi_\pm  S \to \chi_\pm S$ using a semi-relativistic Fokker-Planck-like equation \cite{Binder:2016pnr}, expressed as
\begin{align}
\gamma_i&=  \frac{1}{6m_i T n_i}
 \int  \prod_{\substack{ j=1,2 \\  k=3,4}}
 d\Pi_{j}  \, d\Pi_{k}  f_j (T_j)   (2\pi)^4 \delta^{(4)} (p_{1} +p_{2} - p_{3} - p_{4} )  |M|_{\chi_\pm S \to \chi_\pm S}^2  (-1) (p_1 - p_4)^2 \nonumber\\
 & \simeq   \frac{1}{6 \pi^2 g_i m_i^2 T  K_2(m_i/T_{\rm DM})}
  \int_{ (m_i+m_S)^2 }^{\infty} ds \, \frac{(s- (m_i +m_S)^2) (s-(m_i-m_S)^2)}{4 \sqrt{s}} K_1 \left(\frac{\sqrt{s}}{T} \right) \nonumber\\
 & \hskip4.5cm  \times \int \frac{d\sigma}{d\Omega} m_\pm v_\pm^2 (1-\cos\theta) d\Omega \,,
 \label{app:gamma-chi}
\end{align}
with the distributions approximated by
\begin{align}
f_{i} =  e^{-(E_{i}-\mu_i)/T_i} (1\pm f_i) \simeq e^{-(E_{i}-\mu_i)/T_i}  =\frac{n_i}{n_i^{\rm eq}(T_i)} e^{-E_{i}/T_i}   \,,
\end{align}
where $n_S = n_S^{\rm eq}(T)$, $T_S=T$, and $T_- \equiv T_{\rm DM}$. 

\begin{figure}[ht!]
\begin{center}
\includegraphics[width=0.40\textwidth]{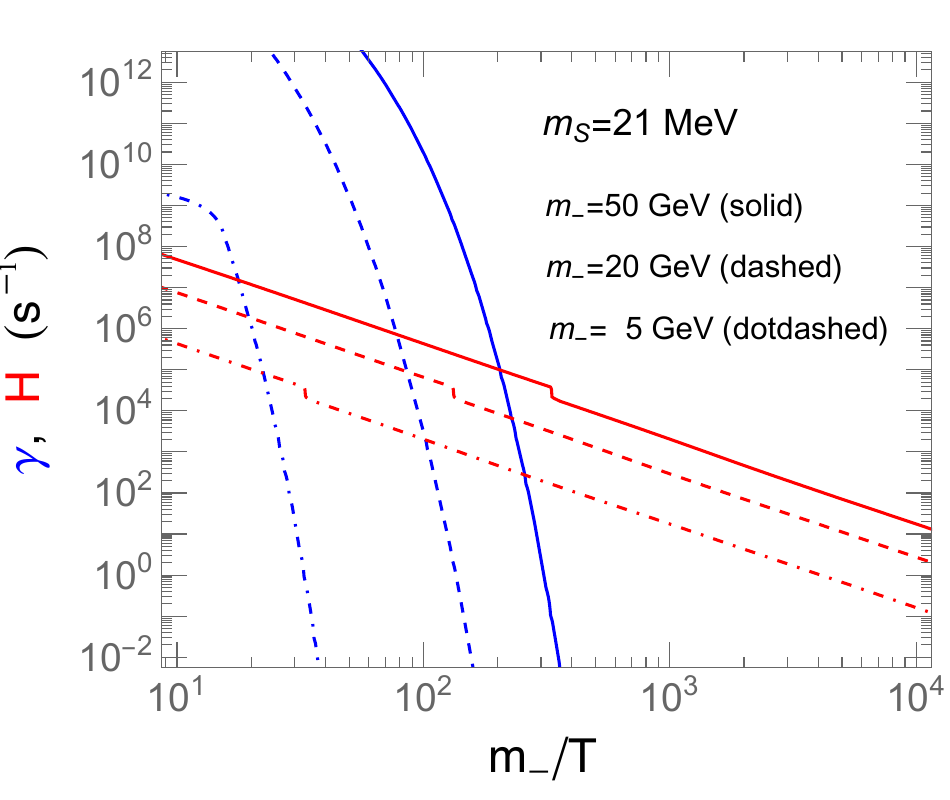}\hskip0.5cm
\includegraphics[width=0.40\textwidth]{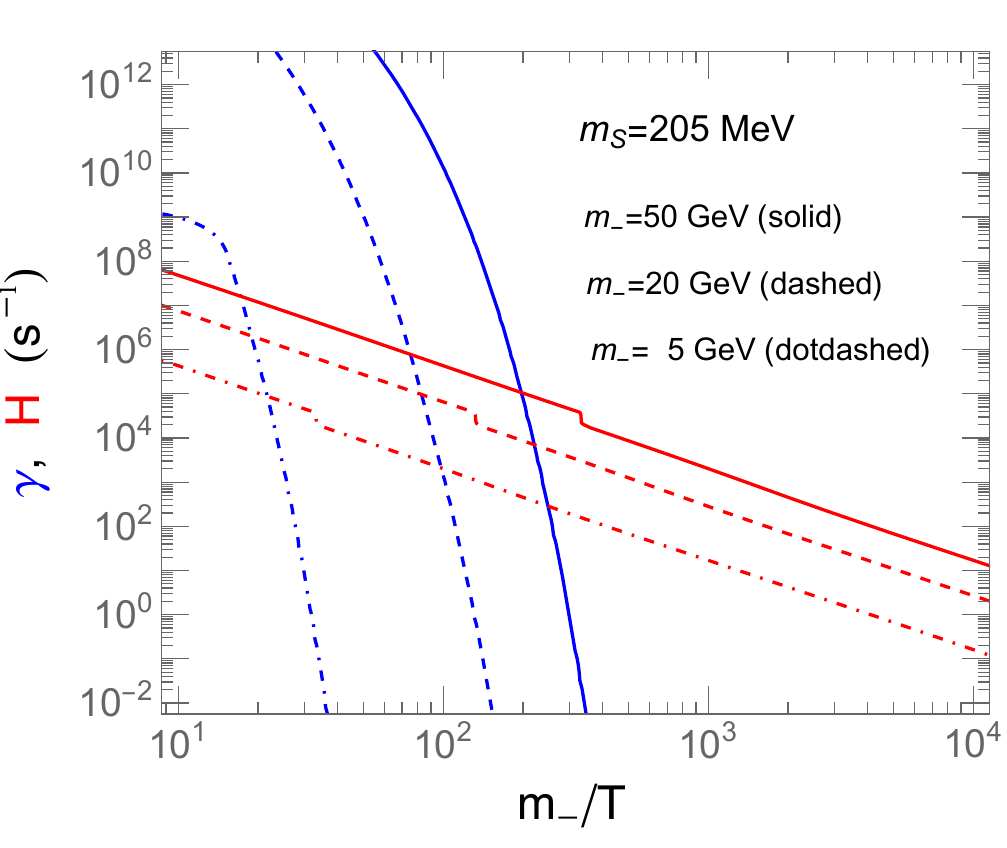} \\
\includegraphics[width=0.40\textwidth]{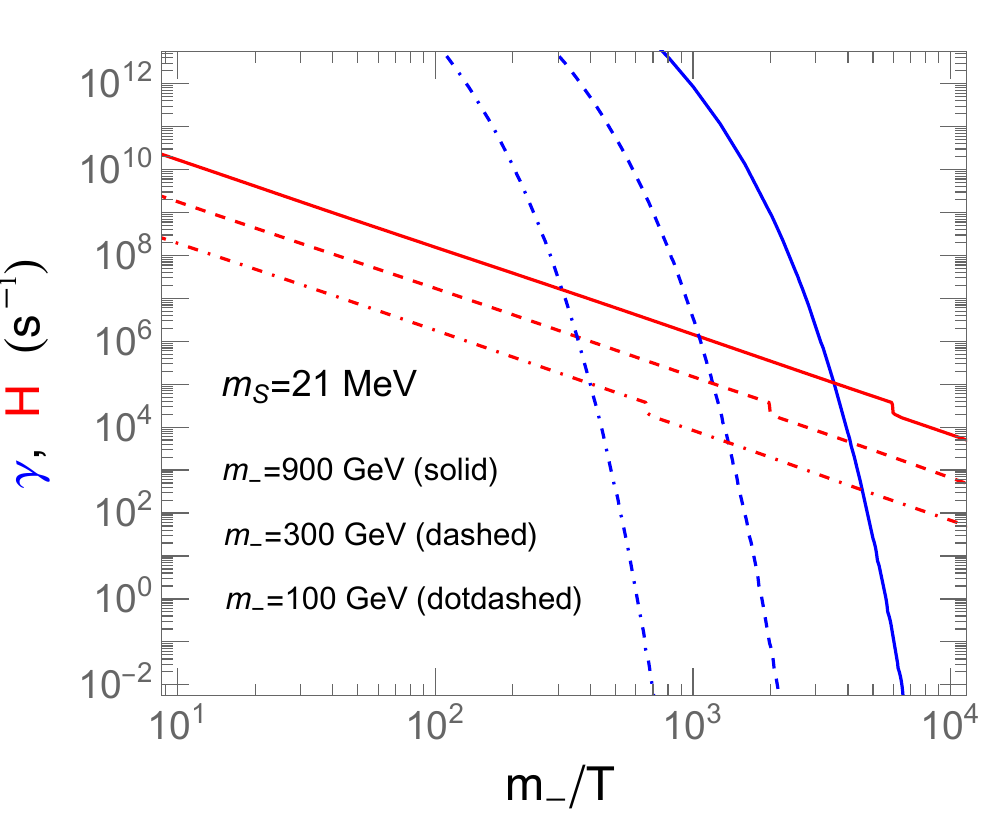}\hskip0.5cm
\includegraphics[width=0.40\textwidth]{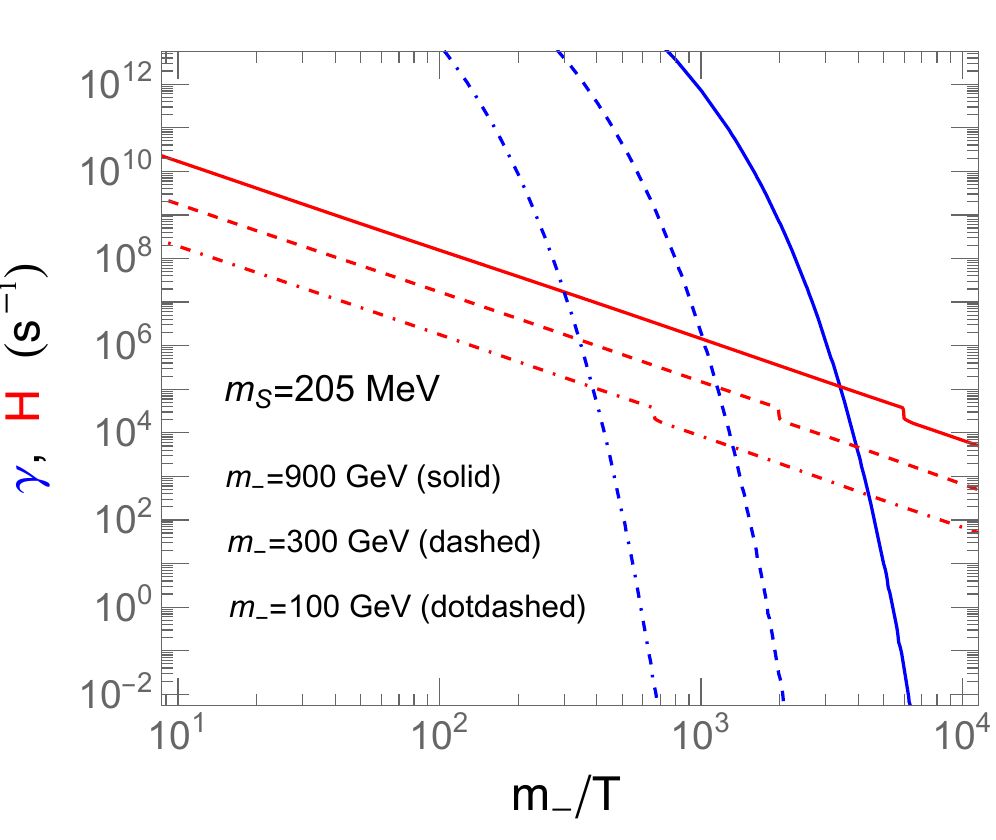} 
\caption{Comparison of kinetic energy injection rates (blue), $\gamma\equiv \gamma_-$, from the thermal bath to dark matter via elastic scattering, $\chi_- \, S \to \chi_- \, S$, and the cooling rate due to Hubble expansion (red), $H$. Results are shown for different values of $m_-$, with $m_S$ fixed at 21 MeV on the left and 205 MeV on the right. When $\gamma < H$, dark matter is kinetically decoupled from the thermal bath. All the decoupling temperatures are roughly between 210 and 250 MeV.
}
\label{fig:elastic-rate}
\end{center}
\end{figure}

The squared amplitude, $|M|_{\chi_\pm S \to \chi_\pm S}^2$, is summed over initial and final internal degrees of freedom,  and
\begin{align}
& \int \frac{d\sigma}{d\Omega} m_\pm v_\pm^2 (1-\cos\theta) d\Omega= \frac{f^4}{48 \pi  (s- m_{\pm}^2)^2} 
\bigg[  \frac{1}{s^3 (m_{\pm}^2 s- \big( m_{\pm}^2 - m_{S}^2 )^2 \big)}
\nonumber\\
&
 \Big( -2 m_{\pm}^2 (m_{S}^2-m_{\pm}^2)^6+s^5 (5 m_{S}^4-62 m_{S}^2 m_{\pm}^2+231 m_{\pm}^4)
 \nonumber\\
 & -s (m_{S}^4-22 m_{S}^2 m_{\pm}^2-3 m_{\pm}^4) (m_{S}^2-m_{\pm}^2)^4+4 s^4 (4 m_{S}^6-45 m_{S}^4 m_{\pm}^2+47 m_{S}^2 m_{\pm}^4-102 m_{\pm}^6)\nonumber\\
 & +2 m_{\pm}^2 s^3 (96 m_{S}^6-109 m_{S}^4 m_{\pm}^2+18 m_{S}^2 m_{\pm}^4+59 m_{\pm}^6)\nonumber\\
 & +s^2 (4 m_{S}^6-87 m_{S}^4 m_{\pm}^2-70 m_{S}^2 m_{\pm}^4+57 m_{\pm}^6) (m_{S}^2-m_{\pm}^2)^2+m_{\pm}^2 s^6  \Big)
 \nonumber\\
 & + \frac{6 (s - m_{\pm}^2)}{m_{S}^4-2 m_{S}^2 (s+ m_{\pm}^2)+( s- m_{\pm}^2)^2}
 \big(-4 m_{S}^6+m_{S}^4 (s-33 m_{\pm}^2)+m_{S}^2 (6 m_{\pm}^4+60 m_{\pm}^2 s-2 s^2)\nonumber\\
 & +31 m_{\pm}^6-13 m_{\pm}^4 s-19 m_{\pm}^2 s^2+s^3 \big)  \log \frac{  s ( s- 2 m_{S}^2 - m_{\pm}^2 )}{ m_{\pm}^2 s- (m_{\pm}^2 - m_{S}^2)^2} 
\bigg]\,,
 \end{align}
which is Lorentz invariant even though the right-hand side is expressed in the center-of-mass frame with $\theta$ and $v_\pm$ representing the scattering angle and velocity of $\chi_\pm$, respectively. 
In Fig.~\ref{fig:elastic-rate}, we show the kinetic energy injection rate from the thermal bath into dark matter via elastic scattering $\chi_\pm \, S \leftrightarrow \chi_\pm \, S$, compared to the cosmic cooling rate caused by Hubble expansion. Dark matter becomes kinetically decoupled from the thermal bath when $\gamma_- < H$. We find that $\chi_-$ with a mass of several GeV to hundreds of GeV kinetically decouples from $S$ at a temperature roughly between 210 and 250 MeV. Meanwhile, this decoupling temperature is insensitive to $m_S$, which ranges from several MeV to hundreds of MeV.

\section{Constraints from neutrino telescopes on the neutrino excess caused by dark matter annihilations}\label{app:SK}

The $\chi_-\chi_-$ pair can annihilate into $SS$ or $Z^\prime Z^\prime$, which subsequently undergo cascade decays into muon or tau neutrinos. In this analysis, we restrict our focus to the kinematic regime $2m_{Z^\prime} < m_S < m_\mu$.
The differential neutrino flux, arising from DM annihilation, can be expressed as
\begin{eqnarray} 
\label{eq:gammaflux}
\frac{d \Phi_\nu}{dE} \simeq \frac{1}{8\pi m_-^2} 
\left[\sum_{i =S, Z^\prime} \langle \sigma v\rangle_{ii} 
\Big( \frac{dN_\nu}{dE} \Big)_{ii} \right]
 \,  \frac{1}{\Delta\Omega}
 \underbrace{ \int_{\Delta\Omega}  \int_{\rm l.o.s.} ds \rho^2(r(s,\psi))d\Omega }_{\text{J-factor}}  ,
\end{eqnarray}
where $\langle \sigma v \rangle_{ii}\equiv \langle \sigma v\rangle_{\chi_\mp \chi_\mp \to i \, i }$. We have $\langle \sigma v \rangle_{SS} \simeq 9 \langle \sigma v \rangle_{Z^\prime Z^\prime} $ in the present model. The J-factor is calculated by integrating the DM density squared along the line of sight (l.o.s.) and over the solid angle $\Delta\Omega$ that covers the source region. The term $(dN_\nu / dE)_{ii}$ represents the neutrino spectrum per DM annihilation.

We use the original Navarro-Frenk-White (NFW) profile  with $\gamma=1$ \cite{Navarro:1995iw,Navarro:1996gj},
 \begin{equation}
 \label{app:gNFW}
 \rho(r)=\displaystyle \rho_{\odot} \left(\frac{r}{r_\odot}\right)^{-\gamma} \left(\frac{1+r/r_s}{1+r_\odot /r_s}\right)^{\gamma-3} \,,
 \end{equation}
to describe the DM density distribution in the Galactic halo, where $r_s=20$~kpc is the scale radius, $r$ is the distance to the GC.
We take the DM density $\rho_\odot=0.35$~GeV in the location of the Sun $r_\odot=8.5$~kpc in the analysis. The variation of $\gamma$ can change the inner slope of the profile.

For the DM annihilation process $\chi_- \chi_- \to S S$, the differential muon-neutrino flux arises from a two-step annihilation process \cite{Elor:2015tva, Yang:2017zor, Yang:2018fje, Yang:2020vxl}. Initially, two DM particles annihilate into a pair of scalar mediators. Each scalar then decays into two $Z^\prime $ particles, with each $Z^\prime $ decay resulting in two neutrinos. 
Since the neutrinos produced in DM annihilation go through intermediate steps, the spectrum from $Z^\prime $ decays is boosted and becomes broader,
\begin{align}
\left( \frac{d N_\nu}{dE} \right)_{SS}= \frac{16/5}{m_-  \sqrt{1-\epsilon_2^2} \sqrt{1-\epsilon_1^2}} \int^{t_{\rm 2, max}}_{t_{\rm 2, min}} \frac{d x_1}{x_1}  \,,
\label{eq: spectrum_SS}
\end{align}
where
 \begin{align}
\epsilon_2  &=\frac{m_S}{m_-}, \quad \epsilon_1 = \frac{2m_{Z^\prime }}{m_S}, 
\\
 t_{\rm 2, max} &= {\rm min} \bigg[\frac{1}{2} \Big(1+\sqrt{1-\epsilon_1^2}\Big), \frac{2 E}{\epsilon_2^2 m_-} \Big(1+\sqrt{1-\epsilon_2^2}\Big) \bigg], \quad
  t_{\rm 2, min} = \frac{2 E}{\epsilon_2^2 m_-} \Big(1-\sqrt{1-\epsilon_2^2}\Big),    
\\
 0\leq E & \leq \frac{m_-}{4} \Big(1+\sqrt{1-\epsilon_1^2} \Big) \Big(1+\sqrt{1-\epsilon_2^2}\Big), 
 \quad  x_1 = \frac{2E_1}{m_S}, 
\end{align}
with $E$, and $E_1$ as the neutrino energies in the center-of-mass frame of the DM pair and the $S$ rest frame, respectively.  
The factor 16/5 accounts for two distinct effects. First, a single DM annihilation eventually produces eight neutrinos. Second, taking into account neutrino oscillations during propagation, the expected flavor ratio at Earth is approximately $1:2:2$ in the $L_\mu - L_\tau$ model \cite{ Maltoni:2004ei, Palomares-Ruiz:2007trf}. 

For $\chi_- \chi_- \to Z^\prime Z^\prime$, the differential muon-neutrino flux arises from a one-step annihilation process \cite{Elor:2015tva, Yang:2017zor, Yang:2018fje, Yang:2020vxl}. The two DM particles annihilate into a pair of the $Z^\prime$ gauge bosons, and then each $Z^\prime$ decays into two neutrinos, resulting in
\begin{align}
\left( \frac{d N_\nu}{dE} \right)_{Z^\prime Z^\prime}
= 2 \int^{t_{\rm max}}_{t_{\rm min}} \frac{d x_0}{x_0\sqrt{1-\epsilon^2}}  \frac{d N_\nu^{Z^\prime}}{dx_0}  
= \frac{8}{5 \sqrt{1-\epsilon^2}} \,,
\label{eq: spectrum_ZpZp}
\end{align}
where
 \begin{align}
 \epsilon  &=\frac{m_{Z^\prime}}{m_-},  \quad
 t_{\rm max} = {\rm min} \bigg[1, \frac{2 E}{m_- \epsilon^2} \Big(1+\sqrt{1-\epsilon^2} \Big) \bigg], \quad
  t_{\rm min} = \frac{2 E}{m_- \epsilon^2} \Big(1-\sqrt{1-\epsilon^2}\Big), \\
 0 & \leq E  \leq \frac{m_-}{2} \Big(1+\sqrt{1-\epsilon^2} \Big) ,  \quad x_0 = \frac{2E_0}{m_{Z^\prime}} \,,  
 \end{align}
with $E_0$ being the neutrino energy in the $Z^\prime$ rest frame.
Here,the muon neutrino spectrum of a $Z^\prime$ decay into two neutrinos in the $L_\mu -L_\tau$ model is given by
 $d N_\nu^{Z^\prime}/ dx_0 = 2\times \frac{2}{5}\delta(x_0 - 1)$, where the factor 2 accounts for two neutrinos are produced per decay, while the factor $2/5$ is the flux ratio
resulting from ocillation observed on the Earth \cite{Palomares-Ruiz:2007trf}.

The upper limit on $\langle\sigma v\rangle = \langle \sigma v \rangle_{SS} + \langle \sigma v \rangle_{Z^\prime Z^\prime}$ can be derived from the relation,
\begin{align}
\mu^{90 \%} &=T_{\rm live} \Delta \Omega\int dE \, A_{\rm eff}(E)  \frac{d \Phi_\nu}{dE}\,,
\end{align}
where $\mu^{90 \%}$ is the 90\% C.L. upper limit on the expect signal from DM annhilations, $\Delta\Omega$ denotes the solid angle of the search region, $A_{\rm eff}$ is the effective area, and $T_{\rm live}$ represents the live time.

Using these inputs, we derive the upper bounds on $\langle \sigma v \rangle$ from various experiments as follows.

\noindent{\underline{\bf  Super-Kamiokande }}

The upper limit for muon-neutrino excess is determined using the on-off source approach of Super-Kamiokande (SK) \cite{Super-Kamiokande:2020sgt, Frankiewicz}, based on analyses of SK-I, II, III, and IV data \cite{Super-Kamiokande:2015qek}. The maximum neutrino excess is expected in the on-source region, centered around the Galactic center, while the off-source region of the same size is located 180$^\circ$ away in right ascension of equatorial coordinates. Therefore, the difference in neutrino event counts ($\Delta N^{\rm sig}$) between these regions directly indicates DM annihilation signals.

The 90\% C.L. upper limit for $\Delta N^{\rm sig}_{90} (\equiv \mu^{90\%})$ is provided separately for $\mu$-like event samples categorized as FC SubGeV, FC MultiGeV, PC, and UP-$\mu$ \cite{Frankiewicz}, which correspond to the measured neutrino energy ranges of approximately $0.1-1$~GeV, $1-5$~GeV, $5-50$~GeV, and greater than 50 GeV, respectively \cite{Super-Kamiokande:2015qek}.  
For the SK, the live time is given by
\begin{align}
T_{\rm live} = \frac{N_{\rm atm}}{ \int dE A_{\rm eff}(E) {\frac{d\Phi_{\rm atm}}{dE} } \Delta\Omega} \,,
\end{align}
where $\Delta\Omega=2\pi(1-\cos\Psi)$ denotes the solid angle of the on-source region for each event sample, $N_{\rm atm}$ is the number of atmospheric muon neutrinos in the off-source region for each sample (see Table 8.2 of Ref.~\cite{Frankiewicz}), which are mainly background events, and the corresponding flux is $d\Phi_{\rm atm}/ dE$ (shown in Fig. 7 of Ref.~\cite{Super-Kamiokande:2015qek}).

The effective area depends on the fiducial volume of the water Cherenkov detector and the scattering cross section. The latter is roughly proportional to the incident neutrino energy \cite{Super-Kamiokande:2005mbp}. Assuming the effective area scales as $\sim E^\beta$, then $\beta$=0 to 1 has little effect on the upper limit of $\langle\sigma v\rangle$. This variation is only slightly noticeable at higher $m_{-}$. When $\beta=0$, the limit at $m_-=300$~GeV is 1.3 times weaker than with $\beta=1$, but at $m_-=1000$~GeV, it is 2.7 times weaker. Here, we use $\beta=1$.

The data show neutrinos from dark matter in the Milky Way are not significantly above atmospheric background. 
Using the 90\% C.L. upper limit for $\Delta N^{\rm sig}_{90}$, derived from $\mu$-like event samples in Ref.~\cite{Frankiewicz}, we establish an upper bound on the dark matter self-annihilation cross section, applicable to the typical velocity range of 140–280 km/s near the Galactic center.

\noindent{\underline{\bf  ANTARES}}

In the indirect detection of dark matter, ANTARES  \cite{ANTARES:2015vis, Arguelles:2019ouk} did not observe a signal significantly higher than the background from data collected between 2007 and 2012 (total effective time of approximately $T_{\rm live}=1321$~days), thus deriving upper limits for neutrino flux and velocity-averaged annihilation cross sections.

ANTARES adopts two reconstruction algorithms,  $\Lambda$Fit and QFit, to analyze the data, where $\mu^{90\%}$ that we used is the 90\% C.L. upper limit on the $\nu_\mu+ \bar\nu_\mu$ flux at Earth. We find that $\Lambda$Fit provides a stronger constraint than QFit in the region of dark matter masses greater than 250 GeV. Therefore, only the results for $\Lambda$Fit are shown in the Fig.~\ref{fig:s1-xs-ss}. The $\mu^{90\%}$ of the $\Lambda$Fit and the angular separation ($\Psi$) of the solid angle can be obtained from Table 3 of Ref.~\cite{ANTARES:2015vis}, while the effective area can be obtained from its Fig.~1.

\noindent{\underline{\bf IceCube}}

Using $T_{\rm live}=9.28$~years of data (from May 2011 to May 2021), IceCube \cite{IceCube:2025fcn} has significantly improved the detection sensitivity of neutrinos in the low-energy region (10 GeV - 1000 GeV). The accumulation of data has made it possible to more accurately remove the atmospheric neutrino background.
In the analysis, we take an opening angle $\Psi \sim 100^\circ$. As illustrated in Fig. 6 of Ref.~\cite{IceCube:2025fcn}, the signal contribution becomes negligible beyond $\Psi \sim 100^\circ$. Since the annihilation flux scales with the squared density ($\rho^2 \propto r^{-2 \sim -4}$), the signal drops rapidly with distance from the Galactic Center, making the contribution to the J-factor from the region $\Psi \gtrsim 100^\circ$ insignificant. 

We approximate the 90\% C.L. upper limit $\mu^{90\%} = \hat{n}_s (1+ 1.282/ \text{z-score})$ for the $\nu_\mu \bar\nu_\mu$ channel using the reported best-fit signal events $\hat{n}_s$ and the corresponding z-score given in Table VI of Ref.~\cite{IceCube:2025fcn}.
To characterize the detector response, we derive the effective area ($A_{\text{eff}}$) by directly inverting the reported 90\% C.L. upper limits.


\bibliographystyle{plain} 
\bibliography{majoranaidmr5n} 

\providecommand{\href}[2]{#2}\begingroup\raggedright\begin{thebibliography}{100}


\bibitem{Tucker-Smith:2001myb}
D.~Tucker-Smith and N.~Weiner,
``Inelastic dark matter,''
Phys. Rev. D \textbf{64}, 043502 (2001)
[arXiv:hep-ph/0101138 [hep-ph]].

\bibitem{Tucker-Smith:2004mxa}
D.~Tucker-Smith and N.~Weiner,
``The Status of inelastic dark matter,''
Phys. Rev. D \textbf{72}, 063509 (2005)
[arXiv:hep-ph/0402065 [hep-ph]].


\bibitem{Batell:2009vb}
B.~Batell, M.~Pospelov and A.~Ritz,
``Direct Detection of Multi-component Secluded WIMPs,''
Phys. Rev. D \textbf{79}, 115019 (2009)
[arXiv:0903.3396 [hep-ph]].

\bibitem{Chang:2010en}
S.~Chang, N.~Weiner and I.~Yavin,
``Magnetic Inelastic Dark Matter,''
Phys. Rev. D \textbf{82}, 125011 (2010)
[arXiv:1007.4200 [hep-ph]].

\bibitem{Okada:2019sbb}
N.~Okada and O.~Seto,
``Inelastic extra $U(1)$ charged scalar dark matter,''
Phys. Rev. D \textbf{101}, no.2, 023522 (2020)
[arXiv:1908.09277 [hep-ph]].

\bibitem{DallaValleGarcia:2024zva}
G.~Dalla Valle Garcia,
``A minimalistic model for inelastic dark matter,''
Phys. Lett. B \textbf{862}, 139320 (2025)
[arXiv:2411.02147 [hep-ph]].

\bibitem{Wang:2025cth}
X.~G.~Wang and A.~W.~Thomas,
``A Viable New Model for Dark Matter,''
[arXiv:2510.24114 [hep-ph]].

\bibitem{Arkani-Hamed:2008hhe}
N.~Arkani-Hamed, D.~P.~Finkbeiner, T.~R.~Slatyer and N.~Weiner,
``A Theory of Dark Matter,''
Phys. Rev. D \textbf{79}, 015014 (2009)
[arXiv:0810.0713 [hep-ph]].

\bibitem{Pospelov:2008jd}
M.~Pospelov and A.~Ritz,
``Astrophysical Signatures of Secluded Dark Matter,''
Phys. Lett. B \textbf{671}, 391-397 (2009)
[arXiv:0810.1502 [hep-ph]].

\bibitem{Finkbeiner:2009mi}
D.~P.~Finkbeiner, T.~R.~Slatyer, N.~Weiner and I.~Yavin,
``PAMELA, DAMA, INTEGRAL and Signatures of Metastable Excited WIMPs,''
JCAP \textbf{09}, 037 (2009)
[arXiv:0903.1037 [hep-ph]].

\bibitem{Gustafson:2024aom}
R.~A.~Gustafson, G.~Herrera, M.~Mukhopadhyay, K.~Murase and I.~M.~Shoemaker,
``Cosmic-ray cooling in active galactic nuclei as a new probe of inelastic dark matter,''
Phys. Rev. D \textbf{111}, no.12, L121303 (2025)
[arXiv:2408.08947 [hep-ph]].

\bibitem{Berlin:2025fwx}
A.~Berlin, J.~W.~Foster, D.~Hooper and G.~Krnjaic,
``dSphobic Dark Matter,''
[arXiv:2504.12372 [hep-ph]].

\bibitem{Hooper:2025fda}
D.~Hooper, G.~Krnjaic, D.~Rocha and S.~Roy,
``Gamma-Rays and Gravitational Waves from Inelastic Higgs Portal Dark Matter,''
[arXiv:2507.22975 [hep-ph]].

\bibitem{Berlin:2023qco}
A.~Berlin, G.~Krnjaic and E.~Pinetti,
``Reviving MeV-GeV indirect detection with inelastic dark matter,''
Phys. Rev. D \textbf{110}, no.3, 035015 (2024)
[arXiv:2311.00032 [hep-ph]].

\bibitem{Krnjaic:2025zjl}
G.~Krnjaic, D.~McKeen, R.~Mizuta, G.~Mohlabeng, D.~E.~Morrissey and D.~Tuckler,
``X-rays from Inelastic Dark Matter Freeze-in,''
[arXiv:2509.19428 [hep-ph]].

\bibitem{Izaguirre:2015zva}
E.~Izaguirre, G.~Krnjaic and B.~Shuve,
``Discovering Inelastic Thermal-Relic Dark Matter at Colliders,''
Phys. Rev. D \textbf{93}, no.6, 063523 (2016)
[arXiv:1508.03050 [hep-ph]].

\bibitem{Berlin:2018jbm}
A.~Berlin and F.~Kling,
``Inelastic Dark Matter at the LHC Lifetime Frontier: ATLAS, CMS, LHCb, CODEX-b, FASER, and MATHUSLA,''
Phys. Rev. D \textbf{99}, no.1, 015021 (2019)
[arXiv:1810.01879 [hep-ph]].

\bibitem{Voronchikhin:2025eqm}
I.~V.~Voronchikhin and D.~V.~Kirpichnikov,
``Examining scalar portal inelastic dark matter with lepton fixed target experiments,''
[arXiv:2505.04290 [hep-ph]].

\bibitem{Krnjaic:2024ols}
G.~Krnjaic, D.~Rocha and I.~R.~Wang,
``Discovering Dark Matter with the MUonE Experiment,''
Phys. Rev. Lett. \textbf{134}, no.16, 161801 (2025)
[arXiv:2409.00170 [hep-ph]].

\bibitem{Garcia:2024uwf}
G.~D.~V.~Garcia, F.~Kahlhoefer, M.~Ovchynnikov and T.~Schwetz,
``Not-so-inelastic Dark Matter,''
JHEP \textbf{02}, 127 (2025)
[arXiv:2405.08081 [hep-ph]].

\bibitem{Abdullahi:2023tyk}
A.~M.~Abdullahi, M.~Hostert, D.~Massaro and S.~Pascoli,
``Semi-Visible Dark Photon Phenomenology at the GeV Scale,''
Phys. Rev. D \textbf{108}, no.1, 015032 (2023)
[arXiv:2302.05410 [hep-ph]].

\bibitem{CarrilloGonzalez:2021lxm}
M.~Carrillo Gonz{\'a}lez and N.~Toro,
``Cosmology and signals of light pseudo-Dirac dark matter,''
JHEP \textbf{04}, 060 (2022)
[arXiv:2108.13422 [hep-ph]].

\bibitem{Heeba:2023bik}
S.~Heeba, T.~Lin and K.~Schutz,
``Inelastic freeze-in,''
Phys. Rev. D \textbf{108}, no.9, 095016 (2023)
[arXiv:2304.06072 [hep-ph]].

\bibitem{Brahma:2023psr}
N.~Brahma, S.~Heeba and K.~Schutz,
``Resonant pseudo-Dirac dark matter as a sub-GeV thermal target,''
Phys. Rev. D \textbf{109}, no.3, 035006 (2024)
[arXiv:2308.01960 [hep-ph]].

\bibitem{Roy:2025zvo}
S.~Roy,
``Dark Matter and Electroweak Baryogenesis with Spontaneous $CP$ Violation in the Early Universe,''
[arXiv:2509.19982 [hep-ph]].

\bibitem{Tulin:2017ara}
S.~Tulin and H.~B.~Yu,
``Dark Matter Self-interactions and Small Scale Structure,''
Phys. Rept. \textbf{730}, 1-57 (2018)
[arXiv:1705.02358 [hep-ph]].

\bibitem{Adhikari:2022sbh}
S.~Adhikari, A.~Banerjee, K.~K.~Boddy, F.~Y.~Cyr-Racine, H.~Desmond, C.~Dvorkin, B.~Jain, F.~Kahlhoefer, M.~Kaplinghat and A.~Nierenberg, \textit{et al.}
``Astrophysical tests of dark matter self-interactions,''
Rev. Mod. Phys. \textbf{97}, no.4, 045004 (2025)
[arXiv:2207.10638 [astro-ph.CO]].


\bibitem{Kaplinghat:2015aga}
M.~Kaplinghat, S.~Tulin and H.~B.~Yu,
``Dark Matter Halos as Particle Colliders: Unified Solution to Small-Scale Structure Puzzles from Dwarfs to Clusters,''
Phys. Rev. Lett. \textbf{116}, no.4, 041302 (2016)
[arXiv:1508.03339 [astro-ph.CO]].



\bibitem{He:1990pn}
X.~G.~He, G.~C.~Joshi, H.~Lew and R.~R.~Volkas,
``NEW Z-prime PHENOMENOLOGY,''
Phys. Rev. D \textbf{43}, 22-24 (1991)

\bibitem{He:1991qd}
X.~G.~He, G.~C.~Joshi, H.~Lew and R.~R.~Volkas,
``Simplest Z-prime model,''
Phys. Rev. D \textbf{44}, 2118-2132 (1991)

\bibitem{Altmannshofer:2016jzy}
W.~Altmannshofer, S.~Gori, S.~Profumo and F.~S.~Queiroz,
``Explaining dark matter and B decay anomalies with an $L_\mu - L_\tau$ model,''
JHEP \textbf{12}, 106 (2016)
[arXiv:1609.04026 [hep-ph]].

\bibitem{Kamada:2018zxi}
A.~Kamada, K.~Kaneta, K.~Yanagi and H.~B.~Yu,
 ``Self-interacting dark matter and muon $g-2$ in a gauged $U(1)_{L_{\mu} - L_{\tau}}$ model,''
JHEP \textbf{06}, 117 (2018)
[arXiv:1805.00651 [hep-ph]].

\bibitem{Foldenauer:2018zrz}
P.~Foldenauer,
``Light dark matter in a gauged $U(1)_{L_\mu-L_\tau}$ model,''
Phys. Rev. D \textbf{99}, no.3, 035007 (2019)
[arXiv:1808.03647 [hep-ph]].

\bibitem{Escudero:2019gzq}
M.~Escudero, D.~Hooper, G.~Krnjaic and M.~Pierre,
``Cosmology with A Very Light L$_{\mu}$ {\ensuremath{-}} L$_{\tau}$ Gauge Boson,''
JHEP \textbf{03}, 071 (2019)
[arXiv:1901.02010 [hep-ph]].

\bibitem{Asai:2020qlp}
K.~Asai, S.~Okawa and K.~Tsumura,
``Search for $ \mathrm{U}{(1)}_{L_{\mu }-{L}_{\tau }} $ charged dark matter with neutrino telescope,''
JHEP \textbf{03}, 047 (2021)
[arXiv:2011.03165 [hep-ph]].

\bibitem{Borah:2021jzu}
D.~Borah, M.~Dutta, S.~Mahapatra and N.~Sahu,
``Muon $(g-2)$ and XENON1T excess with boosted dark matter in ${L_{\mu }-{L}_{\tau }}$ model,"
Phys. Lett. B \textbf{820}, 136577 (2021)
[arXiv:2104.05656 [hep-ph]].

\bibitem{Holst:2021lzm}
I.~Holst, D.~Hooper and G.~Krnjaic,
``Simplest and Most Predictive Model of Muon g-2 and Thermal Dark Matter,''
Phys. Rev. Lett. \textbf{128}, no.14, 141802 (2022)
[arXiv:2107.09067 [hep-ph]].

\bibitem{Drees:2021rsg}
M.~Drees and W.~Zhao,
``$U(1)_{L\ensuremath{\mu}\ensuremath{-}L\ensuremath{\tau}}$ for light dark matter, $g_\mu-2$, the 511 keV excess and the Hubble tension,''
Phys. Lett. B \textbf{827}, 136948 (2022)
[arXiv:2107.14528 [hep-ph]].

\bibitem{Hapitas:2021ilr}
T.~Hapitas, D.~Tuckler and Y.~Zhang,
``General kinetic mixing in gauged $U(1)_{L\ensuremath{\mu}-L\ensuremath{\tau}}$ model for muon g-2 and dark matter,''
Phys. Rev. D \textbf{105}, no.1, 016014 (2022)
[arXiv:2108.12440 [hep-ph]].

\bibitem{Figueroa:2024tmn}
P.~Figueroa, G.~Herrera and F.~Ochoa,
``Direct detection of light dark matter charged under a L{\ensuremath{\mu}}-L{\ensuremath{\tau}} symmetry,''
Phys. Rev. D \textbf{110}, no.9, 095018 (2024)
[arXiv:2404.03090 [hep-ph]].

\bibitem{Asai:2021wzx}
K.~Asai, K.~Hamaguchi, N.~Nagata, S.~Y.~Tseng and J.~Wada,
``Probing the L{\ensuremath{\mu}}-L{\ensuremath{\tau}} gauge boson at the MUonE experiment,''
Phys. Rev. D \textbf{106}, no.5, L051702 (2022)
[arXiv:2109.10093 [hep-ph]].

\bibitem{Zu:2021odn}
L.~Zu, X.~Pan, L.~Feng, Q.~Yuan and Y.~Z.~Fan,
``Constraining $U(1)_{L\ensuremath{\mu}-L\ensuremath{\tau}}$ charged dark matter model for muon g -2 anomaly with AMS-02 electron and positron data,''
JCAP \textbf{08}, no.08, 028 (2022)
[arXiv:2104.03340 [hep-ph]].

\bibitem{Aliberti:2025beg}
R.~Aliberti, T.~Aoyama, E.~Balzani, A.~Bashir, G.~Benton, J.~Bijnens, V.~Biloshytskyi, T.~Blum, D.~Boito and M.~Bruno, \textit{et al.}
``The anomalous magnetic moment of the muon in the Standard Model: an update,''
Phys. Rept. \textbf{1143}, 1-158 (2025)
[arXiv:2505.21476 [hep-ph]].



\bibitem{Muong-2:2023cdq}
D.~P.~Aguillard \textit{et al.} [Muon g-2],
``Measurement of the Positive Muon Anomalous Magnetic Moment to 0.20~ppm,''
Phys. Rev. Lett. \textbf{131}, no.16, 161802 (2023)
[arXiv:2308.06230 [hep-ex]].

\bibitem{Muong-2:2024hpx}
D.~P.~Aguillard \textit{et al.} [Muon g-2],
``Detailed report on the measurement of the positive muon anomalous magnetic moment to 0.20~ppm,''
Phys. Rev. D \textbf{110}, no.3, 032009 (2024)
[arXiv:2402.15410 [hep-ex]].

\bibitem{Muong-2:2006rrc}
G.~W.~Bennett \textit{et al.} [Muon g-2],
``Final Report of the Muon E821 Anomalous Magnetic Moment Measurement at BNL,''
Phys. Rev. D \textbf{73}, 072003 (2006)
[arXiv:hep-ex/0602035 [hep-ex]].


\bibitem{Aoyama:2020ynm}
T.~Aoyama, N.~Asmussen, M.~Benayoun, J.~Bijnens, T.~Blum, M.~Bruno, I.~Caprini, C.~M.~Carloni Calame, M.~C\`e and G.~Colangelo, \textit{et al.}
``The anomalous magnetic moment of the muon in the Standard Model,''
Phys. Rept. \textbf{887}, 1-166 (2020)
[arXiv:2006.04822 [hep-ph]].

\bibitem{Davies:2025pmx}
C.~Davies,
``Muon g-2,''
PoS \textbf{LATTICE2024}, 019 (2025)
[arXiv:2503.03364 [hep-lat]].


\bibitem{LZ:2024zvo}
J.~Aalbers \textit{et al.} [LZ],
``Dark Matter Search Results from 4.2{\,}{\,}Tonne-Years of Exposure of the LUX-ZEPLIN (LZ) Experiment,''
Phys. Rev. Lett. \textbf{135}, no.1, 011802 (2025)
[arXiv:2410.17036 [hep-ex]].

\bibitem{pdg2022}
R.L. Workman et al. (Particle Data Group), Prog. Theor. Exp. Phys. 2022, 083C01 (2022).

\bibitem{Balan:2025uke}
S.~Balan, T.~Bringmann, F.~Kahlhoefer, J.~Matuszak and C.~Tasillo,
``Sub-GeV dark matter and nano-Hertz gravitational waves from a classically conformal dark sector,''
JCAP \textbf{08}, 062 (2025)
[arXiv:2502.19478 [hep-ph]].

\bibitem{Holdom:1985ag}
B.~Holdom,
``Two U(1)'s and Epsilon Charge Shifts,''
Phys. Lett. B \textbf{166}, 196-198 (1986)


\bibitem{Schmitz:2009}
K. Schmitz, Kinetic Mixing in Field Theory (DESY Workshop
Seminar, Winter Semester 2009/2010, 2009)

\bibitem{Gherghetta:2019coi}
T.~Gherghetta, J.~Kersten, K.~Olive and M.~Pospelov,
``Evaluating the price of tiny kinetic mixing,''
Phys. Rev. D \textbf{100}, no.9, 095001 (2019)
[arXiv:1909.00696 [hep-ph]].



\bibitem{Jegerlehner:2009ry}
F.~Jegerlehner and A.~Nyffeler,
``The Muon g-2,''
Phys. Rept. \textbf{477}, 1-110 (2009)
[arXiv:0902.3360 [hep-ph]].


\bibitem{CCFR:1991lpl}
S.~R.~Mishra \textit{et al.} [CCFR],
``Neutrino Tridents and W Z Interference,''
Phys. Rev. Lett. \textbf{66}, 3117-3120 (1991)

\bibitem{Altmannshofer:2014pba}
W.~Altmannshofer, S.~Gori, M.~Pospelov and I.~Yavin,
``Neutrino Trident Production: A Powerful Probe of New Physics with Neutrino Beams,''
Phys. Rev. Lett. \textbf{113}, 091801 (2014)
[arXiv:1406.2332 [hep-ph]].


\bibitem{BaBar:2016sci}
J.~P.~Lees \textit{et al.} [BaBar],
``Search for a muonic dark force at BABAR,''
Phys. Rev. D \textbf{94}, no.1, 011102 (2016)
[arXiv:1606.03501 [hep-ex]].


\bibitem{Bauer:2018onh}
M.~Bauer, P.~Foldenauer and J.~Jaeckel,
``Hunting All the Hidden Photons,''
JHEP \textbf{07}, 094 (2018)
[arXiv:1803.05466 [hep-ph]].

\bibitem{Dreiner:2013tja}
H.~K.~Dreiner, J.~F.~Fortin, J.~Isern and L.~Ubaldi,
``White Dwarfs constrain Dark Forces,''
Phys. Rev. D \textbf{88}, 043517 (2013)
[arXiv:1303.7232 [hep-ph]].

\bibitem{Bellini:2011rx}
G.~Bellini, J.~Benziger, D.~Bick, S.~Bonetti, G.~Bonfini, M.~Buizza Avanzini, B.~Caccianiga, L.~Cadonati, F.~Calaprice and C.~Carraro, \textit{et al.}
``Precision measurement of the 7Be solar neutrino interaction rate in Borexino,''
Phys. Rev. Lett. \textbf{107}, 141302 (2011)
[arXiv:1104.1816 [hep-ex]].

\bibitem{Planck:2018vyg}
N.~Aghanim \textit{et al.} [Planck],
``Planck 2018 results. VI. Cosmological parameters,''
Astron. Astrophys. \textbf{641}, A6 (2020)
[erratum: Astron. Astrophys. \textbf{652}, C4 (2021)]
[arXiv:1807.06209 [astro-ph.CO]].

\bibitem{Kolb:1990}
E. W. Kolb and M. S. Turner, The Early Universe (Addison-Wesley, Redwood City, CA, 1990).

\bibitem{Vagnozzi:2019ezj}
S.~Vagnozzi,
``New physics in light of the $H_0$ tension: An alternative view,''
Phys. Rev. D \textbf{102}, no.2, 023518 (2020)
[arXiv:1907.07569 [astro-ph.CO]].

\bibitem{DiValentino:2021izs}
E.~Di Valentino, O.~Mena, S.~Pan, L.~Visinelli, W.~Yang, A.~Melchiorri, D.~F.~Mota, A.~G.~Riess and J.~Silk,
``In the realm of the Hubble tension{\textemdash}a review of solutions,''
Class. Quant. Grav. \textbf{38}, no.15, 153001 (2021)
[arXiv:2103.01183 [astro-ph.CO]].


\bibitem{CHARM:1985anb}
F.~Bergsma \textit{et al.} [CHARM],
``Search for Axion Like Particle Production in 400-{GeV} Proton - Copper Interactions,’’
Phys. Lett. B \textbf{157}, 458-462 (1985)

\bibitem{Winkler:2018qyg}
M.~W.~Winkler,
``Decay and detection of a light scalar boson mixing with the Higgs boson,’’
Phys. Rev. D \textbf{99}, no.1, 015018 (2019)
[arXiv:1809.01876 [hep-ph]].

\bibitem{Bernardi:1985ny}
G.~Bernardi, G.~Carugno, J.~Chauveau, F.~Dicarlo, M.~Dris, J.~Dumarchez, M.~Ferro-Luzzi, J.~M.~Levy, D.~Lukas and J.~M.~Perreau, \textit{et al.}
``Search for Neutrino Decay,''
Phys. Lett. B \textbf{166}, 479-483 (1986)

\bibitem{Bernardi:1986hs}
G.~Bernardi, G.~Carugno, J.~Chauveau, F.~Dicarlo, M.~Dris, J.~Dumarchez, M.~Ferro-Luzzi, J.~M.~Levy, D.~Lukas and J.~M.~Perreau, \textit{et al.}
``Anomalous Electron Production Observed in the {CERN} Ps Neutrino Beam,''
Phys. Lett. B \textbf{181}, 173-177 (1986)

\bibitem{Bernardi:1987ek}
G.~Bernardi, G.~Carugno, J.~Chauveau, F.~Dicarlo, M.~Dris, J.~Dumarchez, M.~Ferro-Luzzi, J.~M.~Levy, D.~Lukas and J.~M.~Perreau, \textit{et al.}
``FURTHER LIMITS ON HEAVY NEUTRINO COUPLINGS,''
Phys. Lett. B \textbf{203}, 332-334 (1988)

\bibitem{Gorbunov:2021ccu}
D.~Gorbunov, I.~Krasnov and S.~Suvorov,
``Constraints on light scalars from PS191 results,''
Phys. Lett. B \textbf{820}, 136524 (2021)
[arXiv:2105.11102 [hep-ph]].
  
\bibitem{BNL-E949:2009dza}
A.~V.~Artamonov \textit{et al.} [BNL-E949],
``Study of the decay $K^+\to\pi^+\nu \bar\nu$ in the momentum region $140 < P_\pi < 199$ MeV/c,’’
Phys. Rev. D \textbf{79}, 092004 (2009)
[arXiv:0903.0030 [hep-ex]].

\bibitem{NA62:2021zjw}
E.~Cortina Gil \textit{et al.} [NA62],
``Measurement of the very rare K$^{+}${\textrightarrow}$ {\pi}^{+}\nu \overline{\nu} $ decay,''
JHEP \textbf{06}, 093 (2021)
[arXiv:2103.15389 [hep-ex]].

\bibitem{NA62:2020pwi}
E.~Cortina Gil \textit{et al.} [NA62],
``Search for $\pi^0$ decays to invisible particles,''
JHEP \textbf{02}, 201 (2021)
[arXiv:2010.07644 [hep-ex]].

\bibitem{Beacham:2019nyx}
J.~Beacham, C.~Burrage, D.~Curtin, A.~De Roeck, J.~Evans, J.~L.~Feng, C.~Gatto, S.~Gninenko, A.~Hartin and I.~Irastorza, \textit{et al.}
``Physics Beyond Colliders at CERN: Beyond the Standard Model Working Group Report,''
J. Phys. G \textbf{47}, no.1, 010501 (2020)
[arXiv:1901.09966 [hep-ex]].

\bibitem{NA62:2017rwk}
E.~Cortina Gil \textit{et al.} [NA62],
``The Beam and detector of the NA62 experiment at CERN,’’
JINST \textbf{12}, no.05, P05025 (2017)
[arXiv:1703.08501 [physics.ins-det]].

\bibitem{Alekhin:2015byh}
S.~Alekhin, W.~Altmannshofer, T.~Asaka, B.~Batell, F.~Bezrukov, K.~Bondarenko, A.~Boyarsky, K.~Y.~Choi, C.~Corral and N.~Craig, \textit{et al.}
``A facility to Search for Hidden Particles at the CERN SPS: the SHiP physics case,’’
Rept. Prog. Phys. \textbf{79}, no.12, 124201 (2016)
[arXiv:1504.04855 [hep-ph]].

\bibitem{SHiP:2018yqc}
C.~Ahdida \textit{et al.} [SHiP],
``The experimental facility for the Search for Hidden Particles at the CERN SPS,’’
JINST \textbf{14}, no.03, P03025 (2019)
[arXiv:1810.06880 [physics.ins-det]].

\bibitem{Bondarenko:2019vrb}
K.~Bondarenko, A.~Boyarsky, T.~Bringmann, M.~Hufnagel, K.~Schmidt-Hoberg and A.~Sokolenko,
``Direct detection and complementary constraints for sub-GeV dark matter,’’
JHEP \textbf{03}, 118 (2020)
[arXiv:1909.08632 [hep-ph]].

\bibitem{LZ:2018qzl}
D.~S.~Akerib \textit{et al.} [LZ],
``Projected WIMP sensitivity of the LUX-ZEPLIN dark matter experiment,''
Phys. Rev. D \textbf{101}, no.5, 052002 (2020)
[arXiv:1802.06039 [astro-ph.IM]].


\bibitem{Edsjo:1997bg}
J.~Edsjo and P.~Gondolo,
``Neutralino relic density including coannihilations,''
Phys. Rev. D \textbf{56}, 1879-1894 (1997)
[arXiv:hep-ph/9704361 [hep-ph]].

\bibitem{Cerdeno:2011tf}
D.~G.~Cerdeno, T.~Delahaye and J.~Lavalle,
Nucl. Phys. B \textbf{854}, 738-779 (2012)
[arXiv:1108.1128 [hep-ph]].

\bibitem{Gondolo:1990dk}
  P.~Gondolo and G.~Gelmini,
    ``Cosmic abundances of stable particles: Improved analysis,''
  Nucl.\ Phys.\ B {\bf 360}, 145 (1991).
  
\bibitem{Yang:2022zlh}
K.~C.~Yang,
``Freeze-out forbidden dark matter in the hidden sector in the mass range from sub-GeV to TeV,''
JHEP \textbf{11}, 083 (2022)
[arXiv:2209.10827 [hep-ph]].

\bibitem{Cirelli:2024ssz}
M.~Cirelli, A.~Strumia and J.~Zupan,
``Dark Matter,''
[arXiv:2406.01705 [hep-ph]].

\bibitem{Anand:2013yka}
N.~Anand, A.~L.~Fitzpatrick and W.~C.~Haxton,
``Weakly interacting massive particle-nucleus elastic scattering response,''
Phys. Rev. C \textbf{89}, no.6, 065501 (2014)
[arXiv:1308.6288 [hep-ph]].

\bibitem{Yang:2017zor}
K.~C.~Yang,
``Search for Scalar Dark Matter via Pseudoscalar Portal Interactions: In Light of the Galactic Center Gamma-Ray Excess,''
Phys. Rev. D \textbf{97}, no.2, 023025 (2018)
[arXiv:1711.03878 [hep-ph]].

\bibitem{PandaX-4T:2021bab}
Y.~Meng \textit{et al.} [PandaX-4T],
``Dark Matter Search Results from the PandaX-4T Commissioning Run,’’
Phys. Rev. Lett. \textbf{127} (2021) no.26, 261802
[arXiv:2107.13438 [hep-ex]].

\bibitem{XENON:2023cxc}
E.~Aprile \textit{et al.} [XENON],
``First Dark Matter Search with Nuclear Recoils from the XENONnT Experiment,''
Phys. Rev. Lett. \textbf{131}, no.4, 041003 (2023)
[arXiv:2303.14729 [hep-ex]].

\bibitem{Kawasaki:2000en} 
  M.~Kawasaki, K.~Kohri and N.~Sugiyama,
  ``MeV scale reheating temperature and thermalization of neutrino background,''
  Phys.\ Rev.\ D {\bf 62}, 023506 (2000)
  [astro-ph/0002127].

\bibitem{Iengo:2009xf}
R.~Iengo,
``Sommerfeld enhancement for a Yukawa potential,''
[arXiv:0903.0317 [hep-ph]].

\bibitem{Iengo:2009ni}
R.~Iengo,
``Sommerfeld enhancement: General results from field theory diagrams,''
JHEP \textbf{05}, 024 (2009)
[arXiv:0902.0688 [hep-ph]].

\bibitem{Binder:2017lkj}
T.~Binder, M.~Gustafsson, A.~Kamada, S.~M.~R.~Sandner and M.~Wiesner,
``Reannihilation of self-interacting dark matter,''
Phys. Rev. D \textbf{97}, no.12, 123004 (2018)
[arXiv:1712.01246 [astro-ph.CO]].

\bibitem{Kahlhoefer:2017umn}
F.~Kahlhoefer, K.~Schmidt-Hoberg and S.~Wild,
``Dark matter self-interactions from a general spin-0 mediator,''
JCAP \textbf{08}, 003 (2017)
[arXiv:1704.02149 [hep-ph]].

\bibitem{Kamada:2020buc}
A.~Kamada, H.~J.~Kim and T.~Kuwahara,
``Maximally self-interacting dark matter: models and predictions,''
JHEP \textbf{12}, 202 (2020)
[arXiv:2007.15522 [hep-ph]].

\bibitem{Kamada:2023iol}
A.~Kamada, T.~Kuwahara and A.~Patel,
``Quantum theory of dark matter scattering,''
JHEP \textbf{11}, 105 (2023)
[arXiv:2303.17961 [hep-ph]].

\bibitem{Super-Kamiokande:2020sgt}
K.~Abe \textit{et al.} [Super-Kamiokande],
``Indirect search for dark matter from the Galactic Center and halo with the Super-Kamiokande detector,''
Phys. Rev. D \textbf{102}, no.7, 072002 (2020)
[arXiv:2005.05109 [hep-ex]].

\bibitem{Frankiewicz}
K. Frankiewicz,  ``Indirect Search for Dark Matter with the
Super-Kamiokande Detector", Ph.D. thesis, National Centre For Nuclear Research, Poland, Apr. 2018.

\bibitem{Super-Kamiokande:2015qek}
E.~Richard \textit{et al.} [Super-Kamiokande],
``Measurements of the atmospheric neutrino flux by Super-Kamiokande: energy spectra, geomagnetic effects, and solar modulation,''
Phys. Rev. D \textbf{94}, no.5, 052001 (2016)
[arXiv:1510.08127 [hep-ex]].

\bibitem{ANTARES:2015vis}
S.~Adrian-Martinez \textit{et al.} [ANTARES],
``Search of Dark Matter Annihilation in the Galactic Centre using the ANTARES Neutrino Telescope,''
JCAP \textbf{10}, 068 (2015)
[arXiv:1505.04866 [astro-ph.HE]].

\bibitem{Arguelles:2019ouk}
C.~A.~Arg{\"u}elles, A.~Diaz, A.~Kheirandish, A.~Olivares-Del-Campo, I.~Safa and A.~C.~Vincent,
``Dark matter annihilation to neutrinos,''
Rev. Mod. Phys. \textbf{93}, no.3, 035007 (2021)
[arXiv:1912.09486 [hep-ph]].


\bibitem{IceCube:2025fcn}
R.~Abbasi \textit{et al.} [IceCube],
``Search for GeV-scale Dark Matter from the Galactic Center with IceCube-DeepCore,''
[arXiv:2511.00918 [astro-ph.HE]].

\bibitem{Spergel:1999mh}
D.~N.~Spergel and P.~J.~Steinhardt,
``Observational evidence for selfinteracting cold dark matter,''
Phys. Rev. Lett. \textbf{84}, 3760-3763 (2000)
[arXiv:astro-ph/9909386 [astro-ph]].

\bibitem{Tulin:2013teo}
S.~Tulin, H.~B.~Yu and K.~M.~Zurek,
``Beyond Collisionless Dark Matter: Particle Physics Dynamics for Dark Matter Halo Structure,''
Phys. Rev. D \textbf{87}, no.11, 115007 (2013)
[arXiv:1302.3898 [hep-ph]].

\bibitem{Sakurai:1994}
J. J. Sakurai, Modern Quantum Mechanics (Addison-Wesley, Reading, MA, 1994).

\bibitem{Kahlhoefer:2013dca}
F.~Kahlhoefer, K.~Schmidt-Hoberg, M.~T.~Frandsen and S.~Sarkar,
``Colliding clusters and dark matter self-interactions,''
Mon. Not. Roy. Astron. Soc. \textbf{437}, no.3, 2865-2881 (2014)
[arXiv:1308.3419 [astro-ph.CO]].

\bibitem{Colquhoun:2020adl}
B.~Colquhoun, S.~Heeba, F.~Kahlhoefer, L.~Sagunski and S.~Tulin,
``Semiclassical regime for dark matter self-interactions,''
Phys. Rev. D \textbf{103}, no.3, 035006 (2021)
[arXiv:2011.04679 [hep-ph]].


\bibitem{Correa:2020qam}
C.~A.~Correa,
``Constraining velocity-dependent self-interacting dark matter with the Milky Way{\textquoteright}s dwarf spheroidal galaxies,''
Mon. Not. Roy. Astron. Soc. \textbf{503}, no.1, 920-937 (2021)
[arXiv:2007.02958 [astro-ph.GA]].

\bibitem{Turner:2020vlf}
H.~C.~Turner, M.~R.~Lovell, J.~Zavala and M.~Vogelsberger,
``The onset of gravothermal core collapse in velocity-dependent self-interacting dark matter subhaloes,''
Mon. Not. Roy. Astron. Soc. \textbf{505}, no.4, 5327-5339 (2021)
[arXiv:2010.02924 [astro-ph.GA]].

\bibitem{Dave:2000ar}
R.~Dave, D.~N.~Spergel, P.~J.~Steinhardt and B.~D.~Wandelt,
``Halo properties in cosmological simulations of selfinteracting cold dark matter,''
Astrophys. J. \textbf{547}, 574-589 (2001)
[arXiv:astro-ph/0006218 [astro-ph]].

\bibitem{Sagunski:2020spe}
L.~Sagunski, S.~Gad-Nasr, B.~Colquhoun, A.~Robertson and S.~Tulin,
``Velocity-dependent Self-interacting Dark Matter from Groups and Clusters of Galaxies,''
JCAP \textbf{01}, 024 (2021)
[arXiv:2006.12515 [astro-ph.CO]].

\bibitem{Elbert:2014bma}
O.~D.~Elbert, J.~S.~Bullock, S.~Garrison-Kimmel, M.~Rocha, J.~O\~norbe and A.~H.~G.~Peter,
``Core formation in dwarf haloes with self-interacting dark matter: no fine-tuning necessary,''
Mon. Not. Roy. Astron. Soc. \textbf{453}, no.1, 29-37 (2015)
[arXiv:1412.1477 [astro-ph.GA]].

\bibitem{Ren:2018jpt}
T.~Ren, A.~Kwa, M.~Kaplinghat and H.~B.~Yu,
``Reconciling the Diversity and Uniformity of Galactic Rotation Curves with Self-Interacting Dark Matter,''
Phys. Rev. X \textbf{9}, no.3, 031020 (2019)
[arXiv:1808.05695 [astro-ph.GA]].

\bibitem{Harvey:2015hha}
D.~Harvey, R.~Massey, T.~Kitching, A.~Taylor and E.~Tittley,
``The non-gravitational interactions of dark matter in colliding galaxy clusters,''
Science \textbf{347}, 1462-1465 (2015)
[arXiv:1503.07675 [astro-ph.CO]].

\bibitem{Bradac:2008eu}
M.~Bradac, S.~W.~Allen, T.~Treu, H.~Ebeling, R.~Massey, R.~G.~Morris, A.~von der Linden and D.~Applegate,
``Revealing the properties of dark matter in the merging cluster MACSJ0025.4-1222,''
Astrophys. J. \textbf{687}, 959 (2008)
[arXiv:0806.2320 [astro-ph]].

\bibitem{Randall:2008ppe}
S.~W.~Randall, M.~Markevitch, D.~Clowe, A.~H.~Gonzalez and M.~Bradac,
``Constraints on the Self-Interaction Cross-Section of Dark Matter from Numerical Simulations of the Merging Galaxy Cluster 1E 0657-56,''
Astrophys. J. \textbf{679}, 1173-1180 (2008)
[arXiv:0704.0261 [astro-ph]].

\bibitem{Sieber:2021fue}
H.~Sieber, D.~Banerjee, P.~Crivelli, E.~Depero, S.~N.~Gninenko, D.~V.~Kirpichnikov, M.~M.~Kirsanov, V.~Poliakov and L.~Molina Bueno,
``Prospects in the search for a new light Z' boson with the NA64{\ensuremath{\mu}} experiment at the CERN SPS,''
Phys. Rev. D \textbf{105}, no.5, 052006 (2022)
[arXiv:2110.15111 [hep-ex]].

\bibitem{Kahn:2018cqs}
Y.~Kahn, G.~Krnjaic, N.~Tran and A.~Whitbeck,
``M$^{3}$: a new muon missing momentum experiment to probe (g {\ensuremath{-}} 2)$_{\mu}$ and dark matter at Fermilab,''
JHEP \textbf{09}, 153 (2018)
[arXiv:1804.03144 [hep-ph]].

\bibitem{Cirelli:2013ufw}
M.~Cirelli, E.~Del Nobile and P.~Panci,
``Tools for model-independent bounds in direct dark matter searches,''
JCAP \textbf{10}, 019 (2013)
[arXiv:1307.5955 [hep-ph]].

\bibitem{Graham:2010ca}
P.~W.~Graham, R.~Harnik, S.~Rajendran and P.~Saraswat,
``Exothermic Dark Matter,''
Phys. Rev. D \textbf{82}, 063512 (2010)
[arXiv:1004.0937 [hep-ph]].

\bibitem{Li:2022acp}
J.~Li, L.~Su, L.~Wu and B.~Zhu,
``Spin-dependent sub-GeV inelastic dark matter-electron scattering and Migdal effect. Part I. Velocity independent operator,''
JCAP \textbf{04}, 020 (2023)
[arXiv:2210.15474 [hep-ph]].

\bibitem{Potekhin:2013qqa}
A.~Y.~Potekhin, A.~F.~Fantina, N.~Chamel, J.~M.~Pearson and S.~Goriely,
``Analytical representations of unified equations of state for neutron-star matter,''
Astron. Astrophys. \textbf{560}, A48 (2013)
[arXiv:1310.0049 [astro-ph.SR]].

\bibitem{Goriely:2010bm}
S.~Goriely, N.~Chamel and J.~M.~Pearson,
``Further explorations of Skyrme-Hartree-Fock-Bogoliubov mass formulas. XII: Stiffness and stability of neutron-star matter,''
Phys. Rev. C \textbf{82}, 035804 (2010)
[arXiv:1009.3840 [nucl-th]].

\bibitem{Zhang:2020wov}
N.~B.~Zhang and B.~A.~Li,
``Constraints on the muon fraction and density profile in neutron stars,''
Astrophys. J. \textbf{893}, 61 (2020)
[arXiv:2002.06446 [astro-ph.HE]].

\bibitem{Garani:2018kkd}
R.~Garani, Y.~Genolini and T.~Hambye,
``New Analysis of Neutron Star Constraints on Asymmetric Dark Matter,''
JCAP \textbf{05}, 035 (2019)
[arXiv:1812.08773 [hep-ph]].

\bibitem{Feng:2015hja}
J.~L.~Feng, J.~Smolinsky and P.~Tanedo,
``Dark Photons from the Center of the Earth: Smoking-Gun Signals of Dark Matter,''
Phys. Rev. D \textbf{93}, no.1, 015014 (2016)
[erratum: Phys. Rev. D \textbf{96}, no.9, 099901 (2017)]
[arXiv:1509.07525 [hep-ph]].

\bibitem{Blennow:2015hzp}
M.~Blennow, S.~Clementz and J.~Herrero-Garcia,
``Pinning down inelastic dark matter in the Sun and in direct detection,''
JCAP \textbf{04}, 004 (2016)
[arXiv:1512.03317 [hep-ph]].

\bibitem{Lundberg:2004dn}
J.~Lundberg and J.~Edsjo,
``WIMP diffusion in the solar system including solar depletion and its effect on earth capture rates,''
Phys. Rev. D \textbf{69}, 123505 (2004)
[arXiv:astro-ph/0401113 [astro-ph]].

\bibitem{Garani:2019fpa}
R.~Garani and J.~Heeck,
``Dark matter interactions with muons in neutron stars,''
Phys. Rev. D \textbf{100}, no.3, 035039 (2019)
[arXiv:1906.10145 [hep-ph]].

\bibitem{Binder:2016pnr}
T.~Binder, L.~Covi, A.~Kamada, H.~Murayama, T.~Takahashi and N.~Yoshida,
``Matter Power Spectrum in Hidden Neutrino Interacting Dark Matter Models: A Closer Look at the Collision Term,''
JCAP \textbf{11}, 043 (2016)
[arXiv:1602.07624 [hep-ph]].
 
\bibitem{Navarro:1996gj} 
  J.~F.~Navarro, C.~S.~Frenk and S.~D.~M.~White,
  ``A Universal density profile from hierarchical clustering,''
  Astrophys.\ J.\  {\bf 490}, 493 (1997)
  [astro-ph/9611107].

\bibitem{Navarro:1995iw} 
  J.~F.~Navarro, C.~S.~Frenk and S.~D.~M.~White,
  ``The Structure of cold dark matter halos,''
  Astrophys.\ J.\  {\bf 462}, 563 (1996)
  [astro-ph/9508025].
  
\bibitem{Elor:2015tva}
G.~Elor, N.~L.~Rodd and T.~R.~Slatyer,
``Multistep cascade annihilations of dark matter and the Galactic Center excess,''
Phys. Rev. D \textbf{91}, 103531 (2015)
[arXiv:1503.01773 [hep-ph]].

\bibitem{Yang:2018fje}
K.~C.~Yang,
``Hidden Higgs portal vector dark matter for the Galactic center gamma-ray excess from the two-step cascade annihilation, and muon g {\ensuremath{-}} 2,''
JHEP \textbf{08}, 099 (2018)
[arXiv:1806.05663 [hep-ph]].

\bibitem{Yang:2020vxl}
K.~C.~Yang,
``A potentially detectable gamma-ray line in the Fermi Galactic center excess {\textemdash} in light of one-step cascade annihilations of secluded (vector) dark matter via the Higgs portal,''
JHEP \textbf{07}, 148 (2020)
[arXiv:2001.04946 [hep-ph]].

\bibitem{Maltoni:2004ei}
M.~Maltoni, T.~Schwetz, M.~A.~Tortola and J.~W.~F.~Valle,
``Status of global fits to neutrino oscillations,''
New J. Phys. \textbf{6}, 122 (2004)
[arXiv:hep-ph/0405172 [hep-ph]].

\bibitem{Palomares-Ruiz:2007trf}
S.~Palomares-Ruiz and S.~Pascoli,
``Testing MeV dark matter with neutrino detectors,''
Phys. Rev. D \textbf{77}, 025025 (2008)
[arXiv:0710.5420 [astro-ph]].
 
\bibitem{Super-Kamiokande:2005mbp}
Y.~Ashie \textit{et al.} [Super-Kamiokande],
``A Measurement of atmospheric neutrino oscillation parameters by SUPER-KAMIOKANDE I,''
Phys. Rev. D \textbf{71}, 112005 (2005)
[arXiv:hep-ex/0501064 [hep-ex]].

  
\end{thebibliography}\endgroup


\end{document}